\documentclass[11pt]{article}
\usepackage{amsthm}
\usepackage{amsmath}
\usepackage[round]{natbib}
\usepackage[colorlinks,citecolor=blue,urlcolor=blue,filecolor=blue,backref=page]{hyperref}
\usepackage{graphicx}

\usepackage{array,epsfig,enumerate,dsfont,caption,afterpage}
\usepackage{amssymb,latexsym}
\usepackage{amsfonts}
\usepackage{comment}
\usepackage{subcaption}
\usepackage{pdflscape}
\usepackage[mathscr]{euscript}

\newcommand{\bfX}{{\bf X}}
\newcommand{\bfZ}{{\bf Z}}
\newcommand{\bfx}{{\bf x}}
\newcommand{\bfy}{{\bf y}}
\newcommand{\bl}{{\bf l}}
\newcommand{\bfz}{{\bf z}}
\newcommand{\bfP}{{\bf P}}

\newcommand{\bbeta}{\boldsymbol{\beta}}
\newcommand{\Bgamma}{\boldsymbol{\beta}_{\bgm}}
\newcommand{\bmu}{\boldsymbol{\mu}}
\newcommand{\bgm}{\boldsymbol{\gamma}}
\newcommand{\bgmo}{\boldsymbol{\gamma}_{0}}
\newcommand{\bgms}{\boldsymbol{\gamma}^{\star}}
\newcommand{\bgmp}{\boldsymbol{\gamma}^{\prime}}
\newcommand{\bvep}{\boldsymbol{\varepsilon}}

\newcommand{\Sg}{\boldsymbol{\Sigma}}
\newcommand{\diag}{\mathrm{Diag}}

\newcommand{\mA}{\mathcal{A}}
\newcommand{\mI}{\mathcal{I}}
\newcommand{\mM}{\mathcal{M}}
\newcommand{\mG}{\mathcal{G}}
\newcommand{\mC}{\mathcal{C}}
\newcommand{\mE}{\mathcal{E}}
\newcommand{\mQ}{\mathcal{Q}}
\newcommand{\mT}{\mathcal{T}}
\newcommand{\mR}{\mathcal{R}}
\newcommand{\Tggp}{\mT_{\bgm, \bgmp}}

\newcommand{\bXgm}{\bfX_{\bgm}}

\newcommand{\pg}{p_{\bgm}}
\newcommand{\I}{{\bf I}}

\newcommand{\Pgm}{{\bf P_{\bgm}}}
\newcommand{\Pgmp}{{\bf P_{\bgmp}}}

\newcommand{\Pgmo}{{\bf P_{\bgm_{0}}}}
\newcommand{\Pgmco}{{\bf P_{\bgm\cup\bgm_{0}}}}

\newcommand{\kn}{\kappa_{n}}

\newcommand{\gn}{g_{n}}
\newcommand{\ps}{p_{n}^{\star}}
\newcommand{\bfQ}{{\bf Q}}

\newcommand{\argmax}{\mathrm{argmax}}
\newcommand{\argmin}{\mathrm{argmin}}
\newcommand{\var}{\mathrm{var}}

\newtheorem{result}{Result}
\newtheorem{theorem}{Theorem}
\newtheorem{lemma}{Lemma}
\newtheorem{remark}{Remark}

\title{Bayesian Variable Selection for High- dimensional Settings with Grouped Covariates}
\author{Pranay Agarwal \and Subhajit Dutta \and Minerva Mukhopadhyay}
\date{}
\begin{document}

\maketitle







\begin{abstract}
Consider the normal linear regression setup when the number of covariates $p$ is much larger than the sample size $n$, and the covariates form correlated groups. The response variable $y$ is not related to an entire group of covariates in \emph{all or none} basis, rather the sparsity assumption persists within and between groups.
We extend the traditional $g$-prior setup to this framework.
Variable selection consistency of the proposed method is shown under fairly general conditions, assuming the covariates to be random and allowing the true model to grow with both $n$ and $p$.
For the purpose of implementation of the proposed $g$-prior method to high-dimensional setup, we propose two procedures.
First, a group screening procedure, termed as \emph{group SIS (GSIS)}, and secondly, a novel stochastic search variable selection (SSVS) algorithm, termed as \emph{group informed variable selection algorithm (GiVSA)}, which uses the known group structure efficiently to explore the model space without discarding any covariate based on an initial screening.
The screening consistency of GSIS, and the theoretical mixing time of GiVSA are studied using the canonical path ensemble approach of \cite{YWJ_2016}.
Performance of the proposed prior setup with implementation of GSIS, as well as GiVSA
are validated using various simulated data sets and the residential building data.

\end{abstract}

%


\section{Introduction}\label{sec:Intro}
Consider the problem of variable selection in normal linear regression setup when number of predictors, $p ~(= p_{n})$, increases with the sample size $n$. Let $y$ denote the response variable and $X=\{x_{1}, \ldots, x_{p}\}$ be a set of predictors.
In a normal linear regression model, the conditional distribution of $y$ given $X$ is assumed to be normal with a linear combination of $X$ as expectation and a constant variance, say $\sigma^{2}$.
The problem of variable selection is to find the sparsest subset of $X$ which provides the best fit.

A wide variety of methods are available for variable selection in the high-dimensional normal linear regression setup both in the frequentist as well as Bayesian literature. Frequentist methods can be split into two broad categories, the feature screening methods, which are used as a pre-processing step to reduce the dimension of the design matrix, and penalized least squares methods, where the regression coefficients are shrunk towards zero by imposing penalty terms. The former includes methods like sure independence screening \citep[SIS]{FL_2008}, forward selection \citep{Wang_2009}, iteratively thresholded ridge regression screener \citep[ITRRS]{FL_2008}; while the latter includes least absolute shrinkage and selection operator \citep[LASSO]{Tibshirani_1996} and it's variants like the adaptive LASSO \citep{Zou_2006}, fused LASSO \citep{TSRZK_2005}, clustered LASSO \citep{She_2010}, elastic net \citep{ZH_2005}, smoothly clipped absolute deviation \citep[SCAD]{Xie_2009}, Dantzig selector \citep{CT_2007}. For a comprehensive overview of frequentist advancements of variable selection, the readers are referred to the books by \cite{BV_2011} and \cite{FLZZ_2020}.

There is a vast and elegant literature on Bayesian variable selection in normal linear setup as well. Various paths for variable selection include, but are not limited to, variable selection using shrinkage priors (see \cite{Zellner1986,IR2005}), empirical Bayes variable selection \citep{GF_2000,YL_2005}, Bayesian LASSO (\cite{PC_2008,RG_2018}), variable selection using credible region \citep{BR_2012}, Bayesian model averaging \citep{RMH_1997,FGS_2018}. For a comprehensive review of Bayesian variable selection methods, see \cite{OS_2009}. For more recent developments in Bayesian variable selection, see \cite{TV_2021} and the references therein.

From the perspective of implementation, while frequentist methods rely on screening and/or fast optimization, most of the Bayesian methods rely on stochastic search based on reversible jump Markov chain Monte Carlo (RJMCMC, see \cite{GM1993}). RJMCMC explores the model space by scanning the covariates sequentially, and therefore, takes a longer time in the high-dimensional scenario. Several attempts have been made to reduce the time complexity in Bayesian variable selection methods in various ways, for example, by introducing importance sampling \citep{ZR_2019} or screening \citep{SBJ_2018} in stochastic search, or by moving towards deterministic approaches deviating from stochastic search \citep{RG_2014}.

Further, a common limitation of both RJMCMC and screening based approaches is, these methods loose efficiency under the presence of multicollinearity. Various modified screening algorithms have been proposed to mitigate this issue (see, e.g., \cite{WL_2016,LDR_2022}). Assuming a given group structure, a list of frequentist (see \cite{HBM_2012} for a review) and Bayesian (see, e.g., \cite{ NJG_2020, YN_2020, BMACB_2022}) methods have been proposed to select important groups.
However, these group selection methods are sparse at the group level, but not at the individual level. Pertaining to sparsity at the group as well as individual levels, many bi-level selection methods have also been proposed in both the frequentist (see, e.g., \cite{SFHT_2013,Breheny_2015}) and the Bayesian (see, e.g., \cite{MY_2017,LMPS_2017, CDMPLY_2020}) literature.

In this article, we start by extending the traditional $g$-prior setup \citep{Zellner1986} to the high-dimensional scenario, by appropriately modifying the model prior and the $g$-prior on the regression coefficients (Section \ref{sec:2}). Under a reasonable set of assumptions, variable selection consistency of the proposed prior structure is shown when the covariates are stochastic (Section \ref{sec:3}). Particularly, we show that the posterior probability of the true model converges in probability to {\it one}, considering the joint probability distribution of the response and the predictors.

As discussed above, in the high-dimensional scenario the covariates tend to form correlated groups. Moreover, the group structure is often known as domain knowledge from previous studies. For example, genetic variables are usually divided into groups with same functionality. However, existence of groups, in general, does not imply that all the covariates in a particular group would either be associated, or disassociated with the response. Consequently, in search of differentially expressed genes, information on the group structures is often ignored.
Nevertheless, the information of group structure can expedite the search of the underlying true model to a great extent. One of the main objectives of this work is to establish this gain with proper mathematical justification.

Towards this goal, we first propose a group probability vector in Section \ref{sec:GSIS}. The properties of this group probability vector are investigated.
Denoting the groups containing at least one active covariate as `active' and the remaining groups as `inactive', it is shown that the ratio of the proposed group probability of an inactive group w.r.t. the active group converges to {\it zero} in probability, uniformly over the set of active and inactive groups. This property opens up the possibility of using this group-probabilities for group level screening. We term this group screening method as \emph{group SIS (GSIS)}.

We next turn towards improving the stochastic search variable selection algorithms with the help of group information. Suppose that the group structure and a list of group importance probabilities are given.
Then,
the stochastic search for covariates (consequently, models) can be performed via the groups. Towards this, we incorporate the group probabilities in the RJMCMC algorithm in such a way that the long search for a new covariate, while adding or replacing a covariate in an existing model can be reduced efficiently. Further, we show that efficiency of the group probabilities can be directly translated to efficiency of the RJMCMC~algorithm, without discarding any covariate by an initial screening.

The proposed group informed RJMCMC algorithm for variable selection, denoted by \emph{GiVSA}, is described Section \ref{sec:GiVSA}. A detailed analysis of mixing time of the traditional RJMCMC algorithm for variable selection based on $g$-prior is performed by \cite{YWJ_2016}. Following their analysis we show the gain in mixing time due to GiVSA in order of $n$.

Finally, we validate the performances of the proposed prior setup, the GSIS screening, as well as the GiVSA, in comparison with a variety of relevant competitors, with the help of ample simulation experiments and the residential building data set in Section \ref{sec:NumAn}.
The scope and limitations of the proposed methods, and the future directions are discussed in Section \ref{sec:Discuss}.
The proofs of all the theorems are provided in the appendix (Section \ref{sec:app}), and the proofs of all the lemmas are provided in the Supplementary material.

\subsection{Notations} \label{sec:notations}
Before proceeding further, we list some notations and conventions which are used throughout the paper.
\vspace{-.15 in}
\begin{enumerate}
\itemsep-0.5em
\item Let $a$ and $b$ be two real numbers, then $a\wedge b=\max\{a,b\}$ and $a\vee b= \min\{a,b\}$.
\item Let $\{p_{n}\}$ and $\{n\}$ be sequences of real numbers, then $p_{n} \gg n$ implies $n/p_{n} \rightarrow 0$.
\item Let $\{a_{n}\}$ and $\{b_{n}\}$ be sequences of real numbers, then $a_{n} \lesssim b_{n}$, equivalently, $b_{n}\gtrsim a_{n}$, implies that $a_{n}\leq c b_{n}$ for all sufficiently large $n$, where $c>0$. Further, $a_{n} \sim b_{n}$ implies that both $a_{n} \lesssim b_{n}$ and $b_{n} \lesssim a_{n}$ hold.
\item Let ${\bf A}$ be a square matrix, then $\lambda_{\min}({\bf A})$ and $\lambda_{\max}({\bf A})$ are the minimum and maximum eigenvalues of ${\bf A}$, respectively. Let ${\bf A}$ be any matrix, then $\sigma_{\min}({\bf A})$ and $\sigma_{\max}({\bf A})$ are the minimum and maximum singular values of ${\bf A}$.
\item For any square matrix ${\bf A}$, ${\bf A}^{+}$ denotes the Moore Penrose inverse of ${\bf A}$.
\item Let $\bfX$ be the $n\times p$ design matrix and  $\bgm$ be a non-null subset of $\{1, \ldots, p\}$, then $\bfX_{\bgm}$ is the submatrix of $\bfX$ considering the columns corresponding to $\bgm$. Further, let $\bbeta$ be the vector of regression coefficients, then $\bbeta_{\bgm}$ is the sub-vector of $\bbeta$ consisting of the columns corresponding to $\bgm$.
\item Let $\bgm$ be a (non-null) subset of $\{1,\ldots,p\}$ and $\bmu = \bfX\bbeta$. Then, $\bmu_{\bgm}=\bfX_{\bgm}\bbeta_{\bgm}$.
\end{enumerate}

\section{The Normal Linear Model and Prior Settings} \label{sec:2}

In a normal linear regression model, the response variable $y$ and the set of covariates $\{x_{1}, \ldots, x_{p}\}$ are related as

 {\centering
$ y=\beta_{1}x_{1}+\cdots+\beta_{p} x_{p} +\varepsilon; \qquad \varepsilon\sim N (0, \sigma^{2})$.
 \par}

\noindent Each subset of $\{x_{1},\ldots,x_{p}\}$ leads to a separate model, and
with $p$ covariates, there are at most $2^{p}$ possible models. The space of all models is denoted by $\mM$, which has a one-one relation with all subsets of $\{1,\ldots,p\}$, say $\boldsymbol{\Gamma}$. A typical component of $\mM$, is denoted by $M_{\bgm}$, where $\bgm$ is the corresponding element of $\boldsymbol{\Gamma}$. Under $M_{\bgm}$, we have

\vspace{-.15 in}
\begin{equation}
    y= \sum_{j \in \bgm} \beta_{j} x_{j}+ \varepsilon= \bfx_{\bgm}^{\prime}\bbeta_{\bgm}+\varepsilon. \label{eqn:model}
\end{equation}

\vspace{-.1 in}
There are $n$ independent data points available $(\bfx_{1},y_{1}),\ldots, (\bfx_{n}, y_{n})$ with $p\gg n$. The vector of $n$ observations of the response, and that of the $j$-th predictor are denoted by $\bfy$ and $\bfx_{(j)}$, respectively, for $j=1,\ldots,p$. The $n\times p$ design matrix is denoted by $\bfX$, and that for the subsets of the covariates, corresponding to $\bgm$, is denoted by $\bXgm$.

\vspace{-.1 in}
\paragraph{\bf Prior Setup.} In a Bayesian model, each parameter is associated with a prior quatifying the amount of uncertainty associated with different sets of possible values of the parameter. In Zellner's $g$-prior setup \citep{Zellner1986}, a zero mean Gaussian prior with covariance matrix proportional to $\textstyle\left( \bXgm^{\prime} \bXgm\right)^{-1}$ is assigned on the regression parameter. This can be generalized to the high-dimensional situation (i.e., $p\gg n$) as follows
\vspace{-0.1in}
\begin{eqnarray}
   \bbeta_{\bgm} \mid \bfX,\sigma^{2} \sim N_{|\bgm|} \left( {\bf 0} , \gn \sigma^{2} \left( \bXgm^{\prime} \bXgm\right)^{+} \right) \qquad \mbox{and} \quad \pi(\sigma^{2})\propto \frac{1}{\sigma^{2}}, \label{eqn:priorsetup}
\end{eqnarray}

\vspace{-0.1in}
\noindent where $\gn>0$ is a shrinkage parameter depending on $n$, and $\pi(\sigma^{2})$ is the prior density of $\sigma^2$.

Apart from the priors on the parameters, in a variable selection problem, one also assigns a prior distribution on the class of models $\mM$. In the high-dimensional setup, sparsity is commonly assumed, as among a large pool of covariates
often only a few contribute towards the data generating process.
In such situations, the model prior distribution plays a crucial role in selecting the true model.

The Bernoulli class of priors $P(M_{\bgm})=\theta^{|\bgm|}(1-\theta)^{p-|\bgm|}$ for $\theta\in (0,1)$, suggested by \cite{GM1993}, assigns an independent $\mathrm{Bernoulli}(\theta)$ prior to each covariate.
However, it was observed by \cite{MGC2015} that when $p$ grows with $n$, then the hierarchical uniform~prior

{\centering $\displaystyle\int_{0}^{1} \theta^{|\bgm|}(1-\theta)^{p-|\bgm|} d\theta =(p+1)^{-1} \binom{p}{|\bgm|}^{-1}$
\par}

\noindent outperforms the Bernoulli class of priors for any choice of fixed $\theta\in (0,1)$.

Observe that the hierarchical uniform prior provides equal prior mass on the class of models of each dimension. Let $\mM_{|\bgm|}$ be the class of models of dimension $|\bgm|$, then the total prior mass on $\mM_{|\bgm|}$ is $1/(p+1)$ irrespective of the sparsity of $\mM_{|\bgm|}$. In a sparse setup, however, it is additionally necessary to assign a higher prior mass on models having lower dimensions. Keeping this fact in mind, we choose the following prior
\vspace{-0.05in}
\begin{equation}
P\left( M_{\bgm}  \right) \propto {k_{n}^{-|\bgm|}}\binom{p}{|\bgm|}^{-1}, \label{eqn:modelprior}
\end{equation}

\vspace{-0.05in}
\noindent where $k_{n}\rightarrow \infty$ as $n \rightarrow\infty$.
After this modification, observe that the total prior probability received by the class of models $\mM_{|\bgm|}$ is inversely proportional to $|\bgm|$.

\section{Variable Selection Consistency of the Proposed Prior}\label{sec:3}

Under the normal linear regression model and associated priors, one can calculate the posterior probability of the model $M_{\bgm}$ given $\bfy, \bfX$, as follows
\begin{eqnarray*}
P(M_{\bgm}\mid \bfX,\bfy ) = \frac{ P(M_{\bgm}) m_{\bgm}(\bfy\mid \bfX)  }{\sum_{\bgm^{\prime}} P(M_{\bgmp} ) m_{\bgmp}(\bfy\mid \bfX) },
\end{eqnarray*}

\vspace{-0.1in}
\noindent where
\vspace{-0.1in}

{\centering
$m_{\bgm}(\bfy \mid \bfX) = \int f(\bfy \mid \bfX, \bgm, \bbeta, \sigma^{2} ) \pi(\bbeta|M_{\bgm}, \bfX, \sigma^{2}) \pi\left(\sigma^2\right) d\bbeta d\sigma^{2}$
\par}

\noindent is the integrated likelihood of $\bfy$ given $\bfX$ and $\bgm$. Here, $\pi(\bbeta|M_{\bgm}, \bfX, \sigma^{2})$ denotes the prior density of $\bbeta_{\bgm}$ given $\bfX $, $\sigma^{2}$, and $\pi(\sigma^2)$ is the prior on $\sigma^2$.

For the set of priors in (\ref{eqn:priorsetup}) and the data $({\bf y},\bfX)$, the integrated likelihood of a model $M_{\bgm}$
can be expressed as
\vspace{-0.1in}
$$ m_{\bgm}(\bfy \mid \bfX)  \propto (1+g_{n})^{-\rho_{\bgm}/2} \times \left\{ \left( 1- \frac{g_{n}}{1+g_{n}} \right) \bfy^{\prime} \bfy + \frac{g_{n}}{1+g_{n}} \bfy^{\prime} \left( \I - \Pgm\right) \bfy\right\}^{-n/2},$$
where $\rho_{\bgm}$ is the rank of $ \bXgm$ and $ \bfP_{\bgm}$ is the orthogonal projection matrix on to the column space of $\bfX_{\bgm}$ (see Result \ref{res:marginal} in Section \ref{sec:ML} of the Supplementary material for details).

Variable selection consistency (VSC) is a frequentist notion of optimality. Assuming the existence of a unique true model, one verifies if the posterior probability of the true model converges to one in an appropriate sense.  Let $\bgm_{0}$ denote the subset of $\{1,\ldots,p\}$ corresponding to the true model, $M_{\bgm_{0}}$. Then, VSC deals with showing
\vspace{-.1 in}
\begin{equation}
P(M_{\bgm_{0}} \mid \bfX, \bfy) \xrightarrow{p} 1 \quad \text{as}\quad n\rightarrow\infty, \label{eqn:VSC1}
\end{equation}

\vspace{-.1 in}
\noindent where the probability is with respect to the randomness of both $\bfy$ and $\bfX$.

\subsection{Setup and Assumptions}

Assume that $p$ is growing with $n$ is polynomial order (i.e., $p\lesssim n^{t}$ with $t>0$). Next, consider the following assumptions:

\vspace{-.15 in}
\begin{enumerate}[({A}1)]
\itemsep-0.5pt
    \item There exists a true class of covariates, the index set of which, denoted by $\bgm_{0}$, is non-random and has cardinality $|\bgm_{0}|\sim n^{b}$ for some $b$ satisfying $0<b<1/3$.

\item Let $\bmu= E(\bfy \mid \bfX) = \bfX_{\bgm_{0}} \bbeta_{\bgm_{0}}$. There exists constants $M_{0}>0$ and $c_{0}>0$ such that $P\left[(n |\bgm_{0}|)^{-1}\bmu^{\prime} \bmu < M_{0} \right] > 1- \exp\{-c_{0}n\}$ for sufficiently large $n$.

\item Let $\mM_{2}^{\prime}$ be the class of non-supermodels of the true model $\bgm_{0}$ of dimension $|\bgm| \lesssim n^{\xi}$, where $\xi=1-b-\eta$, $0<\eta<1/4$, i.e.,
$\mM_{2}^{\prime}= \left\{ \bgm :   |\bgm|= \pg\lesssim n^{\xi} ,~ \bgmo \nsubseteq \bgm \right\}$.
Then, there exist constants $\delta>0$ and $c_{1}>0$ satisfying

{\centering
$P\left[  \inf_{\bgm\in \mM_{2}^{\prime}}n^{-1}\bmu^{\prime} (\I - \Pgm) \bmu >  \delta\right] >1 - \exp\{ -n c_{1} \}. $
\par}
\end{enumerate}
Assumption (A1) allows the true model to grow with $n$ at a moderate rate.
Assumption (A2) provides a probabilistic upper bound to the conditional expectation of ${\bf y}$ given ${\bf X}$. Assumption (A3) provides a probabilistic lowerbound to the Kullback-Leiber divergence of the true model and any wrong model of small dimension. Assumptions (A2) and (A3) are basically conditions on the random design matrix $\bfX$ and satisfied by common class of distributions like the sub-Gaussian family of distributions (see, e.g., \cite{Vershynin_notes}). The following lemma justifies this claim.

\begin{lemma} \label{lm:1}
Let the distribution of the covariates, $\bfX$, be scaled sub-Gaussian with mean zero and scale matrix $\Sg$ and finite sub-Gaussian parameter $\kappa$. Assume that the submatrix of $\Sg$ corresponding to the true covariates $\bgmo$, say $\Sg_{\bgmo}$, satisfies $0< \tau_{\min} \leq \lambda_{\min} (\Sg_{\bgmo}) \leq \lambda_{\max} (\Sg_{\bgmo}) \leq  \tau_{\max}< \infty$, and the vector of true regression coefficients $\bbeta_{\bgm_{0}}=(\beta_{\bgm_{0},1}, \ldots, \beta_{\bgm_{0},|\bgmo|})^{\prime}$ satisfies $m\leq |\beta_{\bgm_{0},j}|<M$ for some $m,M>0$ and each $j=1,\ldots,|\bgmo|$.
Then, assumptions (A2) and (A3) are satisfied.
\end{lemma}
Proof of Lemma \ref{lm:1} is in the Appendix (see Section \ref{sec:Proof_lm1} of the Supplementary material).


\begin{remark}\label{rm:3}
The above lemma provides an insight into assumptions (A2) and (A3). First, it demonstrates that existence of multicollinearity in the population variance covariance matrix does not affect the assumptions (and consequently, VSC) as long as the true covariates are not highly correlated. The eigen restriction in Lemma \ref{lm:1} also indicates that the true set does not include redundant covariates.
\end{remark}

\subsection{Main Result on Variable Selection Consistency}\label{sec:MR}

Let us split the total class of models into three exhaustive sets:

(a) Supermodels:
\quad $\mM_{1}=\left\{ \bgm : ~\mbox{and } {\bgm_{0}} \subset \bgm ; ~ |\bgm| \lesssim \pg^{\star}= o(n / \log n) \right\}$.

(b) Non-supermodels:
\quad $\mM_{2}=\left\{ \bgm : ~\mbox{and } {\bgm_{0}} \nsubseteq \bgm ; ~ |\bgm| \lesssim \pg^{\star}= o(n / \log n) \right\}$.

(c) Large models: $\mM_{3}=\left\{ \bgm : \bgm \notin \mM_{1} \cup \mM_{2}  \right\}$.

We assume prior knowledge on the existence of sparsity in the true model. As a result, the large models in $\mM_{3}$ (with cardinality bigger than $\pg^{\star}$) are assigned zero model prior, consequently, are discarded. Therefore, VSC is shown over the remaining two classes only. 

\begin{theorem}\label{thm:VSC}
(Variable Selection Consistency) Consider the $g$-prior setup in the normal linear regression framework as described in (\ref{eqn:priorsetup}) with $p\sim n^{t}$ (for some $t>0$), and the modified hierarchical uniform model prior distribution restricted to $\mM_{1}\cup\mM_{2}$ as described in (\ref{eqn:modelprior}). Suppose there exist a true model indexed by $\bgm_{0}$
satisfying (A1), which generates $\bfy$.
Then, under assumptions (A2) and (A3), and for any choice of increasing sequences $g_{n}$ satisfying $|\bgmo|=o(g_{n})$, and $k_{n}$ such that
(i) $\log \left( k_{n}^{2} g_{n} \right)  =o(n^{1-b})$ and (ii) $k_{n}^{2} g_{n}\gtrsim  p^{2(1+\lambda)} $ hold with $\lambda> b/t$,
 the VSC, as described in (\ref{eqn:VSC1}), is achieved under the set of models $\mM_{1}$ or $\mM_{2}$.
\end{theorem}
A proof of Theorem \ref{thm:VSC} is provided in the Appendix (see Section \ref{sec:Proof_Thm1}).

\begin{remark}
\label{rm:VSC}
The result in Theorem \ref{thm:VSC} holds even if the covariates are non-stochastic. Non-stochastic analogs of (A2) and (A3), denoted by (A2$^{\prime}$) and (A3$^{\prime}$), respectively, are

(A2$^{\prime}$) $(n|\bgmo|)^{-1}\|\bmu\|^{2} =O(1)$, and

(A3$^{\prime}$) $\inf_{\bgm\in \mM_{2}^{\prime}}n^{-1}\bmu^{\prime} (\I - \Pgm) \bmu > \delta$ for some $\delta>0$.

\noindent The statement of Theorem \ref{thm:VSC}, with (A2)-(A3) replaced by (A2$^{\prime}$)-(A3$^{\prime}$), holds when the covariates are non-stochastic. The proof is similar to that of Theorem \ref{thm:VSC}, hence omitted.
\end{remark}
Theorem \ref{thm:VSC} is not particularly designed for grouped covariates. It includes both grouped and ungrouped situations, as long as the true covariates are not nearly collinear (see the conditions of Lemma \ref{lm:1}). Further, the group information is not also relevant to the VSC unless additional assumptions are made on the distribution of true covariates in the groups.
However, the group information is useful in fast implementation of Theorem \ref{thm:VSC}, as we will observe in further details in the following sections.

\section{Group Information in Variable Selection Algorithm}\label{sec:GroupInfo}

Variable selection consistency ensures that among all the models of reasonable size, the posterior probability of the true model, $P(\mathcal{M}_{\bgmo} |\bfy, \bfX)$, uniformly dominates all other models, when $n$ is sufficiently large. To identify the true model in practice, it is required to calculate the posterior probabilities
of all the models.
The problem, however, is that the models space is huge. Even if one restricts the model size to $\pg^{\star}$, there are about $p^{\pg^{\star}}$ models to compare with. Any exhaustive model search algorithm can not cover this space in finite time.
Therefore, either a two-step procedure with an initial model screening, or a stochastic search variable selection (SSVS) method must be employed.

In particular, we focus on a situation where the covariates with same functionalities form correlated groups. When $p\gg n$, existence of near multicollinearity in the data is a common phenomenon. Although it may not be true that an entire group is either associated or disassociated with the response.
Incorporating group information in the model-prior setup is therefore of little benefit in the context of variable selection.

Nevertheless, if this group information is carefully utilized, then it can be helpful in making the variable selection algorithm faster to a great extent. The group identities of the covariates act as a latent feature escalating the neighborhood search and thereby, ensuring better mixing. Both from the perspective of screening as well as neighborhood search, the traditional methods can be made more efficient
by utilizing the group information appropriately.

In the following subsection, we propose a class of group inclusion probabilities, called  
group SIS (GSIS). From the theoretical properties, it becomes evident that GSIS can efficiently discriminate the groups containing at least one true covariate from other inactive groups.
One may apply GSIS as a pre-processing step, to reduce the dimensionality of the variable selection problem.

In Section \ref{sec:GiVSA}, we propose a group informed SSVS algorithm, termed as GiVSA, which uses group information, along with an efficient group inclusion probability to make the neighborhood search faster. Theoretical properties of GiVSA guarantees a better mixing rate than the traditional SSVS algorithm, which is also demonstrated numerically later in Section \ref{sec:NumAn}.

In this work, we consider the group information to be known. In many situations, the group information of the covariates is known from previous studies. Finding the group structure among the $p$ covariates is a complex problem of covariate clustering, which we do not attempt to solve here. Our objective is to improvise the current screening based, or SSVS variable selection algorithms with the help of available group information.
For simplicity, we assume the covariates to be non-stochastic in this section, although the results can be extended for stochastic covariates under appropriate assumptions.

\subsection{Group SIS (GSIS)}\label{sec:GSIS}

Let the list of covariates be partitioned into $\kn (\leq p)$ non-overlapping groups.
The set of indices of covariates included in the $k$-th group is denoted by $G_{k}\subseteq \{1,\ldots,p\}$
and the submatrix of $\bfX$ containing the covariates corresponding to $G_{k}$ is denoted by $\bfX_{(k)}$ for $k=1, \ldots, \kn$.
We term a group $G_{k}$ to be `active' if it contains at least one active covariate, and `inactive' otherwise. The classes of active and inactive groups are denoted by $\mA$ and $\mI$, respectively.

\vspace{-.1 in}
\paragraph{GSIS group inclusion probability.} Let $\bfP(\bfX_{(k)})$ denote the projection matrix onto the column space of $\bfX_{(k)}$. Define the inclusion probability of the $k$-th group $q_{k}^{d}$ proportional to $\|\bfP(\bfX_{(k)})\bfy \|^{2}/p_{k}^{d}$, where $\|\cdot\|$ denotes the Euclidean norm, $p_{k}=|G_{k}|$ is the cardinality of the $k$-th group, and $0\leq d\leq 1$.

In the following sub-section, we first demonstrate the property of GSIS group inclusion probabilities with $0<d\leq 1$ to discriminate between active and inactive groups. Some additional properties of GSIS group inclusion probabilities, for $d=0$ will be discussed in Section \ref{sec:mGSIS}.

\subsubsection{Properties of GSIS inclusion probability for $0<d\leq 1$}

Before investigating the properties of the proposed GSIS inclusion probabilities, let us first state the assumptions under which the theoretical results are applicable.

\vspace{-.15 in}
\begin{enumerate}[(B1)]
\itemsep-0.5pt
    \item The maximum group size $ p_{g}^{\star} \lesssim n^{t^{\star}}<n$ for some $t^{\star}\leq 1$.
    \item
    Let $G_{(-k)}$ be the set of indices of the covariates that are absent in $G_{k}$. Then
    \vspace{-.1 in}
     $$ \max \left\{ \max_{k\in \mA} \sigma_{\max}\left(n^{-1}\bfX_{G_{k}\cap \bgmo}^{\prime} \bfX_{G_{(-k)} \cap \bgmo} \right), ~\max_{k\in \mI} \sigma_{\max}\left(n^{-1}\bfX_{\bgmo}^{\prime} \bfX_{(k)} \right)\right\}\leq \eta_{n},$$

     \vspace{-.1 in}
 \noindent   where $\eta_{n}^{2}|\bgmo| \rightarrow 0$ as $n\rightarrow \infty$.

   \item The regression coefficients of the active covariates are strictly larger than a constant, $\beta_{\min}^{\star}>0$, i.e., $|\beta_{0,j}|>\beta_{\min}^{\star}$ for each $j$.
\end{enumerate}

\vspace{-.15 in}
The aforementioned setup includes the ungrouped situation when $\kn=p$, although the group size is restricted to $n$.
Assumption (B2\*) ensures separability of the true set of covariates. The first part asserts that the active covariates have low correlation uniformly across groups. The second part ensures low correlation among the true covariates and the covariates of inactive groups.
Note that there is no restriction on the distribution of the active covariates in the groups.
Under the above assumptions, the following theorem provides an uniform lower bound of the ratio of probabilities of an active and an inactive group.
\begin{theorem}[GSIS Group Probability]\label{res:4}
Let $\bfX_{(k)}$ be the submatrix of $\bfX$ consisting of the members of $k$-th group for $k=1, \ldots, \kn$, and let $q_{k}^{d} \propto \| \bfP({\bf X}_{(k)}) \bfy\|^{2}/p_{k}^{d}$, where $d$ is a constant in $(0,1]$, be the GSIS group inclusion probability of the $k$-th group. Then, under assumptions (A2$^{\prime}$), (A3$^{\prime}$), (B1), (B2\*) and (B3\*), if $n^{-1}\lambda_{\min} \big(\bfX_{\bgmo}^{\prime}\bfX_{\bgmo}\big)\geq \tau_{\min}^{\star}$, then with probability at least $1 - p^{-1} $

 {\centering
 $\inf_{k\in \mA,k^{\prime} \in \mI}~ q_{k^{\prime}}^{d ~-1} q_{k}^{d} \gtrsim  \min\big\{ \eta_{n}^{-2} |\bgmo|^{-1}, n \big(p_{g}^{\star(1-d)}\log p \big)^{-1} \big\} \min_{k \in \mA} \xi_{k} $,
\par}

\noindent where $\xi_{k} = |G_{k}\cap \bgmo|/p_{k}^{d}$.
\end{theorem}
A proof of Theorem \ref{res:4} is provided in the Appendix (see Section \ref{sec:GSIS_proofs}).

From Theorem \ref{res:4} it can be observed that efficiency of the GSIS group probabilities depends on $\min_{k \in \mA} \xi_{k}$, i.e., the minimum of ratio of number of active covariates in a group and an appropriate power of the group size. Even if the group sizes are large, if the active groups are small or dense in the active covariates, then the GSIS shows a clear separation between active and inactive groups.

Another factor which determines the efficiency of GSIS inclusion probabilities is the rate of decay of $\eta_{n}$. A high rate of decay of $\eta_{n}$ can be regarded as an indicator of low correlation among the set of true covariates between active groups, and that among $\bfX_{\bgmo}$ and the inactive groups. Clearly, low correlation between groups makes any group selection method efficient and can be regarded as a basic requirement for efficiency.

The choice of $d$ may depend on maximum group size. If the group sizes are small so that $p_{g}^{\star}\sim n^{t^\star}$ with $t^{\star} < 1$, then a small choice of $d$ suffices. On the other hand if the group sizes are heterogeneous, and some groups are large, then a choice of $d$ close to $1$ is desirable.

\subsubsection{Properties of GSIS inclusion probability for $d = 0$}\label{sec:mGSIS}
When $d=0$, then GSIS group probabilities does not employ any penalization to the group size. Consequently, larger groups receive higher probabilities.
The purpose of introducing group probabilities is not only hard screening of groups, which is a stringent move, as we will see in Section \ref{sec:Mixing_proof}, the group probabilities can also be used appropriately in neighborhood search in an SSVS algorithm. In order to explore the model space properly, one needs to assign higher probability to larger groups, where the special case of GSIS inclusion probability with $d=0$ is more appropriate.
When the group sizes are small compared to $n$, i.e., $p_{g}^{\star}/n \rightarrow 0$, then the special case with $d=0$ achieves some additional properties, which we will discuss here.

Consider the following two additional assumptions:

\vspace{-.15 in}
\begin{enumerate}[(B1)]
\itemsep-0.5pt
\setcounter{enumi}{3}
    \item The highest eigenvalue of $ n^{-1}\bfX^{\prime} \bfX \lesssim n^{r}$ for some $r<1$.
        \item $\min_{k\in\{1,\ldots,\kn\}} \lambda_{\min} \left( n^{-1} \bfX_{(k)}^{\prime} \bfX_{(k)}\right) \geq \tau_{\min}^{\star}  $.
\end{enumerate}

\vspace{-.15 in}
The highest eigenvalue of the design matrix $\bfX$ is allowed to grow at the rate $n^{1+r}$ in assumption (B4\*), which is quite reasonable.
Although the covariates within group are correlated, the possibility of perfect correlation is excluded by assumption (B5\*).

\begin{theorem}\label{lm:GroupPrior}
Let $\bfX_{(k)}$ be the submatrix of $\bfX$ consisting of the members of $k$-th group for $k=1, \ldots, \kn$, and
let $q_{k} \propto \| \bfP({\bf X}_{(k)}) \bfy\|^{2}$ be the GSIS group inclusion probability of the $k$-th group. Then, under assumptions (A2$^{\prime}$)-(A3$^{\prime}$), (B1) with $t^{\star}<1$, and (B2)-(B5) the following results hold with probability at least $1 - p e^{ - c (\log p)^{2}} $ for some constant $c>0$:

\vspace{-.1 in}
\begin{enumerate}[(a)]
\itemsep-0.5pt
    \item $\inf_{k \in \mA} q_{k} \gtrsim |\bgmo|^{-1} n^{-r}\beta_{\min}^{\star} \tau_{\min}^{\star~2}\min_{k\in \mathcal{A}} |\bgmo \cap G_{k}|$,
    \item  $ \sup_{k\in \mI} q_{k} \lesssim \left( |\bgmo| \beta_{\min}^{\star} \tau_{\min}^{\star}  \right)^{-1} \left\{ \eta_{n}^{2}  |\bgmo|/ \tau_{\min}^{\star~2} + 2 \log p \sqrt{|\bgmo|/n} + 5n^{-(1-t^{\star})} \right\}$,
\item $\inf_{k\in \mA,k^{\prime} \in \mI}~ q_{k^{\prime}}^{-1} q_{k} \gtrsim  \eta_{n}^{-2} |\bgmo|^{-1} \wedge n^{1-t^{\star}} $.
\end{enumerate}
\end{theorem}

\vspace{-.1 in}
A proof of Theorem \ref{lm:GroupPrior} is provided in the Appendix (see Section \ref{sec:GSIS_proofs}).

\begin{remark}
Part (c) of Theorem \ref{lm:GroupPrior} indicates that when $p_{g}^{\star}/n \rightarrow 0$ as $n \rightarrow\infty$, the GSIS group inclusion probabilities with $d=0$ can clearly ditinguish between active an inactive groups.
Further, from the proof of Theorem \ref{lm:GroupPrior}, it is also clear that when all the active covariates are from a single group, then the uniform lower bound of the ratio of group probabilities can be sharpened to $\eta_{n}^{-2} \wedge n^{1-t^{\star}}$.
\end{remark}

\begin{remark}
The rate of increase of maximum group size in assumption (B1) and that of the highest eigenvalue of $n^{-1} \bfX^{\prime} \bfX $ in assumption (B4\*) can be same, i.e., $t^{\star}$ can be equal to $r$. Suppose that the design matrix is of the form of a sum of a block diagonal matrix $\boldsymbol{\Sigma}$ and a noise matrix $\bf N$ such that $\sigma_{\max} ({\bf N}) \rightarrow 0$ as $n\to \infty$ at an appropriate rate and the blocks of $\boldsymbol{\Sigma}$ are of equi-correlation structure, then conditions (B1) and (B4\*) hold with $t^{\star}=r$.
\end{remark}

\subsection{Group-informed Variable Selection Algorithm (GiVSA)}\label{sec:GiVSA}

Stochastic search variable selection (SSVS) algorithms are primarily neighborhood based Metropolis Hastings random walks which move through the discrete state space of models. The general algorithm can be described as follows:
\vspace{-.1 in}
\begin{enumerate}[Step 1.]
\itemsep-0.5pt
    \item Suppose at the $m$-th iteration the current stage of the random walk is at the model $\bgm$.
    Let $N(\bgm)$ be the neighborhood of $\bgm$, and $T(\bgm, \cdot)$ be the proposal transition function based on $\bgm$ over $N(\bgm)$. Choose a proposal state $\bgmp$ randomly according to $T(\bgm, \cdot)$.
    \item Accept $\bgmp$ with probability $\alpha$, where

    {\centering
$    \alpha (\bgm, \bgmp)= \min \displaystyle\left\{1, \frac{P(M_{\bgmp}| \bfy)  T(\bgmp,\bgm)}{P(M_{\bgm}| \bfy) T(\bgm,\bgmp)} \right\}.$
\par}
\end{enumerate}
This random walk over models induces a irreducible, aperiodic and positive recurrent Markov chain $\mC$ with transition probability given by
\begin{eqnarray}
    P_{\mathrm{MH}} ( \bgm \to \bgmp ) &=&  \begin{cases} T(\bgm,\bgmp) \alpha(\bgm,\bgmp) & \mbox{if}~ \bgmp \in N(\bgm), \\
    0 &  \mbox{if}~ \bgmp \notin N(\bgm)\cup \{\bgm\}, \\
    1-\sum_{\bar{\bgm}\neq \bgm} P_{\mathrm{MH}} ( \bgm \to \bar{\bgm} ) & \mbox{if}~ \bgmp=\bgm.\end{cases}  \label{eqn:P_MH}
    \end{eqnarray}
The target distribution of $\mC$ is the posterior distribution $ P(M_{\bgm}|\bfy)$. Reversibility of $ \mC$ can be verified by the detailed balance condition $ P(\mM_{\bgm}|\bfy)P_{\mathrm{MH}} ( \bgm \to \bgmp ) = P(\mM_{\bgmp}|\bfy) P_{\mathrm{MH}} ( \bgmp \to \bgm )$ for all $\bgm, \bgmp \in \mM$.

The MCMC algorithm we analyze is based on three local moves: (A) addition of an inactive covariate, (R) removal of an active  covariate and (S) replacing (swapping) an active covariate by an inactive covariate.
Note that `S' is a composite move, composed of an `R' move and an `A' move. The transition function $T(\bgm,\cdot)$ assigns the probabilities $q_{A}(\bgm)$, $q_{R}(\bgm)$ and $q_{S}(\bgm)$ to each of these three moves, respectively. 
Based on the selected move, the neighborhood is determined. If an `A' move is selected, then the neighborhood, say $N_{A}(\bgm)$, consists of all the $p-|\bgm|$ models which can be obtained by adding one inactive covariate to the current model. If an `R' move is selected, then the neighborhood $N_{R}(\bgm)$ consists of all the $|\bgm|$ models which can be obtained by dropping one covariate from the current model. Finally, if an `S' move is selected then all the $|\bgm|(p-|\bgm|)$ models, which can be obtained by replacing an active covariate of the current model with an inactive covariate, constitutes the neighborhood $N_{S}(\bgm)$. The neighborhood class of $\bgm$ is the union of these three types of neighborhoods, $N(\bgm)=N_{A}(\bgm)\cup N_{R}(\bgm) \cup N_{S}(\bgm)$.
Similarly, $T(\bgm,\bgmp)= q_{A}(\bgm) T_{A}(\bgm,\bgmp)+ q_{R}(\bgm) T_{R}(\bgm,\bgmp)+ q_{S}(\bgm) T_{S}(\bgm,\bgmp)$, where $T_{A}$, $T_{R}$ and $T_{S}$ are uniform proposal transition functions restricted to the corresponding neighborhoods. 

The true model is assumed to be non-null. Therefore, for any model $\bgm$ of dimension $1$, we have $q_{A}(\bgm)=q_{S}(\bgm)=1/2$ and $q_{R}(\bgm)=0$. Consequently, $N_{R}(\bgm)=\phi$. Further, suppose we restrict the search to models with dimension at most $p_{n}^{\star}$, then for any model $\bgm$ with dimension $p_{n}^{\star}$, we have $q_{S}(\bgm) =q_{R}(\bgm)=1/2$, and $q_{A}(\bgm)=0$.
For any other model $\bgm$, we have $q_{A}(\bgm)=q_{R}(\bgm)=q_{S}(\bgm)=1/3$.
As $\bgmp \in N_{A}(\bgm)$ if and only if $\bgm \in N_{R}(\bgmp)$, and $\bgmp \in N_{S}(\bgm)$ if and only if $\bgm\in N_{S}(\bgmp)$, the detailed balance condition is satisfied by the proposed MCMC.

\paragraph{Improvisation under the availability of group structure.}
Suppose it is known that the covariates are partitioned in $\kn$ mutually exclusive and exhaustive groups, and a group inclusion probability $q_{k}$ for $k=1,\ldots,\kn$ is available.
Then, for an `A' move, first a group is selected with probability $q_{k}$, and thereafter an inactive covariate within the chosen group is selected following a uniform distribution over the additive neighbors within that group. However, in the `R' move, the proposal distribution remains uniform over $N_{R}(\bgm)$ as before. Since an `S' move is a composition of `A' and `R' moves, the change in proposal distribution corresponding to `A' move is reflected in that of `S' move as well.
We will show theoretically and via simulations that this simple change in implementation reduces the mixing time significantly.


Let $\bgmp$ be obtained from $\bgm$ by an `A' move, and $\bgmp\setminus \bgm=\{j\}$. Then,

{\centering
$ T_{A}(\bgm, \bgmp)
= \displaystyle\sum_{k=1}^{\kn} \frac{q_{k} I(j \in G_{k}) }{p_{k}-|G_{k} \cap \bgm|},$
\par}

\noindent where $I(\cdot)$ is the indicator function
and $p_{k}$ is the size of the $k$-th group.
Let $\bgmp$ be obtained from $\bgm$ by an `R' move. Then,

{\centering
 $ T_{R}(\bgm, \bgmp)= |N_{R}(\bgm)|^{-1}$.
 \par}

\noindent Finally, let $\bgmp$ be obtained from $\bgm$ by an `S' move. Define $\bgm^{\prime\prime}= \bgmp\cap \bgm$ and $\bgmp\setminus\bgm=\{j\} $. Then,

{\centering
 $ T_{S}(\bgm, \bgmp)=  T_{R} (\bgm,\bgm^{\prime\prime}) T_{A}(\bgm^{\prime\prime}\bgmp) =
 |N_{R}(\bgm)|^{-1}  \displaystyle\sum_{k=1}^{\kn} \frac{I(j \in G_{k}) q_{k}}{p_{k}-|G_{k} \cap \bgm^{\prime\prime}|}.$
 \par}

The Markov chain transition probabilities in this case can be summarized as in equation (\ref{eqn:P_MH}) with $T(\bgm,\bgmp) \alpha(\bgm,\bgmp)$ as follows:
     \begin{eqnarray}
    T(\bgm,\bgmp) \alpha(\bgm,\bgmp)   =  \begin{cases} 3^{-1} \min\left\{\sum_{k=1}^{\kn} \frac{q_{k} I(l \in G_{k}) }{p_{k}-|G_{k} \cap \bgm|} , \frac{ (|\bgm|+1) P(M_{\bgmp}| \bfy) }{P(M_{\bgm}| \bfy) }  \right\} & \mbox{if}~ \bgmp \in N_{A}(\bgm) \\
 3^{-1} \min \left\{ |\bgm|^{-1} , \frac{  P(M_{\bgmp}| \bfy) }{P(M_{\bgm}| \bfy) }  \sum_{k=1}^{\kn} \frac{q_{k} I(j \in G_{k}) }{p_{k}-|G_{k} \cap \bgmp|} \right\}   &\mbox{if}~ \bgmp \in N_{R}(\bgm) \\
      (3|\bgm|)^{-1} \min \left\{   \sum_{k=1}^{\kn} \frac{I(l \in G_{k}) q_{k}}{p_{k}-|G_{k} \cap (\bgm\setminus \{j\})|}, \right. &\\
 \left. \qquad \qquad \qquad  \frac{  P(M_{\bgmp}| \bfy) }{P(M_{\bgm}| \bfy) }    \sum_{k=1}^{\kn} \frac{I(j \in G_{k}) q_{k}}{p_{k}-|G_{k} \cap (\bgm\setminus \{l\})|} \right\} & \mbox{if}~ \bgmp \in N_{S}(\bgm),\end{cases}  \label{eqn:P_MH2}
\end{eqnarray}
where both in cases of A and S moves, the $l$-th component is added to $\bgm$; and in cases of R and S moves, the $j$-th component is removed from $\bgm$.
As all the group probabilities are positive, the detailed balance condition under this modification can be verified as before.

\vspace{-.15 in}

\paragraph{Mixing Time.} Let $\bgm\in \mM$ be the initial state, ${\bf S}$ be any subset of the model space $\mM$ and $P(\bgm \to \bgmp)$ be the transition probability of a Markov chain. Define $P\left( \bgm, {\bf S}\right)=\sum_{\bgmp \in {\bf S}}P (\bgm \to \bgmp)$, and the total variation distance of the Markov chain to the target distribution $\pi$ after $t$ iterations is given by

{\centering
$\Delta_{\bgm}(t)= \max_{{\bf S} \subset \mM} \left| P^{t} (\bgm,{\bf S})- \pi({\bf S}) \right|. $
\par}

\noindent The $\upsilon$-mixing time is given by

{\centering
$ \tau_{\upsilon} = \max_{\bgm \in \mM}  \min\left\{ t\in \mathbb{N} \mid   \Delta_{\bgm} (t^{\prime}) \leq \upsilon ~\mbox{for all~} t^{\prime} \geq t\right\}.$
\par}

\cite{YWJ_2016} provides the rate of mixing time in order of $(n,p)$ for the traditional MCMC algorithm (ungrouped version). Following the same line of proof,
in Theorem \ref{thm:Mixing_rate} we will show that the grouped version of MCMC algorithm provides a better mixing time compared to the traditional one.


\paragraph{\bf Setup and Assumptions.}
 Let the space of the models with dimension at most $\ps$ be $\mM^{\star}=\left\{ \bgm\in \mM : |\bgm| \leq \ps\right\}$, where $\ps \sim n^{s}$ for some $s$ satisfying $b<s<1-b$.
 We restrict the search of models to $\mM^{\star}$ only.
 Further, we assume that the covariates are standardized so that $\|{\bf x}_{(j)}\|=\sqrt{n}$ for each $j=1,\ldots,p$.
Below we describe the assumptions under which our result on mixing time is valid.

\vspace{-.15 in}
\begin{enumerate}[(C1)]
\itemsep-0.5pt
    \item $\min_{\bgm}\left(n^{-1}\bfX_{\bgm\cup\bgmo}^{\prime} \bfX_{\bgm\cup\bgmo} \right)\geq \tau_{\min}^{\star}$ for $\bgm\in\mM^{\star}$.
    \item {}[Separability Assumption] Suppose $\bgm\in \mM_{2} \cap \mM^{\star}$, i.e., $\bgm$ is a non-supermodel of $\bgmo$ and $\Pgm$ is the projection matrix corresponding to $\bXgm$, then there exists some $\alpha_{0}\in(0,1)$ such that

{\centering
$\sup_{\bgm } \lambda_{\max} (\Pgmo\bfP_{\bgm} \Pgmo) \leq \alpha_{0}.$
\par}

\item $\lambda_{\max} \left( n^{-1} \bfX^{\prime}_{\bgmo} \bfX_{\bgmo} \right) \leq \tau_{\max}^{\star}$.
\item Assume $\beta_{\min}^{\star}<|\beta_{0,j}|<\beta_{\max}^{\star}$ for all $j$. This together with the lower bound of eigenvalues of $n^{-1} \bfX_{\bgmo}^{\prime}\bfX_{\bgmo}$ in (C1) yields $\|\bmu\|^{2} \sim n|\bgmo|$.
\end{enumerate}

\begin{remark}
\label{rm:2}
Assumptions (C1)-(C4) are similar but stronger versions of previously made assumptions (A2$^{\prime}$),(A3$^{\prime}$), (B3\*) and (B5\*).
Like $\mM^{\star}$, a restricted model space $\mM_{1}\cup \mM_{2}$ is considered in Theorem \ref{thm:VSC}. However, $\mM_{1}\cup \mM_{2}$ contains all models of dimensions at most $O(n/\log n)$, whereas $\mM^{\star}$ contains models of dimension $O(n^{s})$ for $b<s<1-b$.

Assumption (C1) is related to assumption (B5\*) in the previous section. It is intuitive that linear dependence of the covariates within group is more than that between groups. Therefore, restriction on the within group lowest eigenvalue, induces a restriction on the eigenvalues of all subsets of $\bfX$ of small dimension.
However, it is difficult to prove that (B5\*) implies (C1). If the maximum group size $p_g^{\star} \leq \ps$, then (C1) implies (B5\*), although, there is no assumption regarding the relation between $p_g^{\star}$ and $\ps$.

The separability assumption (C2) ensures that

{\centering
$ \inf_{\bgm\in\mM_{2}\cap \mM^{\star}}  \bmu^{\prime} (I-\Pgm)\bmu >(1-\alpha_{0}) \|\bmu\|^{2}.$
\par}

\noindent This, along with (C4), imply assumptions (A2$^{\prime}$) and (A3$^{\prime}$) in Theorem \ref{thm:VSC} (see Remark \ref{rm:VSC}).
Further, assumption (C4) implies (B3\*).

Finally, assumption (C3) ensures that the true model does not include redundant covariates. A similar condition is shown to be sufficient for assumptions (A2) and (A3) in Lemma \ref{lm:1} (see Remark \ref{rm:3}).
\end{remark}

\begin{remark}\label{rm:1}
Since the result on mixing time is much stronger than variable selection consistency (Theorem \ref{thm:VSC}) or consistency of group probabilities (Theorem \ref{res:4}-\ref{lm:GroupPrior}), it is natural that a stronger set of assumptions would be required. Definition of mixing time considers all possible initial values, and a total variation distance between the approximating and target distributions over all possible subsets of $\mM^{\star}$. In practice, if a good initial model is chosen, then the number of iterations needed to reach a $\upsilon$-neighborhood of the true model can be quite less, even if weaker versions of these assumptions hold. 
\end{remark}

\begin{theorem}\label{thm:Mixing_rate}
Consider the $g$-prior setup in the normal linear regression framework as described in (\ref{eqn:priorsetup}), where $p\sim n^{t}$ (for some $t>0$), and the modified hierarchical uniform model prior probability restricted to $\mM^{\star}$ as described in (\ref{eqn:modelprior}). Suppose there exist a true model $M_{\bgm_{0}}$ satisfying (A1) which generate $\bfy$, and the covariates are partitioned into $\kn$ groups satisfying assumptions (B1) with $t^{\star}<1$ and (B2).
Let the groups be split in two classes, namely, the class of active groups ($\mA$), groups containing at least one active covariate, and the class of inactive groups ($\mI$), groups containing no active covariate, and $\{q_{k},~k=1,\ldots,\kn\}$ be a group inclusion probability distribution on the $\kn$ groups.
Then, under assumptions (C1)-(C4) if $g_{n}$ and $k_{n}$ satisfies
(i$^{\prime}$) $k_{n}^{2} g_{n} \gtrsim p^{2(1+\lambda)\ps}$ with $\lambda>0$ and
(ii$^{\prime}$) $\exp\{n/\ps\} = o\left( k_{n}\sqrt{g_{n}}\right)$,
then the $\upsilon$-mixing time corresponding to the proposed GiVSA algorithm satisfies

{\centering
    $ \tau_{\upsilon} \lesssim \ps \max\left\{ \ps, \left(p w_{n} q_{\mI}^{\star} \right)^{-1} p_g^{\star}, q_{\mA}^{\star-1} p_g^{\star} \right\} \times \left[ \max\left\{ n\log n , \ps \log (k_{n}\sqrt{g_{n}})\right\} -\log \upsilon\right] $
\par}

  \noindent  with probability tending to one, where
 $q_{\mA}^{\star} = \min_{k \in \mA} q_{k}$, $q_{\mI}^{\star} = \min_{k \in \mI} q_{k}$ and
    $w_{n}$ is a sequence such that $w_{n}\uparrow \infty$ as $n\rightarrow \infty$.
\end{theorem}
A proof of Theorem \ref{thm:Mixing_rate} is provided in Appendix (see Section \ref{sec:Mixing_proof}).

\begin{remark}
\label{rm:4}
(Implication of Theorem \ref{thm:Mixing_rate})
Suppose that the group probability distribution $\{q_{k},~k=1,\ldots,\kn\}$ contains some information regarding the marginal utility of the groups so that $q_{\mA}^{\star}\gtrsim (n|\bgmo|)^{-1}$ and $q_{\mI}^{\star}\gtrsim (w_{n}p)^{-1}$. If $\ps \gtrsim \sqrt{n}$, then any choice of $(k_n,g_{n})$ satisfying (i$^{\prime}$) also satisfies the condition (ii$^{\prime}$) of Theorem \ref{thm:Mixing_rate}. In that case, we have $\tau_{\upsilon} \lesssim n^{1+s+b+t^{\star}} \left\{ n^{2s}\log n - \log \upsilon\right\}$. Comparing this with \citet[Theorem 2]{YWJ_2016}, who also use the $g$-prior method for variable selection, the mixing rate for the ungrouped case is $n^{t+2s} \left\{ n\log n - \log \upsilon\right\}$. Note that $p\sim n^{t}$ and $t$ can be any positive number, whereas $b+s$ and $t^{\star}$ are less than $1$ each. To summarize, the gain in mixing time is clearly visible, especially when $t$ is larger than $2$.
\end{remark}

\begin{remark}
\label{rm:5}
In view of Remark \ref{rm:4}, it can questioned if there exists a probability distribution on the groups which satisfies the conditions $q_{\mA}^{\star}\gtrsim (n|\bgmo|)^{-1}$ and $q_{\mI}^{\star}\gtrsim (w_{n}p)^{-1}$. The answer is: there are various ways of assigning group probabilities based on marginal utilities. In particular, GSIS (see Section \ref{sec:GSIS}) provides one such probability distribution. It is shown in Theorem \ref{lm:GroupPrior} that $q_{\mA}^{\star}$ for the prior prescribed in GSIS with $d=0$ is at least of order $\left(n^{r}|\bgmo|\right)^{-1}$ with probability tending to one, where $r<1$.
The other condition $q_{\mI}^{\star}\gtrsim (w_{n}p)^{-1}$ is not a desirable property for any good group probability distribution.
It is required in Theorem \ref{thm:Mixing_rate}, as the mixing time considers the worst possible initial setup. However, it is always possible to discard the groups having probability much less than $p^{-1}$. Intuitively, this is as good as discarding groups with probability less than that of a single covariate when an uniform distribution over the covariates is assigned.
\end{remark}

\begin{remark}
\label{rm:6}
Finally, observe that the choices of $k_{n}$ and $g_{n}$ in Theorem \ref{thm:Mixing_rate} are consistent with those in Theorem \ref{thm:VSC}. Suppose $\ps\sim n^{s}$ with $s<0.5$. Then, any $(k_{n},g_{n})$ satisfying (ii$^{\prime}$) also satisfies (i$^{\prime}$). Obviously, this obeys condition (ii) of Theorem \ref{thm:VSC} on $(k_{n},g_{n})$. In particular, if we choose $k_{n}\sqrt{g_{n}} \sim \exp\left\{ n^{1-s}\right\}\log n$, then condition (i) of Theorem \ref{thm:VSC} is also satisfied as $s>b$.

Further, if $s>0.5$, then the condition (i$^{\prime}$) satisfies condition (ii$^{\prime}$) of Theorem \ref{thm:Mixing_rate} and condition (ii) of Theorem \ref{thm:VSC}. In particular, if we choose $k_{n}\sqrt{g_{n}} = p^{(1+\lambda)\ps}$ for some $\lambda>0$, then condition (i) of Theorem \ref{thm:VSC} is also satisfied as $s<(1-b)$.
This, along with Remarks \ref{rm:3} and \ref{rm:2}, suggests that under the conditions of Theorem \ref{thm:Mixing_rate}, the conclusion of Theorem \ref{thm:VSC} is also valid.
\end{remark}

\section{Numerical Analysis}\label{sec:NumAn}

In this section, we validate the performance of our method using some simulated and real data sets. The purpose of our investigation is three fold:
(i) validating the performance of the proposed set of priors associated with GiVSA in variable selection, compared to other state-of-the-art classical and Bayesian variable selection methods;
(ii) studying the performance of GSIS in group selection compared to other classical and Bayesian methods group selection methods; and
(iii) demonstrating the time gain due to GiVSA.

\subsection{Description of the proposed method and competitors}\label{sec:PVSA}

\subsubsection{Proposed variable selection algorithm}

In Section \ref{sec:GiVSA}, we discuss the method of incorporating group information in the basic RJMCMC algorithm based on random walk. For practical purposes, we use the paired-move multiple-try stochastic search (\emph{pMTM}) sampler proposed by \citet{CQT_2016}, which enjoys better mixing properties than the basic single or double-flip RJMCMC.
The proposed modification using group information is demonstrated through the numerical results of the pMTM sampler with and without the group information.

\vspace{-.15 in}
\paragraph{The pMTM algorithm}
The idea of pMTM algorithm (see \citet[Section 3.2]{CQT_2016}) is discussed below in a nutshell, with simple moves ($A$ or $R$) only.
Let $\bgms$ be the present state. Define a weight function $\phi:\{0,1\}\times \mathbb{R}^{+}\times\{A,R\}\to [0,1]$ as

{\centering
$\phi_{j}(m)=\phi(\gamma_{j},\nu_{j};m)=(1-\gamma_{j}) f(\nu_{j}) I_{\{m=A\}} + \gamma_{j} g(\nu_{j}) I_{\{m=R\}},$
\par}

\noindent where $\gamma_{j}$ is the state of the $j$-th covariate in $\bgms$ (i.e., $\gamma_{j}=1$ if the $j$-th covariate is present in $\bgms$ and $\gamma_{j}=0$ otherwise) and $\nu_{j}$ is an importance score of the $j$-th covariate for $j=1,\ldots,p$, while $f,g:\mathbb{R}^{+}\to [0,1]$ are functions determining the probabilities of inclusion and removal of predictors.

First, a move $m$ is selected with probability $\omega_{m}(|\bgms|)$, $m=A,R$.
A random forward neighborhood $N_{m}(\bgms)$ is then created by taking Bernoulli draws with probabilities $\phi_{j}(m)$ for $j=1,\ldots,p$. Next, a proposal model $\bgmp$ is selected from the random forward neighborhood $N_{m}(\bgms)$ as per the proposal transition function $T_{m}(\bgms,\cdot)$ supported on $N_{m}(\bgms)$. To maintain the detailed balance condition, a random backward neighborhood of the selected model $\bgmp$, $N^{\prime}_{rm}(\bgmp)$, is created similarly (here $rm$ is the reverse move). Let $ T^{\prime}_{rm}(\bgmp,\cdot)$ be the corresponding proposal transition probability. Then, $\bgmp$ is accepted with probability
$$ \alpha(\bgms,\bgmp)=\min\left\{ 1, \frac{P(M_{\bgmp}| \bfy) \times \omega_{rm}(|\bgmp|)\times \phi_{j}(rm) \times T_{rm}(\bgmp,\bgms)}{P(M_{\bgms}| \bfy) \times \omega_{m}(|\bgms|) \times \phi_{j}(m) \times T_{m}(\bgms,\bgmp)}\right\},$$
where $j$ is the index of the covariate at which $\bgms$ and $\bgmp$ differ.

\vspace{-.15 in}
\paragraph{$g$-prior (GiVSA)} When the group structure is known, to create the forward neighborhood, we first select a group with respect to group selection probability $\{ q_{k}; k=1, \ldots,\kn\}$, and then select a covariate in the group. Therefore,

{\centering
$\phi_{j}(A)=P(j\text{-th covariate is included in } N_{A}(\bgms))= (1-\gamma_{j}) \sum_{k=1}^{\kn} q_{k} I(j \in G_{k}),$
\par}

\noindent whereas the probability of being included in the backward neighborhood remains
$\phi_{j}(R)=\gamma_{j}$ for $j=1,\ldots,p$.

\vspace{-.15 in}
\paragraph{$g$-prior (MiVSA)} When the group structure is not known (or, not used), then the forward neighborhood is created using the marginal utility function. We consider the absolute value of marginal correlation coefficient, $|r_{j}|$, as the marginal utility function, which leads to $\phi_{j}(A)\propto |r_{j}|$ for $j=1,\ldots,p$. We call this method as marginally informed variable selection algorithm (MiVSA).

\vspace{-.15 in}
\paragraph{Other specifications} In both the grouped and ungrouped cases, the following choices are considered. The maximum model size is fixed to $\ps=n$. The probabilities $\omega_{m}(|\bgm|)$ are set to $1/3$, except for the terminal cases, i.e., for $|\bgm|=1$ or $|\bgm|=p_n^*$, where in each case we provide equal probabilities to the feasible moves. Informed by \cite{HDW_2007}, the choice of transition probability under move $m$ is taken to be $T_{m}(\bgm,\bgms)= S(\bgms)/\sum_{\bgmp\in N_{m}(\bgm)} S(\bgmp)$, where $S(\bgm)$ is a positive score function. In particular, following \citet[Remark 3.3]{CQT_2016}, we consider $S(\bgms) \propto P(M_{\bgmp}| \bfy)$. This leads to the following acceptance probability for the simple moves
$$  \alpha(\bgms,\bgmp)=\min\left\{ 1, \frac{ \omega_{rm}(|\bgmp|)\times \phi_{j}(rm) \times \sum_{\bgm \in N_{rm}(\bgmp)} P(M_{\bgm}| \bfy)}{ \omega_{m}(|\bgms|) \times \phi_{j}(m) \times \sum_{\bgm \in N_{m}(\bgms)} P(M_{\bgm}| \bfy)}\right\} .$$

\vspace{-.15 in}
\paragraph{Group configuration and group probability vector}
The group configurations are provided for the simulation setups. However, we estimate the group structure (details is provided in Section \ref{sec:RDA}) for the analysis of real data. Group structure estimation is used as a pre-processing step. While making inference, we club all the groups with one member together, and form a single group. Finally, a convex combination of the GSIS group inclusion probabilities with $d=0$ and equal probability $\kn^{-1}$ in $2:1$ ratio is used as $q_{k}$ in GiVSA. GSIS with $d=0$ provides higher weight to larger groups, which ensures that all the covariates have a fair chance of being included in the visited model.

\vspace{-.15 in}
\paragraph{Tuning parameters} There are only two tuning parameters in the entire setup, namely, $g_{n}$ and $k_{n}$. Informed by Theorem \ref{thm:VSC}, we consider the choices $g_{n}=c_{1}p^{2(1+\lambda)}$ and $k_{n}=c_{2}n^{r}$, with $c_{1}=\lambda=1e-2$, $r=43e-2$ in all the numerical results. We only tune the parameter $c_{2}$ based on the required sparsity condition. For all the simulation examples, we set $c_{2}=25e-3$, and for real data, we set $c_{2}=5e-1$.
Values of the tuning parameters are chosen in view of theoretical considerations as well as good practical performance.

\vspace{-.15 in}
\paragraph{Iterations and other information}
We consider $5e3$ iterations, with a burnin period of $3e3$ iterations. As a final recommendation, we consider the model with highest posterior probability among the visited models after burnin, denoted by \emph{HPM}. For GiVSA, we also provide the the post-burnin most frequently visited model, denoted by \emph{mode}.

\subsubsection{Measures for comparison and the competing methods}

\paragraph{(I) Variable selection} The $g$-prior method, along with proposed GiVSA algorithm, is compared with the following state-of-the-world methods.
Frequentist penalized likelihood methods: (i) \emph{LASSO} and
(ii) \emph{SCAD} available in R-package $\mathtt{ncvreg}$;
frequentist bi-level selection methods:
(iii) group MCP (\emph{cMCP}) proposed by \cite{BH_2009},
(iv) \emph{group bridge (gBirdge)} proposed by \cite{HMXZ_2009} available in R-package $\mathtt{grprep}$,
(v) \emph{sparse group lasso (SparseGL)} proposed by \cite{SFHT_2013} available in R-package $\mathtt{sparseg}$, and
(vi) \emph{group exponential lasso (GEL)} proposed by \cite{Breheny_2015};
Bayesian methods with optimization:
(vii) EM approach for Bayesian variable selection (\emph{EMVS}) proposed by \cite{RG_2014} available in R-package \href{https://rdrr.io/cran/EMVS/}{$\mathtt{EMVS}$}, and
(viii) spike and slab LASSO (\emph{SSLasso}) proposed by \cite{RG_2018} available in R-package $\mathtt{SSLASSO}$;
scalable Bayesian shrinkage prior method:
(ix) simplified shotgun stochastic search and screening (\emph{S5}) proposed by \cite{SBJ_2018} available in R-package $\mathtt{BayesS5}$;
Bayesian bi-level selection methods:
(x) Bayesian Group Lasso with Spike and Slab prior (\emph{BGLSS}),
(xi) Bayesian Sparse Group Selection with Spike and Slab (BSGSSS) prior proposed by \cite{LMPS_2017} available in R-package $\mathtt{MBSGS}$, and
(xii) Bayesian bi-level variable selection method (\emph{BIVAS}) proposed by \cite{CDMPLY_2020} available in R-package \href{https://github.com/mxcai/bivas}{$\mathtt{BayesS5}$}; Bayesian screening methods:
(xiii) screening embedded Bayesian variable selection (\emph{SVEN}) proposed by \cite{DSV_2023} and
(xiv) Bayesian iterative screening (BITS) proposed by \cite{WDR_2023} available in R-package $\mathtt{bravo}$.

For each of the competing methods, we choose the tuning parameters as recommended, or the default values have been selected.
For \emph{S5}, we calculate both the product inverse moment (\emph{piMOM}) and product exponential moment (\emph{peMOM}) non-local priors, and both the MAP estimators and the least squares (LS) estimators. Since the performances of MAP and LS are similar, we only report results corresponding to MAP estimators. For \emph{SVEN}, we provide results corresponding to both the MAP model and weighted average model (WAM). Finally, the BSGSSS and BGLSS methods are iterated $5e3$ times with a burnin period of $3e3$ and the number of updates $2e1$.

\vspace{-.15 in}
\paragraph{Measures for comparison} For each simulation setup, we repeat the experiment $N=1e2$ times. Efficiency of variable selection is validated using the following four measures: 

\vspace{-.15 in}
\begin{enumerate}
\itemsep-0.5pt
    \item \emph{Number of selected covariates ($\|\widehat{\bbeta}\|_{0}$)}: Average number of selected covariates.
    \item \emph{True positives (TP):} Average number of true covariates in the selected models.
    \item \emph{False positives (FP):} Average number of inactive (false) covariates in the selected models.
    \item \emph{Jaccard index:} The probability of the event that the chosen model contains all active covariates. It is calculated as the ratio of the number of times all true parameters are included in the chosen model and $N$.
\end{enumerate}

\vspace{-.15 in}
\paragraph{(II) Group selection} When the purpose is solely to select groups with respect to some group utility function, then GSIS with $d>0$ is preferred over GSIS with $d=0$, as it employs group size penalization. In particular, to employ highest group size penalty we take $d=1$. The GSIS algorithm with $d=1$ (\emph{GSIS$_{1}$}) is compared with all the frequentist and Bayesian group and bilevel selection algorithms described above, viz., \emph{cMCP}, \emph{gBridge}, \emph{SparseGL}, \emph{GEL}, \emph{BGLSS}, \emph{BSGSSS} and \emph{BIVAS}. Additionally, we include group LASSO (\emph{gLASSO}) proposed by \cite{YL_2006}, group SCAD (\emph{gSCAD}) and group MCP (\emph{gMCP}), which are variants of \emph{gLASSO} with SCAD and MCP penalties, and are available in R-package $\mathtt{grpreg}$. For the competitors, we consider the $L^{1}$ norm of the group regression coefficient vector as the group~utility.

As hard thresholding based group selection is not the primary goal of this investigation, we do not recommend a threshold beyond which the groups are discarded. Instead, we compare the competing methods with respect to the following two criteria.

Let there be $K$ active groups.
(i) The average number of active groups among the $K$ highest utility groups, is considered as a measure, and is referred to  \emph{true positive} (\emph{TP}).
(ii) Secondly, the number of  active groups among the top $K$ groups as per the average utility over $N$ repetitions is considered, and is termed as  \emph{true positive on average} (\emph{TPa}).

\subsection{Simulation experiments}\label{sec:SE}

\subsubsection{(I) Variable selection}\label{sec:VarSelect}

In all the simulation settings, we set $p=2e3$ and $n=2e2$. The covariates in each case are generated from a symmetric distribution with location parameter ${\bf 0}$ and scale matrix $\boldsymbol{\Sigma}_{X}$. Two choices of the scale matrix $\boldsymbol{\Sigma}_{X}$ are considered: (i) block AR(1) and (ii) block equicorrelation (EC). For the underlying distribution, we take two choices (a) multivariate normal, and (b) multivariate $t$ with degrees of freedom $4$.

We consider a total of $32$ blocks with four distinct sizes. There are $8$ blocks of $4$ different sizes, viz., $25$, $50$, $75$ and $100$. The correlation and assignment of true covariates differ in the following four simulation schemes.

\noindent {\bf Scheme 1} (AR(1)-normal) Gaussian covariates with scale matrix $\boldsymbol{\Sigma}_{X}$ of block AR(1) structure are considered.
The AR(1) probability for the blocks of sizes $25$, $50$, $75$ and $100$ are $0.35$, $0.45$, $0.65$ and $0.75$, respectively.
We consider $11$ true covariates, all from a single block of size $100$. Although all the true covariates are in a single block, they are evenly separated across the block so that the maximum correlation between two active covariates is $0.75^9$.

\noindent {\bf Scheme 2} (AR(1)-$t_{4}$).
This setup is exactly same as Scheme 1, except that the covariates are now generated from a scaled $t$ distribution with degrees of freedom $4$.

\afterpage{
\begin{table}
\renewcommand*{\arraystretch}{0.9}
{\centering
\begin{tabular}{|p{0.155\textwidth} |p{0.06\textwidth}| p{0.06\textwidth} | p{0.06\textwidth} | p{0.075\textwidth} | p{0.06\textwidth}| p{0.06\textwidth} | p{0.06\textwidth} | p{0.075\textwidth} |}
\hline
 & \multicolumn{4}{|c|}{Scheme 1} & \multicolumn{4}{|c|}{Scheme 2}
\\ \hline
Methods $\downarrow$ & $\|\widehat{\bbeta}\|_{0}$  & TP & FP & Jaccard Index & $\|\widehat{\bbeta}\|_{0}$  & TP & FP & Jaccard Index \\
\hline
LASSO   &  73.03   &  10.94   &  62.09   &  0.94   &  83   &  10.61   &  72.39      &  0.71       \\

SCAD   &  60.02   &  10.65   &  49.37    &  0.71   &  61.92   &  10.03   &  51.89      &  0.49  \\

cMCP   &  38.31   &  10.68   &  27.63    &  0.73 &  36.39   &  9.99   &  26.4   &  0.49     \\

gBridge   &  21.47   &  7.8   &  13.67    &  0.7   &  28.1   &  10.19   &  17.91     &  0.89      \\

SparseGL   &  142.64   &  11   &  131.64    &  1   &  122.25   &  11   &  111.25      &  1  \\

GEL  &  86.97   &  11   &  75.97    &  1    &  55.13   &  10.1   &  45.03     &  0.87    \\

EMVS   &  6.52   &  5.98   &  0.54    &  0.16  &  7.06   &  5.81   &  1.25     &  0.11     \\

SSLasso   &  10.37   &  9.89   &  0.48   &  0.45    &  9.47   &  8.94   &  0.53    &  0.35   \\

S5(piMOM)   &  10.58   &  10.24   &  0.34    &  0.63  &  9.2   &  8.55   &  0.65     &  0.37     \\

S5(peMOM)   &  11.05   &  10.6   &  0.45    &  0.73      &  10.44   &  9.79   &  0.65   &  0.5   \\

BGLSS   &  329.5   &  1.1   &  328.4   &  0.1   &  434.75   &  0.66   &  434.09    &  0.06      \\

BSGSSS   &  12.51   &  10.88   &  1.63    &  0.89    &  12.51   &  10.63   &  1.88  &  0.73    \\

BIVAS   &  11.11   &  10.47   &  0.64    &  0.59  &  11.13   &  9.93   &  1.2     &  0.47    \\

SVEN(MAP)   &  5.12   &  4.77   &  0.35   &  0.1   &  4.01   &  3.24   &  0.77   &  0.04 \\

SVEN(WAM)   &  4.89   &  4.62   &  0.27    &  0.09  &  3.87   &  3.21   &  0.66     &  0.04  \\

BITS   &  84.86   &  10.6   &  74.26   &  0.65  &  86.79   &  9.9   &  76.89    &  0.49    \\

MiVSA(HPM)   &  15.15   &  9.78   &  5.37    &  0.46    &  14.61   &  8.96   &  5.65   &  0.37   \\


GiVSA(HPM)   &  10.87   &  10.35   &  0.52    &  0.61   &  9.86   &  8.96   &  0.9    &  0.47    \\

GiVSA(mode)   &  10.8   &  10.44   &  0.36   &  0.65    &  9.98   &  9.45   &  0.53     &  0.54  \\

\hline
\end{tabular}
}
\caption{Results of Scheme 1 and Scheme 2. }
\label{tab:2}

\end{table}

}

\noindent {\bf Scheme 3} (EC-normal). Gaussian covariates with scale matrix $\boldsymbol{\Sigma}_{X}$ of block equi-correlation structure are considered. The $8$ blocks of each size have common probabilities ranging from $0.25$ to $0.95$ with a separation of $0.1$ between two consecutive blocks. 
We consider $12$ true covariates, each in a separate group. There are $3$ covariates in $3$ groups of each size with probabilities $0.35$, $0.45$ and $0.55$, respectively.

\noindent {\bf Scheme 4} (EC-$t_{(4)}$).
This setup is exactly same as Scheme 3, except that the covariates are now generated from a scaled $t_{(4)}$ distribution.

The regression parameter of the active covariates are set to be {\it one} each, and $\sigma^{2}$ is so chosen that the signal to noise ratio $\var(\bmu)/\sigma^{2} = 3$.
Results for Scheme 1 and 2 are provided in Table \ref{tab:2}, and that of Scheme 3 and 4 are provided in Table \ref{tab:1}.

\afterpage{
\begin{table}
\begin{center}
\renewcommand*{\arraystretch}{.9}
\begin{tabular}{|p{0.155\textwidth} |p{0.06\textwidth}| p{0.06\textwidth} | p{0.06\textwidth} | p{0.075\textwidth} | p{0.06\textwidth}| p{0.06\textwidth} | p{0.06\textwidth} | p{0.075\textwidth} |}
\hline
 & \multicolumn{4}{|c|}{Scheme 3} & \multicolumn{4}{|c|}{Scheme 4}
\\ \hline
Methods $\downarrow$ & $\|\widehat{\bbeta}\|_{0}$  & TP & FP & Jaccard Index & $\|\widehat{\bbeta}\|_{0}$  & TP & FP & Jaccard Index \\
\hline

LASSO  &  75.46   &  11.95   &  63.51   &  0.96   &  71.49   &  11.87   &  59.62    &  0.89   \\
SCAD   &  43.37   &  11.94   &  31.43    &  0.95    &  43.88   &  11.86   &  32.02  &  0.89    \\
cMCP   &  36.92   &  11.94   &  24.98   &  0.95   &  30.08   &  11.79   &  18.29    &  0.84  \\
gBridge   &  23.68   &  1.29   &  22.39  &  0  &  21.69   &  1.77   &  19.92
&  0 \\

SparseGL   &  940.21   &  10.9   &  929.31    &  0.44    &  895.16   &  10.36   &  884.8    &  0.38      \\

GEL   &  23.76   &  9.4   &  14.36    &  0.16  & 23.24   &  8.91   &  14.33      &  0.12   \\

EMVS   &  8.6   &  7.75   &  0.85  &  0.17    &  11.01   &  8.63   &  2.38   &  0.08   \\
SSLasso   &  11.84   &  10.75   &  1.09     &  0.58  &  10.52   &  9.81   &  0.71    &  0.4       \\
S5(piMOM)   &  10.01   &  9.5   &  0.51     &  0.52   &  9.32   &  8.65   &  0.67    &  0.42     \\
S5(peMOM)   &  11.47   &  10.91   &  0.56     &  0.69     &  10.74   &  10.08   &  0.66   &  0.55   \\
BGLSS   &  390.25   &  4.03   &  386.22     &  0  &  473   &  5.17   &  467.83    &  0   \\
BSGSSS   &  11.8   &  11.4   &  0.4   &  0.69    &  11.46   &  10.97   &  0.49      &  0.5   \\
BIVAS   &  11.92   &  11.67   &  0.25     &  0.77    &  11.75   &  11.59   &  0.16    &  0.74   \\
SVEN(MAP)   &  1.89   &  1.77   &  0.12    &  0.01   &  1.82   &  1.61   &  0.21  &  0    \\
SVEN(WAM)   &  1.81   &  1.74   &  0.07    &  0.01   &  1.76   &  1.61   &  0.15    &  0 \\
BITS   &  89.55   &  11.68   &  77.87     &  0.82   &  90.99   &  11.52   &  79.47     &  0.7  \\
MiVSA(HPM)   &  12.96   &  9.13   &  3.83     &  0.41   &  12.17   &  8.44   &  3.73    &  0.29  \\
GiVSA(HPM)   &  11.47   &  10.98   &  0.49    &  0.7   &  10.83   &  10.38   &  0.45    &  0.53  \\
GiVSA(mode)   &  11.41   &  11.07   &  0.34   &  0.73  &  10.78   &  10.4   &  0.38    &  0.59    \\
\hline
\end{tabular}
\captionof{table}{Results of Scheme 3 and Scheme 4. }
\label{tab:1}
\end{center}

\end{table}
}

From the simulation results, it is clear that the frequentist methods tend to select more number of covariates than the Bayesian methods. Cosequently, the \emph{FP} of all the frequentist methods are quite high. Among the Bayesian methods, S5, BIVAS and GiVSA have comparable performance. In particular, S5, BIVAS and GiVSA have best overall performance in Scheme 1; S5 (peMOM) and GiVSA (mode) in Scheme 2; BSGSSS, BIVAS and GiVSA (mode) in Scheme 3; and BIVAS in Scheme 4. Among the S5 methods, performance of S5 (peMOM) is marginally better than S5 (piMOM). Among the GiVSA methods, performance of GiVSA (mode) is marginally better than GiVSA(HPM).
Observe that the performance of GiVSA (HPM) is uniformly better than MiVSA (HPM). Although the same prior setup and pMTM algorithm is applied to both, the only difference is in the choices of $\phi_{j}$ (as described in Section \ref{sec:PVSA}). This difference is due to the group information employed via the group probability vector.

\subsubsection{(II) Group Selection}\label{sec:GrpSelect}

For comparison of the group selection algorithms, we consider four simulation settings.
As in Schemes 1 and 2 above, we consider Gaussian and multivariate $t_{(4)}$ covariates with scale matrix $\boldsymbol{\Sigma}_{X}$ of block AR(1) settings.
In each of the simulation schemes, we consider $40$ blocks. There are $10$ blocks in four groups with sizes $20$, $40$, $60$ and $80$, respectively.
The AR(1) probability of the $10$ groups of each size ranges from $0.05$ to $0.95$ with a difference of $0.1$ between two consecutive groups. There are eight  active groups, two active groups of each size with probabilities $0.25$ and $0.75$, respectively.

With respect to sparsity within a group, two types of settings are considered, dense, where the true covariates are densely populated in the active groups, and sparse, where a small proportion of covariates in an active group is active.

The simulation schemes are described below:

\noindent {\bf Scheme 5} (AR(1)-normal-dense)
The underlying distribution of the covariates is multivariate normal with mean ${\bf 0}$ and covariance matrix $\boldsymbol{\Sigma}_{X}$. Further, every third component in an active group is an active covariate. Thus, there are $7$, $14$, $20$ and $27$ active covariates in groups of sizes $20$, $40$, $60$ and $80$, respectively.

\noindent {\bf Scheme 6} (AR(1)-$t_{4}$-dense).
In this setup, the covariates are generated from a scaled $t$ distribution with degrees of freedom $4$, with scale matrix $\boldsymbol{\Sigma}_{X}$. The other specifications are same as in Scheme 5.

\afterpage{

\begin{table}[h]
\begin{small}
    \centering
    \renewcommand*{\arraystretch}{1.2}
    \begin{tabular}{| c | c | c | c | c | c | c | c | c |}\hline
     & \multicolumn{2}{|c|}{Scheme 5} & \multicolumn{2}{|c|}{Scheme 6} & \multicolumn{2}{|c|}{Scheme 7} & \multicolumn{2}{|c|}{Scheme 8} \\ \hline
    Methods & \emph{TP}   & \emph{TPa}  & \emph{TP}   & \emph{TPa} &\emph{TP}   & \emph{TPa} & \emph{TP}   & \emph{TPa} \\
    \hline
gLASSO    &  4.26   &  6  &  3.88   &  8  &  4.43   &  8 &  3.7   &  8   \\
gSCAD  &  4.27   &  6  &  3.8   &  7 &  4.54   &  8  &  3.74   &  8   \\
gMCP   &  3.26   &  7  &  2.62   &  6  &  3.55   &  7   &  2.77   &  8 \\
cMCP   &  3.54   &  7  &  3.19   &  7  &  7.53   &  8  &  4.6   &  8    \\
gBridge &  2.56   &  3  &  2.96   &  4  &  3.34   &  6    &  2.66   &  5    \\
SparseGL &  5.45   &  8  &  5.67   &  8  &  6.41   &  8 & 4.95   &  8   \\
GEL   &  2.56   &  6  &  2.83   &  7  &  6.16   &  8   &  3.16   &  8   \\
BGLSS   &  1.8   &  2  &  1.6   &  2  &  1.7   &  2  &  1.77   &  2   \\
BSGSSS  &  5.29   &  8  &  5.52   &  8 &  7.25   &  8  &  4.03   &  8    \\
BIVAS   &  2.24   &  6   &  2.23   &  7  &  5.99   &  8  &  2.37   &  8   \\
GSIS$_{1}$  &  4.48   &  8  &  4.26   &  8 &  4.81   &  8 &  4.22   &  8    \\
 \hline
    \end{tabular}
    \caption{Results for Scheme 5, 6, 7 and 8}
    \label{tab:7}
\end{small}
\end{table}
}

\noindent {\bf Scheme 7} (AR(1)-normal-sparse)
The covariates are generated from multivariate normal distribution with mean ${\bf 0}$ and covariance matrix $\boldsymbol{\Sigma}_{X}$. Further, every $10$-th component in an active group is an active covariate. Thus, there are $2$, $4$, $6$ and $8$ active covariates in groups of sizes $20$, $40$, $60$ and $80$, respectively.

\noindent {\bf Scheme 8} (AR(1)-$t_{4}$-sparse)
This setup is similar to Scheme 7, except that the covariates are now generated from a scaled $t$ distribution with degrees of freedom $4$, with scale matrix $\boldsymbol{\Sigma}_{X}$.

The other specifications of Schemes 5-8, like choices of $\boldsymbol{\beta}_{\bgmo}$ and $\sigma^{2}$, etc., are as in Schemes 1-4.
Results of Schemes 5-8 are provided in Table \ref{tab:7}.

Across all the simulation schemes, \emph{SparseGL} yields the best overall performance, followed by \emph{BSGSSS}. All the sparse bi-level selection methods, viz., \emph{gLASSO, gSCAD, cMCP, GEL, BIVAS}, work well under the sparse normal setup. However, these methods fail to perform well in terms of \emph{TP} even under the sparse-$t_{(4)}$ setup (Scheme 3), and also in the dense setups with respect to both the measures. \emph{GSIS}$_{1}$ shows consistent performance over dense as well as sparse setups, and both under Gaussian and $t_{(4)}$ covariates. While about $4$-$5$ active groups are listed among the top $8$ groups with respect to GSIS inclusion probability with $d=1$ on an average, the $8$ active groups receive the highest average GSIS probability over the $N$ random repetitions.

\subsubsection{(III) Time comparison}\label{sec:Time}

In Section \ref{sec:GiVSA}, we theoretically demonstrated the advantage of using group information in terms of mixing time for a simple random walk based SSVS algorithm. Now, we demonstrate the same numerically with respect to the computational time.
This analysis is different from the mixing time analysis. While mixing time can be interpreted as the number of iterations required for convergence, here we consider a fixed number of iterations $M= 5e3$ for all the methods. Therefore, this analysis highlights the differences in per iteration computing time (on an average).

For this comparison, we consider two variants of the proposed g-prior method with the pMTM algorithm, viz., GiVSA and MiVSA. As described in Section \ref{sec:PVSA}, the only difference between GiVSA and MiVSA lies in the choice of weight function $\phi_{j}$ in the pMTM algorithm.

For runtime comparison we consider two new situations. First, the case when the number of groups increases as $(n,p)$ increases keeping the group sizes fixed (\emph{Scheme 9}), and second, the case where the group sizes increase as $(n,p)$ increases keeping the number of groups fixed  (\emph{Scheme 10}). In both the schemes, we take seven nearly equidistant choices of $n$ ranging from $150$ to $300$, and $p=an^{t}$, where $a=15e-2$ and $t=2$. In each simulation scheme there are groups of four different sizes. There are twelve active covariates in the true model, each from a separate group, and groups of each size contain three active covariates. Details of the simulation schemes are given below:

\noindent {\bf Scheme 9.} There are groups of $4$ different sizes $25$, $50$, $75$ and $100$.
As $p$ increases, the number of groups of each size increase equally.
The other aspects are similar to Scheme 3 (described above), except that the common correlation is $0.5$ for all the groups.

\noindent {\bf Scheme 10.} There are $60$ groups, where each set of $4$ consecutive groups have sizes in $1:2:3:4$ ratio. This simulation scheme is similar to Scheme 3 (described above), except that the common probabilities of the $15$ groups of each size now ranges from $0.2$ to $0.9$ with an interval of $0.05$. The active covariates of the true model lies in groups of each size with probabilities $0.2$, $0.45$ and $0.7$.

\begin{figure}
    \centering
    \includegraphics[scale=0.5]{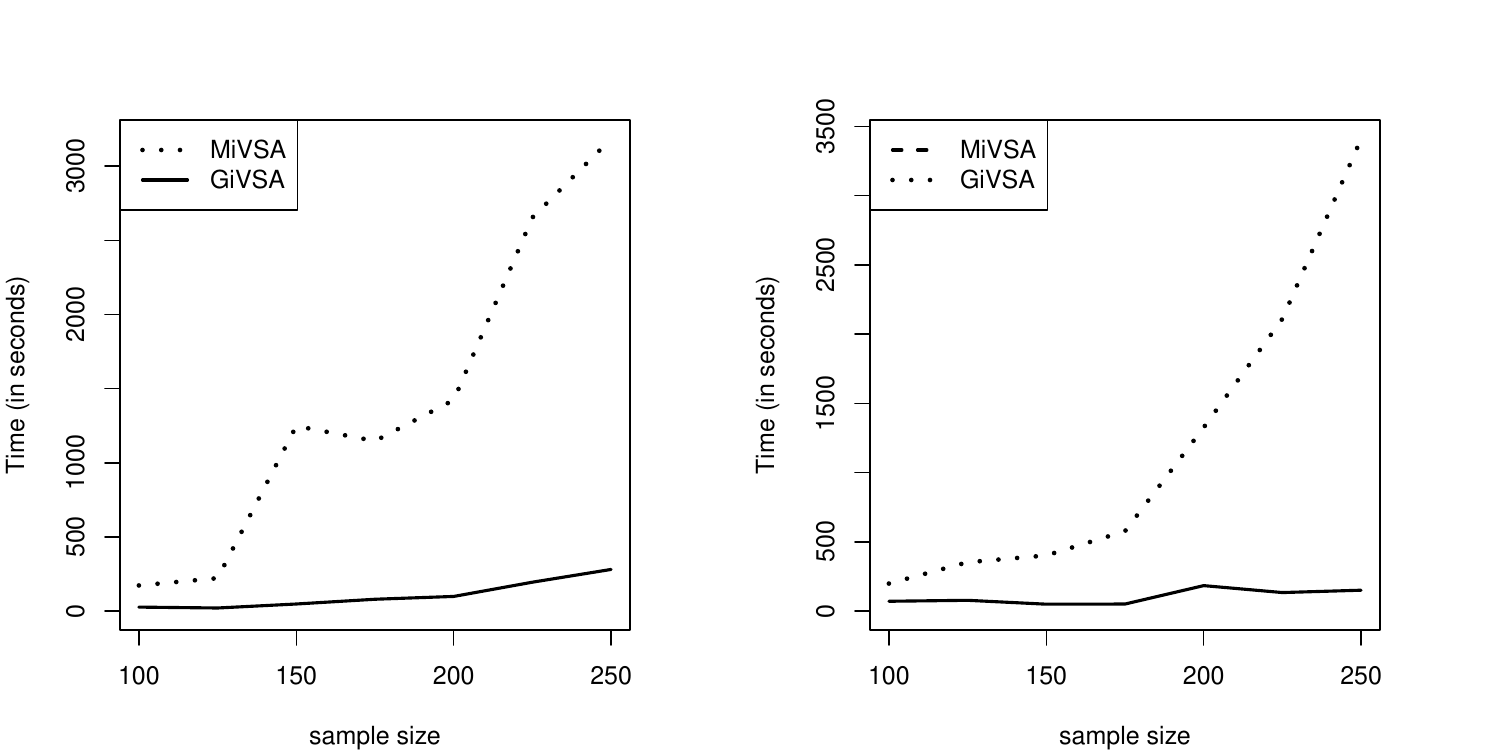}
    \caption{Time (in seconds) required to complete $5e3$ iterations as $(n,p)$ increases for GiVSA and MiVSA, in Scheme 9 (left) and Scheme 10 (right).}
    \label{fig:1}
\end{figure}

Graphs of the computational time for the competing methods are given in Figure \ref{fig:1}. Clearly, the computational time of GiVSA is much smaller than MiVSA in both the situations. If the group inclusion probabilities are not much informative, then the mixing time of GiVSA may not be much different from MiVSA. However, the runtime results suggest that even in that case, the per iteration computation time of GiVSA will be much less than MiVSA resulting a faster convergence.

\subsection{Application to the Residential Building data}\label{sec:RDA}

The Residential Building data set is
available from the UCI machine learning repository.
This data contains the construction costs and sale prices corresponding to $n=372$ real estate single-family residential apartments in Tehran, Iran. The data include $8$ physical and financial regressors, and $19$ economic regressors collected for $5$ time lags before the construction. As is evident from the data structure, the covariates form near collinear groups, which we also observe from a covariance based complete linkage hierarchical clustering result. The clustering result shows existence of $7$ groups of sizes, $72$, $10$, $7$, $5$, $5$, $2$, $2$, respectively (see Figure \ref{fig:2}). Considering this group structure as the underlying truth, we separately regress the construction costs and sale prices.

To validate the performance of the competing methods, we split the data into two parts of sizes $272$ and $100$, respectively. The first part is used as the training set, and the second part is used as the validation set. We repeat this experiment $N=1e2$ times, and the average number of covariates chosen by each model and the mean square prediction errors (MSPE) are reported. The results are provided in Table \ref{tab:6} \footnote{In the real data analysis, we drop EMVS because it produced errors when applied to the real data sets.
Further, MiVSA is also excluded from comparison, because GiVSA and MiVSA are variants of same g-prior method, and it is already evident from simulation results that GiVSA provides better results when a group structure exists and is known as domain knowledge. }.

\afterpage{
\begin{small}
\begin{minipage}[c]{0.55\textwidth}
    \renewcommand*{\arraystretch}{1.2}
\begin{tabular}{|p{0.25\textwidth}  | p{0.1\textwidth} | p{0.1\textwidth} | p{0.1\textwidth} | p{0.1\textwidth} |}\hline
 & \multicolumn{2}{|c|}{Construction} & \multicolumn{2}{|c|}{Sale} \\
 & \multicolumn{2}{|c|}{cost} & \multicolumn{2}{|c|}{price} \\\hline
  Methods $\downarrow$ & $\|\widehat{\bbeta}\|_{0}$  & MSPE & $\|\widehat{\bbeta}\|_{0}$  & MSPE \\
\hline
LASSO   &  24.87   &  2.24 &  30.53    &  4.14  \\
SCAD   &  18.56      &  2.50  &  18.68      &  4.66   \\
cMCP   &  19.57     &  2.32  &  19.21      &  4.31    \\
gBridge   &  19.79   &  2.34  &  33.62    &  3.85  \\
SparseGL   &  66.31   &  2.74  &  76.54   &  5.20 \\
GEL   &  45.92   &   2.80  &  27.63   &   6.61  \\
SSLasso   &  9.61     &  2.38  &  5.92     &  4.18 \\
S5(piMOM)   &  3.07   &  3.38 &  3.94   &  4.12   \\
S5(peMOM)   &  5.7   &  3.56  &  6.5    &  5.88\\
BGLSS   &  12.1   &  2.92  &  15.73      &  6.25   \\
BIVAS   &  7.11   &  2.58 &  4.53      &  4.97     \\
SVEN(MAP)   &  6.26   &  2.25 &  7.83   &  3.98   \\
SVEN(WAM)   &  5.62  &  3.80   &  7.13    &  6.50  \\
BITS   &  16.76     &  2.35 &  18.98   &  4.14  \\
GiVSA(HPM)   &  4.84     &  2.17  &  5.41     &  3.98      \\
GiVSA(mode)   &  4.84    &  2.15    &  5.40     &  4.01  \\
\hline
\end{tabular}
\captionof{table}{Results of residential building data.}
\label{tab:6}
\end{minipage}
\begin{minipage}[c]{0.4\textwidth}
\centering
    \includegraphics[height=6.5 cm, width =5.5 cm]{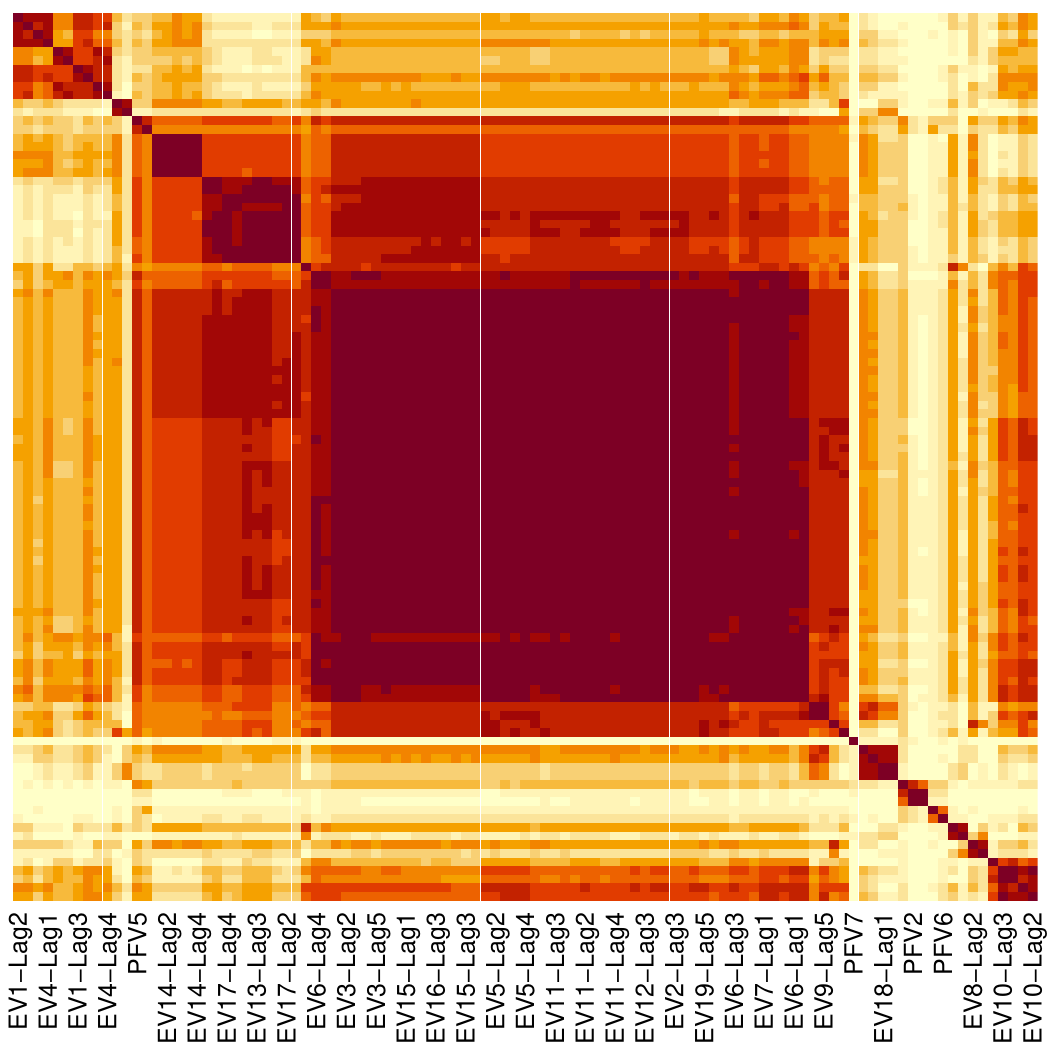}
    \captionof{figure}{Heatmap of the matrix of absolute correlation among the regressors of the Residential building data. The physical and financial variables are abbreviated as \emph{PFV1}-\emph{PFV8}, and the economic variables are abbreviated as \emph{EV1}-\emph{EV19}.}
    \label{fig:2}
\end{minipage}

\end{small}

\vskip10pt
}

For both the responses, GiVSA selects $5$ covariates on an average and yet produces the lowest MSPE among all the methods. For construction costs, MSPE is lowest for GiVSA (mode), followed by GiVSA (HPM), and the number of selected covariates is $4.84$ for both the models, which is the second lowest among the competing models. S5 (piMOM) selects about $3$ covariates on an average, however, the MSPE is almost $1.6$ times higher than GiVSA MSPEs. We observe a similar situation for the sale prices. GiVSA methods attain an average MSPE of $4$, which is the third lowest among the competing models, and $5$ covariates on an average. The lowest MSPE ($3.85$) is achieved by gBridge, which selects about $33$ covariates on an average. GiVSA (HPM) and SVEN (MAP) attains an average MSPE of $3.98$, which is slightly  better than GiVSA (mode). However, SVEN (MAP) selects about $8$ covariates on an average.

\section{Discussion}\label{sec:Discuss}

This paper deals with the classic variable selection problem in the normal linear regression setup for the high-dimensional scenario. While developing a new methodology for variable selection, we have acknowledged different aspects of the high-dimensional regime like multicollinearity, sparsity, and tried to adapt our method accordingly. Under the popular $g$-prior setup, we have provided sufficient conditions under which the variable selection consistency (VSC) holds when the covariates are stochastic and the true model grows with $(n,p)$. Consequently, it is now theoretically guaranteed that under a set of fairly reasonable assumptions (A1)-(A3) the posterior probability of the sparsest true model dominates all other model. While proving VSC for the proposed method, we did not make any assumption on the group structure of the covariates. In fact, the VSC property holds even if the covariates are not grouped, or the group structure in unknown. Under a similar set up, it is also possible to prove VSC for the popular spike and slab prior.

VSC is a necessary property, that any good Bayesian model selection procedure should possess. However, implementation of any method remains challenging due to the huge model space under consideration. In order to make the implementation fast, we use the information of group structure whenever available. On one hand, a group importance probability (GSIS) is imposed which can be helpful in discarding unimportant groups, while on the other hand an improvised SSVS algorithm (GiVSA) is proposed which ensures faster convergence of the RJMCMC to the target posterior distribution.
Ideally, one should use a combination of both the methods while implementing our model. Instead of discarding most of the covariates in a radical screening step, one may discard only the groups with extremely small probabilities, and apply GiVSA on the remaining covariates.

The fundamental difference between GiVSA and the traditional SSVS algorithm is in the proposal transition function when an `addition' (or `swap') move is taken. In an SSVS algorithm, a significant proportion of time is spent on searching for a new covariate that can be included in the model. However, this proposed small change proves to be very effective as we have seen theoretically in Section \ref{sec:GiVSA}. Although, we mainly deal with the traditional RJMCMC based on random walk in this section, the proposed change can be implemented in any RJMCMC method. As an example, we may see the performance of GiVSA on the pMTM algorithm in Section \ref{sec:NumAn}. Apart from the gain in mixing time, which ensures faster convergence, we also demonstrate the gain in per iteration runtime due to the group iteration in Section \ref{sec:SE}.

To conclude, we provide a list of future directions which demands proper investigation. First of all, in this work we consider the group structure to be known. However, finding the groups among a set of covariates is a challenging and interesting problem, especially in the high-dimensional scenario. It is also interesting to investigate if the response variable can be informative in finding the groups of covariates. Another extension of this work can be made for the case of multivariate responses. When there is a vector of responses with a set of common predictors, it will pose a challenging task to identify a few common predictors among a large pool. We plan to work along these directions in the future.

\section{Appendix}\label{sec:app}
This section contains the proof of the theorems. Proofs of all the auxiliary results and lemmas are deferred to the Supplementary material.

In the proofs of all the theorems and lemmas, the notation $c$ is used as a generic symbol for constants. In the proofs of the mathematical results, we often encounter situations where existence of a constant is of importance, but the value of the constant is unimportant. In such cases, we denote the constant by $c$. Therefore, all constants denoted by $c$ are not the same.

\subsection{Proof of Theorem \ref{thm:VSC}}\label{sec:Proof_Thm1}
To show $P(M_{\bgm_{0}} \mid \bfX, \bfy) \xrightarrow{p} 1$ as $n\rightarrow\infty$, it is enough to show that
\begin{eqnarray}
\sum_{\bgm \neq \bgm_{0} } \frac{P(M_{\bgm} \mid \bfX) m_{\bgm}(\bfy\mid \bfX) }{P(M_{\bgm_{0}} \mid \bfX) m_{\bgm_{0}}(\bfy\mid \bfX) } \xrightarrow{p}  0, \notag
\end{eqnarray}
which implies
\begin{eqnarray}
\sum_{\bgm \neq \bgm_{0} }
\binom{p}{|\bgm_{0}|}\binom{p}{|\bgm|}^{-1}\hspace{-.05 in}
\left(  \frac{1}{k_{n}\sqrt{1+g_{n}}}\right)^{|\bgm|- |\bgm_{0}|} \hspace{-.05 in}
\left\{ 1+ \frac{  \bfy^{\prime} \left( \Pgmo - \Pgm\right) \bfy  }{g_{n}^{-1} \bfy^{\prime} \bfy  +  \bfy^{\prime} \left( \I - \Pgmo\right) \bfy } \right\}^{-n/2}\hspace{-.05 in}
\xrightarrow{p}  0.\label{eqn:pp_ratio}
\end{eqnarray}


We split the class of non-supermodels, $\mM_{2}$, into two sub-classes.

(a) Non-supermodels of small dimension: Fix $0<\eta <1/4$. Let $\xi=(1-b-\eta)$, then
 $\mM_{2}^{\prime}=\left\{ \bgm : ~ |\bgm| \lesssim n^{\xi} ~\mbox{and } {\bgm_{0}} \nsubseteq \bgm \right\}$.

(b) Non-supermodels of large dimension: $\mM_{2}^{\prime \prime}=\left\{ \bgm: \bgm \in \mM_{2} ~\text{and}~ \bgm \notin \mM_{2}^{\prime} \right\}$.

\begin{proof}[Proof of VSC over $\mM_{2}^{\prime}$:]
We will find probabilistic upper bounds for each of the quantities in (\ref{eqn:pp_ratio}).

Let $\nu_{n}$ be the probability measure corresponding to $\bfX$. By the assumptions of the theorem, $\nu_{n}$ satisfies assumptions (A2) and (A3). Define
\vspace{-.1 in}
\begin{multline}
A_{n}=\{ {\bf M} : \textstyle(n |\bgm_{0}|)^{-1} \textstyle( {\bf M}_{\bgmo} \bbeta_{\bgmo})^{\prime} {\bf M}_{\bgmo} \bbeta_{\bgmo}  < M_{0}\}  \\ \cap  \{{\bf M}: \inf_{\bgm\in \mM_{2}^{\prime}}n^{-1}({\bf M}_{\bgmo} \bbeta_{\bgmo})^{\prime} \{ \I - \bfP({\bf M}_{\bgm}) \} {\bf M}_{\bgmo} \bbeta_{\bgmo} >  \delta\}
\end{multline}

\vspace{-.15 in}
\noindent Then $\nu_{n}(A_{n})>1-2\exp\{ -n c_{0}\wedge c_{1}\}$.

\noindent(i) Consider the term $ g_{n}^{-1} \bfy^{\prime} \bfy$:
As $\bfy=\bmu+\bvep$, we have $\bfy^{\prime} \bfy= \bmu^{\prime} \bmu + 2\bmu^{\prime} \bvep + \bvep^{\prime} \bvep $, where $\bvep \sim \mathrm{N}_{n} ({\bf 0} , \sigma^2 \I)$.
By assumption (A2), we have $(n|\bgm_{0}|)^{-1}\bmu^{\prime} \bmu \leq M_{0}$ with probability at least~$1-\exp\{-c_{0}n\}$.

Given $\bfX$, $ \bmu^{\prime} \bvep \sim \mathrm{N} (0, \sigma^{2} \bmu^{\prime} \bmu )$ and thus, by assumption (A2) for any $\epsilon>0$
\vspace{-.15 in}
    \begin{eqnarray}
P\left(n^{-1} |\bmu^{\prime}\bvep| > \sigma \epsilon \right) &=& \int P\left(n^{-1} |\bmu^{\prime}\bvep| > \epsilon \sigma \mid \bfX \right) d\nu_{n}(\bfX) \notag \\
&\leq & \int \exp\{-n^{2} \epsilon /\bmu^{\prime} \bmu \} d\nu_{n}(\bfX) \notag \\
&\leq & \int_{A_{n}} \exp\{-n^{2} \epsilon /\bmu^{\prime} \bmu \} d\nu_{n}(\bfX)
+\int_{A_{n}^{c}} \exp\{-n^{2} \epsilon /\bmu^{\prime} \bmu \} d\nu_{n}(\bfX)
\notag \\
&\leq & 2\exp\{ - n c_{0}\wedge c_{1}\} + \exp\{-n |\bgmo|^{-1}\epsilon/M_{0}\} = \exp\{-c n^{1-b}\} \label{eqn:mu_e}
\end{eqnarray}
for some $c>0$ which depends on $\epsilon$.

Finally, $\bvep^{\prime} \bvep = \sum_{i=1}^{n} \varepsilon_{i}^{2}$ with $\varepsilon_{i}^{2}$s i.i.d. from $\sigma^{2} \chi^{2}_{(1)}$ distribution and independent of $\bfX$.
Thus, for any $\epsilon>0$, by \citet[Lemma 1]{LM2000}, there exists a constant $c>0$, such that
\vspace{-.1 in}
\begin{equation}
P \left\{ n^{-1} \bvep^{\prime} \bvep > \sigma^{2} (1+ \epsilon) \right\} = P \left\{  \sigma^{-2}\bvep^{\prime} \bvep > n (1+ \epsilon) \right\} \leq \exp\{ -cn\epsilon\}.  \label{eqn:e_e}
\end{equation}

\vspace{-.1 in}
Therefore, as $g_{n}$ satisfies $|\bgmo| =o(g_{n})$, by assumption (A2), (\ref{eqn:mu_e}) and (\ref{eqn:e_e}), we have $n^{-1} g_{n}^{-1}\bfy^{\prime} \bfy \leq \epsilon$ for any $\epsilon>0$, with probability at least $1- \exp\{ - c n^{1-b} \}$ for some $c>0$.

\vskip5pt

\noindent (ii) Next, consider the term $ \bfy^{\prime}( \I - \Pgmo) \bfy$:
As $( \I - \Pgmo)\bmu = {\bf 0} $, $ \bfy^{\prime} ( \I - \Pgmo) \bfy = \bvep^{\prime} ( \I - \Pgmo) \bvep $.
Given $\bfX$, $\I-\Pgmo={\bf U} {\bf U}^{\prime}$ where ${\bf U}$ satisfies ${\bf U}^{\prime}{\bf U}=\I_{n-|\bgm_{0}|}$. As $\bvep\sim \mathrm{N}_{n} ({\bf 0}, \sigma^{2} \I)$, given $\bfX$, ${\bf v}={\bf U}^{\prime} \bvep\sim \mathrm{N}_{n-|\bgm_{0}| } ({\bf 0}, \sigma^{2} \I)$. Note that the distribution ${\bf v}$ depends on $\bfX$ only through $|\bgm_{0}|$ which is fixed. So, ${\bf v}$ is independent of $\bfX$. Further, as $ {\bf v}^{\prime}{\bf v}= \bvep^{\prime} ( \I - \Pgmo) \bvep \sim \sigma^{2} \chi^{2}_{(n-|\bgmo|)} $, by \citet[Lemma 1]{LM2000},
for any $\epsilon>0$, there exists a constant $c>0$ such that
$P\left\{ \sigma^{-2} \bvep^{\prime} ( \I - \Pgmo) \bvep  >  n (1+\epsilon) \right\} \leq \exp\{ - cn\}$ and $P\left\{ \sigma^{-2} \bvep^{\prime} ( \I - \Pgmo) \bvep  <  n (1-\epsilon) \right\} \leq \exp\{ - cn\}$.

\vskip5pt

\noindent (iii) Next, we consider $ \bfy^{\prime} ( \Pgmo - \Pgm) \bfy $:
Observe that
\vspace{-.1 in}
\begin{eqnarray}
\bfy^{\prime} \left( \Pgmo - \Pgm\right) \bfy
&=&  \bmu^{\prime} \left( \I - \Pgm\right) \bmu + 2 \bmu^{\prime} \left( \I - \Pgm\right) \bvep +  \bvep^{\prime} \left( \Pgmo - \Pgm\right) \bvep  \notag \\
&\geq &  \bmu^{\prime} \left( \I - \Pgm\right) \bmu - 2 \left|\bmu^{\prime} \left( \I - \Pgm\right) \bvep \right| -  \bvep^{\prime}  \Pgm \bvep. \notag
\end{eqnarray}
Given $\bfX\in A_{n}$, for any $\bgm \in \mM_{2}^{\prime}$, $ \bmu^{\prime} ( \I - \Pgm) \bvep \sim \mathrm{N} \left(0, \sigma^{2} \bmu^{\prime} ( \I - \Pgm) \bmu \right)$. Thus, writing $\Delta_{\bgm} :=\bmu^{\prime} ( \I - \Pgm) \bmu  $, for any $\epsilon>0$ we have
    \begin{eqnarray}
&& P \left( \sup_{\bgm \in \mM_{2}^{\prime}} n^{-1} \left|\bmu^{\prime} ( \I - \Pgm) \bvep \right| > \sigma \epsilon \mid \bfX \in A_{n} \right)\notag \\
 && \qquad \leq  \sum_{\bgm \in \mM_{2}^{\prime}} P \left(  \frac{ \left|\bmu^{\prime} ( \I - \Pgm) \bvep \right|}{\sigma \sqrt{\Delta_{\bgm}}}  >\frac{ n \epsilon}{ \sqrt{\Delta_{\bgm}} } \mid \bfX \in A_{n} \right)  \notag \\
 && \qquad  \leq  \sum_{\bgm \in \mM_{2}^{\prime}} P \left\{  \frac{ \left|\bmu^{\prime} ( \I - \Pgm) \bvep \right|}{\sigma \sqrt{\Delta_{\bgm} }}  >\frac{ \sqrt{n} \epsilon}{ \sqrt{|\bgm_{0}| M_{0}}} \mid \bfX \in A_{n} \right\}  \notag \\
 && \qquad  \leq  \sum_{\bgm \in \mM_{2}^{\prime}} \exp \left\{- c n^{1-b}\right\} ~~~\leq ~~ \exp \left\{- c n^{1-b}\right\}  \sum_{|\bgm|:\bgm\in \mM_{2}^{\prime}}  \binom{p}{|\bgm|} \notag \\
 && \qquad  \leq \exp \left\{- c n^{1-b}\right\}  \sum_{|\bgm|:\bgm\in \mM_{2}^{\prime}} p^{|\bgm|} ~~\lesssim~~ n^{tn^{\xi}} \exp \left\{- c n^{1-b}\right\} . \notag
\end{eqnarray}
Here, the constant $c>0$ depends on $M_{0}$ and $\epsilon$.
Further,
\vspace{-.1 in}
\begin{eqnarray}
&& P \left\{ \sup_{\bgm \in \mM_{2}^{\prime}} n^{-1} \left|\bmu^{\prime} ( \I - \Pgm) \bvep \right| > \sigma \epsilon \right\} \notag \\
&&\qquad =  P \left\{ \sup_{\bgm \in \mM_{2}^{\prime}} n^{-1} \left|\bmu^{\prime} ( \I - \Pgm) \bvep \right| > \sigma \epsilon \mid \bfX \in A_{n}  \right\}  P(\bfX\in A_{n}) \notag \\
&& \qquad \qquad \qquad   +
P\left[ \left\{ \sup_{\bgm \in \mM_{2}^{\prime}} n^{-1} \left|\bmu^{\prime} ( \I - \Pgm) \bvep \right| > \sigma \epsilon \right\}  \cap \left\{ \bfX\in A_{n}^{c} \right\} \right]\notag \\
&& \qquad \lesssim   n^{tn^{\xi}} \exp \left\{- c n^{1-b}\right\} + 2\exp\{-c_{0}\wedge c_{1} n\} = n^{tn^{\xi}} \exp \left\{- c n^{1-b}\right\}. \label{eqn:m_I-P_e}
\end{eqnarray}
Note here that the constant $c$ in the last $2$ steps may be different from each other.

Given $\bfX$, $\bvep^{\prime} \Pgm \bvep \sim \sigma^{2} \chi^{2}_{(\pg)}$ distribution, which depends on the rank of $\bXgm$ (say, $\rho_{\bgm}$). As $1 \leq \rho_{\bgm} \lesssim n^{\xi}$, by \citet[Lemma 1]{LM2000}, there exists a constant $c$ free of $\bfX$ such that $P \left(  \bvep^{\prime} \Pgm \bvep > n\sigma^{2} \epsilon \mid \bfX \right) \leq \exp \left\{- c n\right\}$, for any $\epsilon >0$. Therefore,
\vspace{-.1 in}
    \begin{eqnarray}
P \Big( \sup_{\bgm \in \mM_{2}^{\prime}} n^{-1} \bvep^{\prime} \Pgm \bvep  > \sigma^{2} \epsilon \Big) &=& \int  \sum_{\bgm \in \mM_{2}^{\prime}} P \left(  \bvep^{\prime} \Pgm \bvep > n\sigma^{2} \epsilon \mid \bfX \right) d\nu_{n}(\bfX)     \notag \\
&\lesssim & \int n^{tn^{\xi}} \exp \left\{- c n\right\} d\nu_{n}(\bfX) = n^{tn^{\xi}} \exp \left\{- c n\right\} \label{eqn:e_P_e}
\end{eqnarray}
for some $c>0$ depending on $n$.
Combining (\ref{eqn:m_I-P_e}) and (\ref{eqn:e_P_e}) along with assumption (A3), we have

{\centering
$P\left\{ \inf_{\bgm \in \mM_{2}^{\prime}} n^{-1} \bfy^{\prime} (\Pgmo- \Pgm)\bfy  \geq \delta/2 \right\}  \geq  1 - n^{tn^{\xi}} \exp \left\{- c n^{1-b}\right\} \quad \mbox{for some $c>0$.}$  
\par}

\noindent (iv) From (i)-(iii), for any $\epsilon>0$ we get
    \begin{eqnarray}
\frac{  \bfy^{\prime} \left( \Pgmo - \Pgm\right) \bfy  }{g_{n}^{-1} \bfy^{\prime} \bfy  +  \bfy^{\prime} \left( \I - \Pgmo\right) \bfy } = \frac{ n^{-1} \bfy^{\prime} \left( \Pgmo - \Pgm\right) \bfy  }{n^{-1} g_{n}^{-1}  \bfy^{\prime} \bfy  + n^{-1} \bfy^{\prime} \left( \I - \Pgmo\right) \bfy } \geq \frac{\delta}{2(1+\epsilon)} \notag
\end{eqnarray}
with probability at least $1 - n^{tn^{\xi}} \exp \left\{- c n^{1-b}\right\}$, for some $c>0$ depending on $\epsilon>0$.
\vskip5pt

\noindent (v) Combining all the components we get
\begin{eqnarray}
&&\sup_{\bgm \in \mM_{2}^{\prime}}\frac{P(M_{\bgm} \mid \bfy, \bfX)}{P(M_{\bgm_{0}} \mid \bfy, \bfX)} \notag \\
&& \quad=   \sum_{|\bgm|: \bgm \in \mM_{2}^{\prime} }
\binom{p}{|\bgm|} \frac{\binom{p}{|\bgm_{0}|}}{\binom{p}{|\bgm|}}
\left(\frac{1}{k_{n}\sqrt{1+g_{n}}}\right)^{|\bgm|- |\bgm_{0}|}
\left\{ 1+ \frac{  \bfy^{\prime} \left( \Pgmo - \Pgm\right) \bfy  }{g_{n}^{-1} \bfy^{\prime} \bfy  + \bfy^{\prime} \left( \I - \Pgmo\right) \bfy } \right\}^{-n/2} \notag \\
&& \quad \leq \sum_{|\bgm|: \bgm \in \mM_{2}^{\prime} }   p^{|\bgm_{0}|} \left\{ k_{n}^{2} (1+g_n)\right\}^{(|\bgm_{0}|-|\bgm|)/2} \left(  1+ \frac{\delta}{3} \right)^{-n/2} \notag \\
&& \quad \leq (pk_{n} \sqrt{1+g_{n}})^{|\bgmo|} \left(  1+ \frac{\delta}{3} \right)^{-n/2} \label{eqn:ProbBoundM2}
\end{eqnarray}
with probability at least $1 - n^{tn^{\xi}} \exp \left\{- c n^{1-b}\right\}$ for some $c>0$.
Taking $\log$ of the last expression, we obtain

{\centering
$ |\bgmo| \log p +  \displaystyle\frac{|\bgmo|}{2} \log \{k_{n}^{2}(1+g_{n} )\} - \frac{n}{2} \log \left(1+\frac{\delta}{3}\right).$
\par}

\noindent The above expression converges to $-\infty$ if $\log k_{n}^{2} g_{n} = o(n^{1-b})$. Thus, (\ref{eqn:pp_ratio}) is proved for $\mM_{2}^{\prime}$.
\end{proof}

\begin{proof}[VSC over $\mM_{1}$]
The $-2/n$-th power of the last term in (\ref{eqn:pp_ratio}) is as follows:
\begin{eqnarray}
 1+ \frac{  \bvep^{\prime} \left( \Pgmo - \Pgm\right) \bvep } {g_{n}^{-1}\bfy^{\prime} \bfy  +  \bfy^{\prime} \left( \I - \Pgmo\right) \bfy }
 = 1-  \frac{  \bvep^{\prime} \left(\Pgm-  \Pgmo \right) \bvep } {g_{n}^{-1} \bfy^{\prime} \bfy  +  \bfy^{\prime} \left( \I - \Pgmo\right) \bfy } .
\notag
\end{eqnarray}
We will first provide a probabilistic upper bound of the term $\bvep^{\prime} (\Pgm-  \Pgmo ) \bvep$.
Let $d_{\bgm}=\rho_{\bgm}-\rho_{\bgmo}$, where $\rho_{\bgm}=\rho(\bfX_{\bgm})$ is the rank of $\bXgm$. If $d_{\bgm}=0$, then $ \bvep^{\prime} (\Pgm-  \Pgmo ) \bvep =0$ with probability one. Given $\bfX$, if $d_{\bgm}>0$, then $\bvep^{\prime} (\Pgm-  \Pgmo ) \bvep \sim \sigma^{2}\chi^{2}_{(d_{\bgm})}$. For any $\epsilon>0$ and a sufficiently large $n$ depending on $\epsilon$, by \citet[Lemma 1]{LM2000} we have
\begin{eqnarray*}
&& P\left\{ \sigma^{-2}\sup_{\bgm \in \mM_{1}} \bvep^{\prime} \left(\Pgm-  \Pgmo \right) \bvep  > 2(1+ \epsilon)^{2} d_{\bgm}  \log p \mid \bfX \right\} \\
&& \quad \leq \sum_{\bgm \in \mM_{1}}  P\left[ \sigma^{-2}  \bvep^{\prime} \left(\Pgm-  \Pgmo \right) \bvep  > d_{\bgm} + 2 d_{\bgm}\sqrt{ (1+\epsilon)  \log p} +   2 d_{\bgm} (1+\epsilon) \log p \mid \bfX  \right]\\
&& \quad \leq \sum_{d_{\bgm}=1}^{\pg^{\star}-|\bgm_{0}|} \binom{p-|\bgm_{0}|}{d_{\bgm}}
\exp\{ - (1+\epsilon)d_{\bgm}  \log p \} \\
&& \quad
\leq  \sum_{d_{\bgm}=1}^{\pg^{\star}-|\bgm_{0}|} (p-|\bgm_{0}|)^{d_{\bgm}}
p^{ - (1+\epsilon)d_{\bgm}  } ~ \leq ~  p^{-\epsilon} \times \frac{p}{p-1}.
\end{eqnarray*}
As the last expression depends only on $\epsilon$ and $p$, the same upper bound holds for the unconditional probability.
Therefore, with probability at least $p^{-\epsilon} \times p/(p-1)$, we have

{\centering
$\sigma^{-2}\sup_{\bgm \in \mM_{1}} \bvep^{\prime} \left(\Pgm-  \Pgmo \right) \bvep  \leq 2(1+ \epsilon)^{2} d_{\bgm}  \log p. $
\par}

\noindent Combining the results in (i) and (ii), for any $\epsilon>0$, the final term of (\ref{eqn:pp_ratio}) is bounded above by

\vskip5pt
{\centering
 $   \displaystyle\left\{ 1 - \frac{2(1+ \epsilon)^{2} d_{\bgm} \log p }{n} \right\}^{-n/2}
\leq  \exp \left\{ (1+\epsilon)^{3} d_{\bgm} \log p \right\}
= p ^{(1+\epsilon)^{3} d_{\bgm}} . $
\par}

\noindent The last inequality results from the fact that $ \log (1-x) \geq -x /(1-\delta_{1})$ for any $ x<\delta_{1}<1 $ (see, e.g., \cite{SC2011}). We apply this inequality with $x= 2(1+ \epsilon)^{2} d_{\bgm}\log p /n \leq 3t\pg^{\star} \log n/n$ which is less than $1-(1+\epsilon)^{-1}$ for any pre-assigned $\epsilon>0$, for sufficiently large $n$.

Combining the above facts and noting that $d_{\bgm} \leq |\bgm|-|\bgmo|$, we get
\begin{eqnarray}
\sup_{\bgm \in \mM_{1}}\frac{P(M_{\bgm} \mid \bfy, \bfX)}{P(M_{\bgm_{0}} \mid \bfy, \bfX)}
&\leq & \sum_{\bgm \in \mM_{1} }
\binom{p}{|\bgm_{0}|}\binom{p}{|\bgm|}^{-1}
\left(  \frac{1}{k_{n}\sqrt{1+g_{n}}}\right)^{|\bgm|- |\bgm_{0}|}
p ^{(1+\epsilon)^{3} d_{\bgm}} \notag \\
&\leq & \sum_{|\bgm|-|\bgm_{0}|=1 }^{\pg^{\star} - |\bgm_{0}|}
\binom{p-|\bgm_{0}|}{|\bgm|-|\bgm_{0}|} \binom{p}{|\bgm_{0}|}\binom{p}{|\bgm|}^{-1}
\left(  \frac{p ^{(1+\epsilon)^{3}}}{k_{n}\sqrt{1+g_{n}}}\right)^{|\bgm|- |\bgm_{0}|}
 \notag \\
 &\leq & \sum_{|\bgm|-|\bgm_{0}|=1 }^{\pg^{\star} - |\bgm_{0}|}
\binom{p-|\bgm_{0}|}{|\bgm|-|\bgm_{0}|} \binom{p}{|\bgm_{0}|}\binom{p}{|\bgm|}^{-1}
\left(  \frac{p ^{(1+\epsilon)^{3}}}{k_{n}\sqrt{1+g_{n}}}\right)^{|\bgm|- |\bgm_{0}|}
 \notag \\
 &\leq & \sum_{|\bgm|-|\bgm_{0}|=1 }^{\pg^{\star} - |\bgm_{0}|}
\binom{|\bgm|-|\bgm_{0}|+ |\bgm_{0}|}{|\bgm_{0}|}
\left(  \frac{p ^{(1+\epsilon)^{3}}}{k_{n}\sqrt{1+g_{n}}}\right)^{|\bgm|- |\bgm_{0}|}
 \notag \\
 & \leq & \left( 1- \frac{p ^{(1+\epsilon)^{3}}}{k_{n}\sqrt{1+g_{n}}} \right)^{-|\bgm_{0}|} -1. \label{eqn:ProbBoundM1}
\end{eqnarray}
Here, we use the identity $\sum_{x=0 }^{\infty} \binom{x+r-1}{r-1}y^{x}  = (1-y)^{-r}$ for any $0<y<1$ and the fact $k_{n}^{2} g_{n}\gtrsim p^{2(1+\lambda)}$. As before, we can show that the above identity is dominated above by

{\centering
$ \exp \left\{ (1+\epsilon) p ^{(1+\epsilon)^{3}} k_n^{-1} g_{n}^{-1/2} |\bgm_{0}|\right\} - 1 =  \exp \left\{ (1+\epsilon) n ^{t(1+\epsilon)^{3}+b  -t(1+\lambda) }\right\} - 1,$
\par}

\noindent where $\epsilon>0$ is arbitrary. The above quantity converges to zero if we choose $\lambda$ such that $\lambda > (1+\epsilon)^{3}+b/t -1$.
\end{proof}

\begin{proof}[VSC over $\mM_{2}^{\prime\prime}$:]
Define $\bgm \cup \bgm_{0}$ as the model containing all the covariates which are active in $\bgm $ or $\bgm_{0}$, and $\bgm\setminus \bgm_{0}$ as the model containing only those covariates which are active in $\bgm$, but not in $\bgm_{0}$. Then, $ \bgm\setminus \bgm_{0}  \subseteq \bgm \subseteq \bgm\cup \bgm_{0} $ and $|\bgm \cup \bgm_{0} |= |\bgm_{0}|+ |\bgm \setminus \bgm_{0} |$.

Now, consider the $-2/n$-th power of the last term in (\ref{eqn:pp_ratio}) as follows:
\begin{eqnarray}
\frac{ g_{n}^{-1} \bfy^{\prime} \bfy  +  \bfy^{\prime} \left( \I - \Pgm\right) \bfy } {g_{n}^{-1} \bfy^{\prime} \bfy  +  \bfy^{\prime} \left( \I - \Pgmo\right) \bfy } \geq  \frac{ g_{n}^{-1}  \bfy^{\prime} \bfy  +  \bfy^{\prime} \left( \I - \Pgmco\right) \bfy } { g_{n}^{-1} \bfy^{\prime} \bfy  +  \bfy^{\prime} \left( \I - \Pgmo\right) \bfy }
= 1-  \frac{  \bvep^{\prime} \left(\Pgmco-  \Pgmo \right) \bvep } {g_{n}^{-1} \bfy^{\prime} \bfy  +  \bfy^{\prime} \left( \I - \Pgmo\right) \bfy } .
\notag
\end{eqnarray}
Define $d_{\bgm}=\rho_{\bgm\cup\bgmo}-\rho_{\bgmo}$.
As before, it can be shown that for any $\epsilon>0$ and a sufficiently large $n$ (depending on $\epsilon$), we get
\vspace{-.1 in}
\begin{eqnarray*}
 &&P\left\{ \sigma^{-2}\sup_{\bgm \in \mM_{2}^{\prime\prime}} \bvep^{\prime} \left(\Pgmco-  \Pgmo \right) \bvep  > 2(1+ \epsilon)^{2} d_{\bgm}  \log p \mid \bfX \right\}\\
&& \qquad  \leq  \sum_{d_{\bgm}=1}^{\pg^{\star}-|\bgm_{0}|} \binom{p-|\bgm_{0}|}{d_{\bgm}}
\exp\{ - (1+\epsilon)d_{\bgm} \log p \} \\
&& \qquad \leq   \sum_{d_{\bgm}=1}^{\pg^{\star}-|\bgm_{0}|} (p-|\bgm_{0}|)^{d_{\bgm}}
p^{ - (1+\epsilon)d_{\bgm}   } ~ \leq ~ p^{-\epsilon} \times \frac{p}{p-1}.
\end{eqnarray*}
As the upper bound is free of $\bfX$ and converges to zero for any $\epsilon>0$, we conclude that

{\centering
$ \sigma^{-2}\sup_{\bgm \in \mM_{2}^{\prime\prime}} \bvep^{\prime} \left(\Pgm-  \Pgmo \right) \bvep  \leq 2(1+ \epsilon)^{2} d_{\bgm}  \log p  $
\par}

\noindent with probability at least $1-p^{1-\epsilon}/(p-1)$.

Combining the above facts and noting that $d_{\bgm} \leq |\bgm\setminus\bgmo| $, we get
\begin{eqnarray}
&&\sum_{\bgm \in  \mM_{2}^{\prime\prime}} \frac{P(M_{\bgm} \mid \bfX) m_{\bgm}(\bfy\mid \bfX) }{P(M_{\bgm_{0}} \mid \bfX) m_{\bgm_{0}}(\bfy\mid \bfX) } \notag \\
&& \qquad \leq \sum_{|\bgm \setminus \bgm_{0}|=n^{\xi}-|\bgm_{0}| }^{\pg^{\star} - |\bgm_{0}|}
\binom{p-|\bgm_{0}|}{|\bgm\setminus \bgm_{0}|} \binom{p}{|\bgm_{0}|}\binom{p}{|\bgm|}^{-1}
\left(  \frac{p ^{(1+\epsilon)^{3}}}{k_{n}\sqrt{1+g_{n}}}\right)^{|\bgm \setminus \bgm_{0}|}
 \notag \\
 && \qquad \leq  \sum_{|\bgm \setminus \bgm_{0}|= n^{\xi}-|\bgm_{0}| }^{\pg^{\star} - |\bgm_{0}|}
p^{|\bgm_{0}|}
\left(  \frac{p ^{(1+\epsilon)^{3}}}{k_{n}\sqrt{1+g_{n}}}\right)^{|\bgm \setminus \bgm_{0}|}
 \notag \\
 && \qquad \leq   p^{|\bgm_{0}|} \pg^{\star}
\left(  \frac{p ^{(1+\epsilon)^{3}}}{k_{n}\sqrt{1+g_{n}}}\right)^{n^{\xi}-|\bgm_{0}|}
\label{eqn:ProbBoundM3}
\end{eqnarray}
for sufficiently large $n$ as $ p ^{(1+\epsilon)^{3}}/ (k_{n}\sqrt{1+g_{n}}) \rightarrow 0$. Taking log of the last quantity and expressing $p,g_{n}, k_{n}$ and $|\bgm_{0}|$ in terms of $n$, we get

{\centering
$ \log n + tn^{b}\log n - \left( n^{\xi} - n^{b} \right)\left\{  1+\lambda - (1+\epsilon)^{3} \right\} t \log n , $
\par}

\noindent as $ \xi = 1-b-\eta$, $\eta <1/4$ and $b<1/3$, $\xi > b$. Finally, as $\epsilon>0$ is arbitrary, for any fixed $\lambda>0$ the above quantity converges to $-\infty$, implying variable selection consistency under $\mM_{2}^{\prime\prime}$.
\end{proof}

\vskip10pt

\subsection{Proof of Theorem \ref{res:4} and Theorem \ref{lm:GroupPrior}}\label{sec:GSIS_proofs}
The proofs of Theorem \ref{res:4} and Theorem \ref{lm:GroupPrior} depend on two lemmas, viz.,  Lemma \ref{lm:ProjMat} and \ref{lm:GroupProb} stated below.


\begin{lemma}\label{lm:ProjMat}
Let $ \bfX=[\bfX_{1} : \bfX_{2}]$ and $\bfP$, $\bfP_{k}$ and $\bfP_{(-k)}$ be the projection matrices corresponding to $\bfX$, $\bfX_{k}$ and $\bfX_{(-k)}$, respectively, for $k=1,2$, where $\bfX_{(-1)}=\bfX_{2}$ and $\bfX_{(-2)}=\bfX_{1}$.
\vspace{-.1 in}
\begin{enumerate}[(a)]
    \item Then, $ \bfP=\bfP_{k}+ \boldsymbol{\Psi}$, where $\boldsymbol{\Psi}$ is the orthogonal projection matrix corresponding to the column space of $\bfZ=(\I-\bfP_{k})\bfX_{(-k)}$, for $k=1,2$. 
\item If
  $\min_{k\in\{1,2\}} \lambda_{\min} \left( \bfX^{\prime}_{k} \bfX_{k}\right) \geq n \tau_{\min}^{\star}$~
and~ $\max_{k\in\{1,2\}} \sigma_{\max}\left(\bfX_{k}^{\prime} \bfX_{(-k)} \right) \leq n \eta_{n}$, then

  $\max_{k} \lambda_{\max} \left(\bfP_{k}\bfP_{(-k)}\bfP_{k} \right) \leq \eta_{n}^{2}/\tau_{\min}^{\star~2}$.

    Consequently, $\max_{k} \sigma_{\max}\left(\bfP_{k} \bfP_{(-k)}\right) = \max_{k} \sigma_{\max}\left(\bfP_{(-k)} \bfP_{k}\right) \leq \eta_{n}/\tau_{\min}^{\star}$.
\end{enumerate}

\end{lemma}

\begin{lemma}\label{lm:GroupProb}
Under the setup of Theorem \ref{res:4} the followings are satisfied.
\vspace{-.1 in}
\begin{enumerate}[(a)]
\itemsep-0.25em
\item Let $G_{k}$ be an active group, then {\small $ \bmu^{\prime} \bfP({\bf X}_{(k)}) \bmu \geq n \beta_{\min}^{\star} \tau_{\min}^{\star} |G_{k} \cap \bgmo| \left\{ 1- 2 \eta_{n} \sqrt{|\bgmo|} \beta_{\min}^{\star} \tau_{\min}^{\star~-1}\right\}$}.
\item Let $G_{k}$ be an inactive group then {\small $\bmu^{\prime} \bfP({\bf X}_{(k)}) \bmu \lesssim  \eta_{n}^{2} n |\bgmo|/ \tau_{\min}^{\star~2}$}.
\end{enumerate}
\end{lemma}

The proof of Theorem \ref{lm:GroupPrior} is provided first, and is followed by the proof of Theorem \ref{res:4}.

\vskip5pt
\noindent {\bf Proof of Theorem \ref{lm:GroupPrior}}
\begin{proof}[Proof of part (a)]
Observe that $\|\bfP({\bf X}_{(k)}) \bfy \|^{2} = \bmu^{\prime} \bfP({\bf X}_{(k)}) \bmu + 2 \bmu^{\prime} \bfP({\bf X}_{(k)}) \bvep+ \bvep^{\prime} \bfP({\bf X}_{(k)}) \bvep$.
By Lemma \ref{lm:GroupProb}(a), $\bmu^{\prime} \bfP({\bf X}_{(k)}) \bmu \geq n \beta_{\min}^{\star} \tau_{\min}^{\star} |G_{k} \cap \bgmo| \left( 1- 2 \eta_{n} \sqrt{|\bgmo|} \beta_{\min}^{\star} \tau_{\min}^{\star~-1}\right)$.

Next, consider the second term $\bmu^{\prime} \bfP({\bf X}_{(k)}) \bvep$, which is distributed as a normal distribution with mean $0$ and variance $\sigma^{2}\bmu^{\prime} \bfP({\bf X}_{(k)}) \bmu$. Therefore,
    \vspace{-.1 in}
\begin{eqnarray}
    P \left( \max_{k} \left| \bmu^{\prime} \bfP({\bf X}_{(k)}) \bvep \right| > \sqrt{n |\bgmo|} \log p \right) &\leq & \sum_{k} P \left( \frac{\left| \bmu^{\prime} \bfP({\bf X}_{(k)}) \bvep \right|}{\sigma\sqrt{\bmu^{\prime} \bfP({\bf X}_{(k)}) \bmu}} > \frac{\sqrt{n |\bgmo|} \log p}{\sigma \sqrt{\bmu^{\prime}  \bmu}} \right) \notag \\
    &\leq & \kn \times \exp \left\{- c\frac{n |\bgmo| (\log p)^{2}}{ \bmu^{\prime} \bmu} \right\} \notag \\
    &=& p \times \exp\{ - c(\log p)^{2} \} \rightarrow 0, \notag 
\end{eqnarray}
as there can be at most $p$ groups and $\bmu^{\prime} \bmu \lesssim n|\bgmo|$.

Finally, as $\bvep^{\prime} \bfP({\bf X}_{(k)}) \bvep$ is non-negative with probability at least $1 - p\times \exp\{ - c(\log p)^{2}\}$,
\vspace{-.2 in}
\begin{multline*}
 \min_{k\in \mA} \left\{\|\bfP({\bf X}_{(k)}) \bfy \|^{2}/ |G_{k} \cap \bgmo| \right\} \gtrsim \\
n \beta_{\min}^{\star}\tau_{\min}^{\star} (1-2 \eta_{n} \sqrt{|\bgmo|} \beta_{\min}^{\star} \tau_{\min}^{\star~-1}  - c \log p \sqrt{|\bgmo|/n})
\end{multline*}

Next, consider the denominator $\sum_{k} \|\bfP({\bf X}_{(k)}) \bfy  \|^{2} = \bfy^{\prime} \left\{ \sum_{k} \bfP({\bf X}_{(k)}) \right\} \bfy $.
Consider the matrix $\sum_{k} \bfP({\bf X}_{(k)})=\bfX {\bf S}_{\mathrm{opt}}^{-1} \bfX^{\prime} $, where ${\bf S}_{\mathrm{opt}}=\diag\left\{ \bfX_{1}^{\prime} \bfX_{1}, \ldots, \bfX_{\kn}^{\prime} \bfX_{\kn}\right\}$. For any non-zero $n$-dimensional vector ${\bf z}$, we have
\vspace{-.1 in}
\begin{eqnarray*}
    \bfz^{\prime} \Big\{ \sum_{k} \bfP({\bf X}_{(k)}) \Big\} \bfz \leq \bfz^{\prime} \bfz  \times \lambda_{\max} \left( \bfX  {\bf S}_{\mathrm{opt}}^{-1} \bfX^{\prime} \right) \leq \bfz^{\prime} \bfz  \times \frac{\lambda_{\max} \left( \bfX \bfX^{\prime}  \right)}{\lambda_{\min}  ({\bf S}_{\mathrm{opt}})}.
\end{eqnarray*}
Now, $ \lambda_{\min}  ({\bf S}_{\mathrm{opt}}) = \min_{k} \lambda_{\min}  (\bfX_{k}^{\prime} \bfX_{k}) \geq n \tau_{\min}^{\star}$ and $ \lambda_{\max}(\bfX \bfX^{\prime}) = \lambda_{\max}(\bfX^{\prime} \bfX)\lesssim n^{1+r}$ with $r<1$. Thus, we get

{\centering
$     \displaystyle\bfz^{\prime} \Big\{ \sum_{k} \bfP({\bf X}_{(k)}) \Big\} \bfz \lesssim \|\bfz\|^{2} \tau_{\min}^{\star -1} n^{r}.$
\par}

\noindent By splitting $\bfy$ into $\bmu+\bvep$ and applying assumption (A2$^{\prime}$), (\ref{eqn:mu_e}) and (\ref{eqn:e_e}), we can show that $\|\bfy\|^{2} \lesssim n |\bgmo|$ with probability at least $1-\exp\{- cn^{1-b} \}$.

Combining all the above facts, we get

{\centering
$ \displaystyle \inf_{k\in \mA} q_{k} \gtrsim \frac{ \beta_{\min}^{\star} \tau_{\min}^{\star~2}\min_{k\in \mathcal{A}} |\bgmo \cap G_{k}|}{|\bgmo| n^{r}},$
\par}

\noindent with probability at least $1 - p\times \exp\{ - c(\log p )^{2}\} $.
\end{proof}

\begin{proof}[Proof of part (b)]
As in part (a), we start by splitting $\|\bfP({\bf X}_{(k)}) \bfy \|^{2}$ into three components.
By Lemma \ref{lm:GroupProb}(b), the first component is bounded above because $\bmu^{\prime} \bfP({\bf X}_{(k)}) \bmu \lesssim  \eta_{n}^{2} n |\bgmo|/ \tau_{\min}^{\star~2}$.

By part (a), the second component is also bounded above since $ \max_{k} \left| \bmu^{\prime} \bfP({\bf X}_{(k)}) \bvep \right| \lesssim \sqrt{n |\bgmo|} \log p $ with probability at least $p\times \exp\{ - c(\log p)^{2}\}$.

Finally, consider the third component $\bvep^{\prime} \bfP({\bf X}_{(k)})\bvep$. For a group $G_k$ with $|G_{k}|=p_{k}$ components, $\bvep^{\prime} \bfP({\bf X}_{(k)}) \bvep \sim \sigma^{2} \chi^{2}_{(p_{k})}$.
As $p_{k} \lesssim n^{t^{\star}}$, we have

{\centering
$\displaystyle    P\left(\max_{k} \sigma^{-2} \bvep^{\prime} \bfP({\bf X}_{(k)}) \bvep \geq 5n^{t^{\star}}  \right) \leq \sum_{k} \exp\{-c n^{t^{\star}} \} \leq \kn \exp\{- c n^{t^{\star}} \} \rightarrow 0 \mbox{ as } n \to \infty.$
\par}

Combining these, we get

{\centering
$\max_{k} \|\bfP({\bf X}_{(k)}) \bfy \|^{2} \lesssim n \left\{ \eta_{n}^{2}  |\bgmo|/ \tau_{\min}^{\star~2}+ 2 \log p\sqrt{|\bgmo|/n} + 5n^{-(1-t^{\star})} \right\} $
\par}

\noindent with probability at least $1 - p\times\exp\{ - c(\log p)^{2}\} $ for some constant $c>0$.

\vskip5pt

Next, consider the denominator $ \bfy^{\prime} \left\{ \sum_{k}\bfP({\bf X}_{(k)})\right\} \bfy $.
From part (a),

{\centering
$\|\bfP({\bf X}_{(k)}) \bfy \|^{2}  \gtrsim n \beta_{\min}^{\star}\tau_{\min}^{\star} (1-2 \eta_{n} \sqrt{|\bgmo|} \beta_{\min}^{\star} \tau_{\min}^{\star~-1}  -  c \log p \sqrt{|\bgmo|/n} ) |G_{k} \cap \bgmo|$
\par}

\noindent uniformly over $k \in \mA$, with probability at least $1 - p\times \exp\{ - c(\log p)^{2}\}$. This in turn implies that

{\centering
    $\displaystyle\sum_{k} \| \bfP({\bf X}_{(k)}) \bfy \|^{2} \geq  \sum_{k \in \mA} \|\bfP({\bf X}_{(k)}) \bfy \|^{2} \gtrsim n |\bgmo| \beta_{\min}^{\star} \tau_{\min}^{\star}.$
\par}

Combining all the above facts, we get

{\centering
$\displaystyle \sup_{k\in \mI} q_{k} \lesssim \frac{ \eta_{n}^{2}  |\bgmo|/ \tau_{\min}^{\star~2} + 2 \log p \sqrt{|\bgmo|/n} + 5n^{-(1-t^{\star})}}{ |\bgmo| \beta_{\min}^{\star} \tau_{\min}^{\star} },$
\par}

with probability at least $1 - p\times \exp\{ - c(\log p)^{2}\} $ for some constant $c>0$.
\end{proof}


\begin{proof}[Proof of part (c)]
Observe that
\vspace{-.1 in}
\begin{eqnarray*}
\inf_{k\in \mA,k^{\prime} \in \mI} \frac{ q_{k} }{ q_{k^{\prime}} } & \geq  & \frac{ \inf_{k\in \mA} q_{k} }{ \sup_{k^{\prime} \in \mI} q_{k^{\prime}} }
~ = ~ \frac{ \inf_{k\in \mA} \| \bfP(\bfX_{(k)} \bfy\|^{2} }{ \sup_{k^{\prime} \in \mI} \| \bfP(\bfX_{(k^{\prime})} \bfy\|^{2} } \\
& \geq & \frac{  \beta_{\min}^{\star}\tau_{\min}^{\star} (1-2 \eta_{n} \sqrt{|\bgmo|} \beta_{\min}^{\star} \tau_{\min}^{\star~-1}  - c \log p \sqrt{|\bgmo|/n}) \inf_{k\in \mA} |G_{k} \cap \bgmo|  }{  \left\{ \eta_{n}^{2}  |\bgmo|/ \tau_{\min}^{\star~2} + 2 \log p \sqrt{|\bgmo|/n}  + 5n^{-(1-t^{\star})} \right\} }\\
&\gtrsim & \eta_{n}^{-2} |\bgmo|^{-1} \wedge n^{1-t^{\star}} ,
\end{eqnarray*}
with probability at least $1 - p\times\exp\{ - c(\log p)^{2}\} $ from parts (a) and (b) above. As $\eta_{n}^{2} |\bgmo| \rightarrow 0$, the above ratio increases to infinity.
\end{proof}

\begin{proof}[{\bf Proof of Theorem \ref{res:4}}]
Let $G_{k}$ be an active group.

Writing $ \|\bfP({\bf X}_{(k)}) \bfy \|^{2} = \bmu^{\prime} \bfP({\bf X}_{(k)}) \bmu + 2 \bmu^{\prime} \bfP({\bf X}_{(k)}) \bvep+ \bvep^{\prime} \bfP({\bf X}_{(k)}) \bvep $, we recall by Lemma \ref{lm:GroupProb}(a), $\bmu^{\prime} \bfP({\bf X}_{(k)}) \bmu \geq n \beta_{\min}^{\star} \tau_{\min}^{\star} |G_{k} \cap \bgmo| \left\{ 1- 2 \eta_{n} \sqrt{|\bgmo|} \beta_{\min}^{\star} \tau_{\min}^{\star~-1}\right\}$.

Next, as in the proof of part (a) of Theorem \ref{lm:GroupPrior}, we argue that
    \vspace{-.1 in}
\begin{eqnarray}
    P \left( \max_{k} \left| \bmu^{\prime} \bfP({\bf X}_{(k)}) \bvep \right| > \sqrt{n |\bgmo|} \log p \right) \leq p \times \exp\{ - c(\log p)^{2} \} \rightarrow 0. \label{eqn:mu_P_e}
\end{eqnarray}
Therefore, with probability at least $1 - p\times \exp\{ - c(\log p)^{2}\} $, we have
\vspace{-.05 in}
\begin{equation}
\min_{k\in \mA} q_{k}^{d} \gtrsim n \beta_{\min}^{\star}\tau_{\min}^{\star} (1-2 \eta_{n} \sqrt{|\bgmo|} \beta_{\min}^{\star} \tau_{\min}^{\star~-1}  - c \log p \sqrt{|\bgmo|/n}) \min_{k\in \mA} \xi_{k}, \label{eqn:MGSIS1}
\end{equation}
where $\xi_{k} = |G_{k} \cap \bgmo|/p_{k}^{d}$.

Let $G_{k}$ be an inactive group.
By Lemma \ref{lm:GroupProb}(b), we have $\bmu^{\prime} \bfP({\bf X}_{(k)}) \bmu \lesssim  \eta_{n}^{2} n |\bgmo|/ \tau_{\min}^{\star~2}$.
Next consider the term $\bvep^{\prime} \bfP({\bf X}_{(k)})\bvep/p_{k}^{d}$. As $\bvep^{\prime} \bfP({\bf X}_{(k)}) \bvep \sim \sigma^{2} \chi^{2}_{(p_{k})}$, we have
\begin{eqnarray*}
    P \left(\max_{k \in \mI} \frac{\bvep^{\prime} \bfP({\bf X}_{(k)})\bvep}{\sigma^{2} p_{k}^{d} } \geq 5 p_{g}^{\star(1-d)} \log p \right) &\leq & \sum_{k\in \mI} P \left( \sigma^{-2} \bvep^{\prime} \bfP({\bf X}_{(k)})\bvep  \geq 5 p_{k} \log p \right) \\
    &\leq & \sum_{k\in \mI} \exp\left\{ - 2p_{k} \log p\right\} \leq p^{-1} \rightarrow 0,
\end{eqnarray*}
as $p\rightarrow\infty$.
Finally, consider the term $\bmu^{\prime} \bfP({\bf X}_{(k)})\bvep/p_{k}^{d}$. By Cauchy-Schwartz's inequality
\vspace{-.1 in}
 $$ p_{k}^{-d}\bmu^{\prime} \bfP({\bf X}_{(k)})\bvep \leq \sqrt{\bmu^{\prime} \bfP({\bf X}_{(k)}) \bmu ~ p_{k}^{-2d}\bvep^{\prime} \bfP({\bf X}_{(k)}) \bvep} \lesssim \eta_{n} \sqrt{n |\bgmo| p_{g}^{\star(1-d)}\log p} $$
with probability at least $1-p^{-1}$.
Combining the above facts with (\ref{eqn:mu_P_e}) we get

{\centering
$\max_{k\in \mI} q_{k}^{d}\lesssim n \left\{ \eta_{n}^{2}  |\bgmo|/ \tau_{\min}^{\star~2}+ c \eta_{n} \sqrt{|\bgmo| p_{g}^{\star(1-d)} \log p /n} + 5 p_{g}^{\star(1-d)} n ^{-1}\log p \right\} $
\par}

\noindent with probability at least $1 - p^{-1}$.

Therefore, from (\ref{eqn:MGSIS1}) and the above expression we get
$$ \inf_{k\in \mA,k^{\prime} \in \mI}~ q_{k^{\prime}}^{d ~-1} q_{k}^{d} \gtrsim \min\Big\{ \eta_{n}^{-2} |\bgmo|^{-1}, n \big(p_{g}^{\star(1-d)}\log p \big)^{-1} \Big\} \min_{k \in \mA} \xi_{k} .$$
\end{proof}
\subsection{Proof of Theorem \ref{thm:Mixing_rate}} \label{sec:Mixing_proof}

The idea of the proof of Theorem \ref{thm:Mixing_rate} follows from that of Theorem 2 of \cite{YWJ_2016}. The main difference is in the algorithm itself, and consequently, in the mixing rate. The underlying assumptions of these two theorems are also quite different. Therefore, the whole proof needs to be re-iterated even if some of the definitions, concepts and results are already stated in \cite{YWJ_2016}.
However, we have attempted to minimize repetitions as much as possible.

Let $\Tilde{\bfQ}$ be the transition matrix corresponding to the Metropolis Hastings sampler, stated in (\ref{eqn:P_MH})-(\ref{eqn:P_MH2}) for the restricted model space containing models of size at most $\ps$, and $\bfQ= (\I + \Tilde{\bfQ})/2$. The spectral gap is defined as $\mathrm{Gap}(\bfQ)=1-\max\{ \lambda_{2}(\bfQ), \lambda_{\min}(\bfQ)\}$, where $\lambda_{2} (\bfQ)$ is the second largest eigenvalue of $\bfQ$. It can be shown that
\vspace{-.05 in}
\begin{equation}
    \frac{1-\mathrm{Gap}(\bfQ)}{2\mathrm{Gap}(\bfQ)} \log \left(\frac{1}{2\upsilon} \right)\leq \tau_{\upsilon}  \leq  \frac{\log(1/\min_{\bgm\in \mM^{\star}} P(M_{\bgm}\mid \bfy)) + \log (1/\upsilon)}{\mathrm{Gap}(\bfQ)}. \label{eqn:mixing_time_bound}
\end{equation}
We will find lower bounds of $\min_{\bgm\in \mM^{\star}} P(M_{\bgm}\mid \bfy) $ and $\mathrm{Gap}(\bfQ)$ in order to obtain an upper bound of $\tau_{\upsilon}$.

\begin{lemma}\label{lm:PostRatioLowerbound}
    Under the assumptions of Theorem \ref{thm:Mixing_rate}, we have

    {\centering
    $ \displaystyle\min_{\bgm} P(M_{\bgm}|\bfy) \gtrsim \left\{ \frac{|\bgm_{0}|}{p k_{n} \sqrt{1+\gn}} \right\}^{\ps -|\bgmo|} \times \left(1+c |\bgm_{0}| \right)^{-n/2}. $
\par}

\end{lemma}

\subsubsection{Lower bound of $\mathrm{Gap} (\bfQ) $:}
A lower bound of $ \mathrm{Gap} (\bfQ)$ is provided in \cite{YWJ_2016}, which is based on the canonical path ensemble (CPE) argument of \cite{Sinclair1992}.
Before providing the lower bound, we briefly describe the notion of CPE below.

\vspace{-.2 in}
\paragraph{Canonical Path Ensemble.}
Let $\mG(\mC)=(\mM,\mE)$ be a directed graph generated by the proposed Markov chain $\mC$, where the set of vertices $\mM$ is the state space of the Markov chain and $\mE$ is the set of edges.
An ordered pair $e=(\bgm, \bgmp)$ is included as an edge if and only if the weight of $e$, namely, $\mQ(e)= P(M_{\bgm}|\bfy) P_{\mathrm{MH}}(\bgm \rightarrow \bgmp) >0$.

A canonical path ensemble (CPE) $\mT$ is a collection of paths that contains for each ordered pairs $(\bgm, \bgmp)$ of distinct vertices, a unique simple path (with no repeating vertices) $\Tggp$ in $\mG(\mC)$ which connects $\bgm$ to $\bgmp$. Any path in the CPE $\mT$ is called a canonical path.
It can be shown following \cite{Sinclair1992} that
\vspace{-.1 in}
\begin{equation}
    Gap(\bfQ)\geq \frac{1}{\bar{\rho}(\mT) l(\mT)},\label{eqn:GapUB}
\end{equation}
where $l(\mT)$ is the length of the longest path in the CPE $\mT$, and $ \bar{\rho}(\mT)$ is the path congestion parameter defined as
\vspace{-.1 in}
\begin{equation}
    \bar{\rho}(\mT) = \max_{e\in \mE} \frac{1}{\mQ(e)} \sum_{e \in \Tggp} P(M_{\bgm}|\bfy) P(M_{\bgmp}| \bfy) . \label{eqn:PCP}
\end{equation}
\paragraph{Set operations on simple paths $T_{1}$ and $T_{2}$:}
Next, we define a few notations:
\vspace{-.15 in}
\begin{enumerate}[i.]
\itemsep-0.5pt
    \item If the endpoint of the simple path $T_{1}$ matches with the starting point of another simple path $T_{2}$, then $T_{1} \cup T_{2}$ is defined as the path that connects $T_{1}$ and $T_{2}$ together.
    \item For two simple paths $T_{1}$ and $T_{2}$ with one or more connected overlapping edges, define $T_{1} \cap T_{2}$ as the path of overlapping subsets. By the definition of CPE, there will not exist more than one disconnected overlapping subsets of $T_{1}$ and $T_{2}$.
    \item Let $T_{1}$ and $T_{2}$ be simple paths such that $T_{2} \subset T_{1}$. Then, $T_{1} \setminus T_{2}$ denotes the path obtained by removing all the edges in $T_{2}$ from $T_{1}$.
   \item For any simple path $T_{1}$, the reverse simple path is denoted by $\bar{T_{1}}$.
\end{enumerate}

\paragraph{Memoryless canonical path ensemble.}
Let us assume the central set to be $\bgmo$. A set of canonical paths (say, $\mR$) is said to have memoryless property with respect to $\bgmo$ if the following two criteria are satisfied:
\vspace{-.15 in}
\begin{enumerate}[i.]
\itemsep-0.5pt
    \item For any state $\bgm$ in $\mM\setminus \{\bgmo\}$, there exists a unique path $\mT_{\bgm, \bgmo}$ in $\mR$, connecting $\bgm$ to $\bgmo$.
    \item Let $\bgms$ be an intermediate state in $\mT_{\bgm, \bgmo}$ defined in i., then the unique path $\mT_{\bgms, \bgmo}$ is a sub-path of $\mT_{\bgm, \bgmo}$.
\end{enumerate}
\vspace{-.1 in}
Due to the above characterizations, in order to specify a canonical path from $\bgm \in \mM\setminus \{\bgmo\}$ to $\bgmo$ in a memoryless CPE, it is enough to specify a transition function $\mG: \mM\setminus \{\bgmo\} \to \mM$, that maps the current state to the next state.
One may include the central state $\bgmo$ in the domain of $\mG$ by defining $\mG(\bgmo) = \bgmo$.

The function $\mG: \mM\setminus \{\bgmo\} \to \mM$ is termed as a valid transition function if there exists a memoryless CPE, corresponding to which $\mG$ is a transition function.  Finally, we conclude this part by stating a lemma from \cite{YWJ_2016}, which provides a sufficient condition for a valid transition function $\mG$.

\begin{lemma}\label{lm:lemma1_YWJ}
If a transition function $\mG: \mM\setminus \{\bgmo\} \to \mM$ satisfies that for any state $\bgm \in \mM \setminus \{\bgmo\}$, the Hamming distance between $\mG(\bgm)$ and $\bgmo$ is strictly less than that of $\bgm$ and $\bgmo$, then $\mG$ is a valid transition function.
\end{lemma}

\vspace{-.2 in}
\paragraph{Canonical Path Ensemble for Grouped Covariates.}
First, we define some partitions of the model space as follows:
\vspace{-.15 in}
\begin{enumerate}[i.]
\itemsep-0.5pt
    \item {\it Supermodel and non-supermodel.} A state $\bgm$ is said to be a supermodel if $\bgmo\subseteq \bgm$, i.e., $\bgm \in \mM_{1}\cap \mM^{\star}$ and $\bgm$ is said to be non-supermodel if $\bgm \in \mM_{1}^{c}\cap \mM^{\star}$.
    \item {\it Over-explored or under-explored.} A model $\bgm \in  \mM^{\star}$ is said to be over-explored, if components from all the active groups are present in the model. The model $\bgm \in  \mM^{\star}$ is said to be under-explored otherwise.
    \item {\it Saturated or unsaturated.} A model $\bgm \in  \mM^{\star}$ is said to be saturated if $|\bgm |= \ps$ and it is called unsaturated otherwise.
\end{enumerate}
For each $\bgm\in \mM\setminus \{\bgmo\}$, we provide $\mG(\bgm)$ in terms of two moves: {\it add} (A) and {\it remove} (R). As the {\it swap} (S) move is a composition of R and A, it is not considered separately. 
Transitions from each possible state is given below:
\vspace{-.1 in}
\begin{enumerate}[(i)]
\itemsep-0.5pt
    \item {\it Supermodel.} If $\bgm$ is a supermodel, then $\mG(\bgm)=\bgmp=\bgm\setminus \{l\}$, where $\{l\}$ is the least influential covariate in $\bgm\setminus \bgmo$. In other words,
    $ l = \argmin_{j \in \bgm\setminus \bgmo} \left\| \bfP(\bfx_{(j)})\bmu \right\|$
    and $\bfP(\bfx_{(j)})$ is the projection matrix corresponding to the $j$-th covariate.

\item {\it Saturated, non-supermodel.}
If $\bgm$ is saturated, then $\mG(\bgm)=\bgmp=\bgm\setminus \{l\}$, where $\{l\}$ is the least influential covariate in $\bgm\setminus \bgmo$, i.e.,
~$ l = \argmin_{j \in \bgm\setminus \bgmo}  \|  (\bfP_{\bgm} - \bfP_{\bgm\setminus \{j\}}) \bmu\| . $

Note that after this step either $\bgmp$ will be in the class of over-explored, unsaturated non-supermodels, or the class of under-explored, unsaturated non-supermodels.

\item  {\it Unsaturated, over-explored non-supermodel.}
If $\bgm$ is an unsaturated, over-explored non-supermodel, then among the active groups the best covariate $l$ is included, which is not presently included in the model. Thus $\mG(\bgm)=\bgmp= \bgm\cup\{ l\}$, where
$ l = \argmax_{j \in \bgmo\setminus \bgm} \|\bfP_{\bgm\cup \{j\} } \bmu \| $.

\item {\it Unsaturated, under-explored non-supermodel.}
If $\bgm$ is an unsaturated, under-explored non-supermodel, then among the active groups with no representative in $\bgm$, the best active covariate $l$ in the {\it best} unexplored group $G_{k^{\star}}$ is included. Thus $\mG(\bgm)=\bgmp= \bgm\cup\{ l\}$, where

{\centering
$ l = \argmax_{j \in G_{k^{\star}} \cap \bgmo } \|\bfP_{\bgm\cup \{j\} } \bmu \| \quad \mbox{with} ~~ k^{\star} = \argmax_{k\in \mA: \bgm\cap G_{k}=\phi} \| \bfP(\bfX_{k})\bmu \|.  $
\par}
\end{enumerate}
By Lemma \ref{lm:lemma1_YWJ}, the transition function5 $\mG$ described above is valid and gives rise to a unique memoryless set of canonical paths from any state $\mM\setminus \{\bgmo\} \to \bgmo $.
We complete the CPE by defining the following
\vspace{-.1 in}
\begin{enumerate}[i.]
\itemsep-0.5pt    \item $\mT_{\bgm \cap \bgmp}:= \mT_{\bgm,\bgmo}\cap \mT_{\bgmp,\bgmo}$,
    \quad ii. $\mT_{\bgm \setminus \bgmp}:= \mT_{\bgm,\bgmo}\setminus \mT_{\bgm\cap \bgmp}$,
    \quad iii. $\Tggp:=\mT_{\bgm \setminus \bgmp} \cup \bar{\mT}_{\bgmp \setminus \bgm}$.
    \item[iv.] The model $\bgm$ is called a precedent of $\bgmp$ if $\bgmp$ is on the canonical path $\mT_{\bgm,\bgmo}$.
    \item[v.] A pair of
states $(\bgm,\bgmp)$ is called adjacent if the canonical path $\Tggp$ is $e_{\bgm,\bgmp}$.
\item[vi.] For any $\bgm\in \mM\setminus \{\bgmo\}$, denote the set of all precedent of $\bgm$ as
$ \Lambda(\bgm) := \{ \bgms : \bgm \in \mT_{\bgms,\bgmo} \}$.
\item[vii.] $|\mT|$ denotes the length of the path $\mT$, i.e., the number of models covered by $\mT$.
\end{enumerate}
Next, consider the following lemma by \cite{YWJ_2016} which continues to hold in our case.

\vskip5pt
\begin{lemma}\label{lm:lemma2_YWJ}
\vspace{-.1 in}
\begin{enumerate}[(a)]
\itemsep-0.5pt
    \item For any $\bgm \in \mM^{\star}$, $|\mT_{\bgm,\bgmo}| \leq d_{H} (\bgm,\bgmo) \leq \ps$, where $d_{H}(\bgm,\bgmo)=|\bgm\Delta \bgmo|$ is the Hamming distance between $\bgm$ and $\bgmo$.
    \item For any $\bgm,\bgmp\in \mM^{\star}$, $ |\Tggp| \leq d_{H} (\bgm,\bgmo) + d_{H} (\bgmp,\bgmo) \leq 2 \ps$.
    \item If $\bgm$ and $\bgmp$ are adjacent and $\bgmp$ is the precedent of $\bgm$, then

    {\centering
    $ \left\{ (\bar{\bgm},\bar{\bgm}^{\prime}) :  e_{\bgm,\bgmp} \in \mT_{\bar{\bgm},\bar{\bgm}^{\prime}}   \right\} \subseteq \Lambda(\bgm) \times \mM^{\star}$.
    \par}
\end{enumerate}
\end{lemma}

\noindent{\bf Upper-bounds of $l(\mT)$ and $\bar{\rho}(\mT)$.}

\noindent From Lemma \ref{lm:lemma2_YWJ}, it is clear that $l(\mT) \leq 2 \ps$. Further, from (\ref{eqn:PCP}) and Lemma \ref{lm:lemma2_YWJ}(c) we have
\vspace{-.1 in}
\begin{eqnarray}
 \bar{\rho}(\mT) & = &  \max_{e_{\bgm,\bgmp} \in \mE} \frac{1}{\mQ(e_{\bgm,\bgmp})} \sum_{\bar{\bgm}\in \Delta(\bgm)} \sum_{\bar{\bgm}^{\prime}\in \mM^{\star}} P(M_{\bar{\bgm}}|\bfy) P(M_{\bar{\bgm}^{\prime}}| \bfy) \notag \\
 & = & \max_{e_{\bgm,\bgmp} \in \mE} \frac{1}{\mQ(e_{\bgm,\bgmp})} \sum_{\bar{\bgm}\in \Delta(\bgm) } P(M_{\bar{\bgm}}|\bfy)  \notag  \\
 & = & \max_{e_{\bgm,\bgmp} \in \mE} \frac{1}{P(M_{\bgm}|\bfy) P_{\mathrm{MH}}(\bgm \to \bgmp)} \sum_{\bar{\bgm} \in \Delta(\bgm) } P(M_{\bar{\bgm}}|\bfy) . \label{eqn:PCP2}
 \end{eqnarray}
To obtain an upper bound of $\bar{\rho}(\mT)$, we separately provide a lower bound of $P_{\mathrm{MH}}(\bgm \to \bgmp)$, and an upper bound of $P(M_{\bgm}|\bfy)^{-1} \sum_{\bar{\bgm} \in \Delta(\bgm) } P(M_{\bar{\bgm}}|\bfy)$ uniformly over the class of all $(\bgm,\bgmp)$ such that $e_{\bgm,\bgmp}$ forms an edge.


In order to provide an upper bound of $P(M_{\bgm}|\bfy)^{-1} \sum_{\bar{\bgm} \in \Delta(\bgm) } P(M_{\bar{\bgm}}|\bfy)$, we first derive an upper bound of $P(M_{\bgm}|\bfy)^{-1}  P(M_{\bar{\bgm}}|\bfy)$ over all $(\bgm,\bgmp)$ such that $e_{\bgm,\bgmp}$ forms an edge in Lemma \ref{res:1}, and then provide an upper bound of the sum in Lemma \ref{res:2}.

\begin{lemma}\label{res:1}
Let $\bgm \in \mM^{\star}$ be any model of dimension at most $p_{n}^{\star}$ and $\bgmp= \mG(\bgm)$ in the canonical map. Then, for some sequence $\{\omega_{n}\}$ such that  $\omega_{n}\uparrow\infty$ as $n\to\infty$ the following holds with probability tending to one
    \begin{eqnarray}
\frac{P(M_{\bgm}|\bfy)}{ P(M_{\bgmp}|\bfy)} \lesssim \left\{
\begin{array}{ll} p^{-1} \omega_{n}^{-1} \qquad & \mbox{ if $\bgm$ is a supermodel or saturated non-supermodel,} \\
\exp\{ -cn^{1-b}\} \qquad & \mbox{if $\bgm$ is an unsaturated non-supermodel.} \end{array} \right.  \notag
    \end{eqnarray}
\end{lemma}

\begin{lemma}\label{res:2}
    Let $\bgm \in \mM^{\star}$ be any model, and $ \Lambda(\bgm) := \{ \bgms : \bgm \in \mT_{\bgms,\bgmo} \}$. Then,
    \begin{eqnarray}
\sup_{\bgm} \frac{1}{P(M_{\bgm}|\bfy)}\sum_{\bar{\bgm} \in \Lambda(\bgm) } P(M_{\bar{\bgm}}|\bfy) \leq \left\{
\begin{array}{ll} 2 \qquad \mbox{ if $\bgm$ is a non-supermodel ,} \\
 5 \qquad  \mbox{ if $\bgm$ is a supermodel,} \end{array} \right. \notag
    \end{eqnarray}
    with probability tending to one.
\end{lemma}

Finally, a uniform lower bound to $P_{\mathrm{MH}}(\bgm \to \bgmp)$ is provided in the following lemma.

\begin{lemma}\label{res:3}
Let $\bgm \in \mM^{\star}$ be any model of dimension at most $p_{n}^{\star}$ and $\bgmp= \mG(\bgm)$ in the canonical map. Then, with probability tending to one
    \begin{eqnarray}
\inf_{e_{\bgm,\bgmp} \in \mE} P_{\mathrm{MH}}(\bgm \to \bgmp) \gtrsim \left\{
\begin{array}{llll} \min\left\{ p_{n}^{\star-1}, pw_{n} q_{\mI}^{\star}/p_g^{\star} \right\} \\
 \qquad\quad \mbox{ if $\bgm$ is a supermodel, or saturated non-supermodel,}  \\
q_{\mA}^{\star} \pg^{\star-1} \\
\qquad \quad \mbox{if $\bgm$ is an unsaturated non-supermodel.} \end{array} \right.  \notag
    \end{eqnarray}
\end{lemma}

From the Lemmas \ref{res:1}--\ref{res:3}, we get

{\centering
$\bar{\rho}(\mT) \lesssim \max\left\{ p_{n}^{\star}, \left(p w_{n} q_{\mI}^{\star} \right)^{-1}p_g^{\star}, q_{\mA}^{\star-1} p_g^{\star} \right\}$.
\par}

\noindent Thus $ Gap(\bfQ)\gtrsim p_{n}^{\star-1} \left( \max\left\{ p_{n}^{\star}, \left(p w_{n} q_{\mI}^{\star} \right)^{-1} p_g^{\star}, q_{\mA}^{\star-1} p_g^{\star} \right\}\right)^{-1}$.

Combining the above with Lemma \ref{lm:PostRatioLowerbound} we get
\begin{eqnarray*}
\tau_{\upsilon} &\lesssim & \frac{n\log\left(1+c |\bgm_{0}| \right)/2-(\ps -|\bgmo|) \log ( |\bgm_{0}|)+ (\ps -|\bgmo|) \log \left(p k_{n} \sqrt{1+\gn} \right) -\log \upsilon}{p_{n}^{\star-1} \left( \max\left\{ p_{n}^{\star}, \left(p w_{n} q_{\mI}^{\star} \right)^{-1} p_g^{\star}, q_{\mA}^{\star-1} p_g^{\star} \right\}\right)^{-1}} \\
&\lesssim & \ps \max\left\{ \ps, \left(p w_{n} q_{\mI}^{\star} \right)^{-1}p_g^{\star}, q_{\mA}^{\star-1} p_g^{\star} \right\} \times \left[ \max\left\{ n\log n , \ps \log (k_{n}\sqrt{g_{n}})\right\} -\log \upsilon\right].~\qed
\end{eqnarray*}

\noindent Proofs of Lemmas \ref{lm:PostRatioLowerbound}, \ref{res:1}, \ref{res:2} and \ref{res:3} are provided in Section \ref{sec:Proof_Mixing} of the Supplementary material.




\section{Supplement}
The supplementary material contains proofs of the results and lemmas. In particular, it contains the proofs of Result \ref{res:marginal} and Lemma \ref{lm:1} in Section \ref{sec:3}, proofs of Lemma \ref{lm:ProjMat} and \ref{lm:GroupPrior} related to the theorems in Section \ref{sec:GSIS}, proofs of Lemma \ref{lm:PostRatioLowerbound} and Lemmas \ref{res:1}-\ref{res:3} related to Theorem \ref{thm:Mixing_rate} in Section \ref{sec:GiVSA}.

\subsection{Supplement to the results in Section \ref{sec:3}}\label{sec:ML}

\subsubsection{Expression of the integrated likelihood under the $g$-prior setup}
\begin{result}\label{res:marginal}
    Consider the normal linear model under $\bgm$ as in (\ref{eqn:model}). Given $n$ independent observations from (\ref{eqn:model}) and the prior setup (\ref{eqn:priorsetup}), the integrated likelihood of $M_{\bgm}$ is
    $$ m_{\bgm}(\bfy \mid \bfX)  \propto (1+g_{n})^{-\rho_{\bgm}/2} \times \left\{ \left( 1- \frac{g_{n}}{1+g_{n}} \right) \bfy^{\prime} \bfy + \frac{g_{n}}{1+g_{n}} \bfy^{\prime} \left( \I - \Pgm\right) \bfy\right\}^{-n/2},$$
    where $\rho_{\bgm}$ is the rank of the matrix $\bXgm$.
\end{result}
\begin{proof}
By definition
    \begin{eqnarray*}
 && m_{\bgm}(\bfy \mid \bfX)\\
 && \quad  =   \int f(\bfy \mid \bfX, \bgm, \bbeta, \sigma^{2} ) \pi(\bbeta|M_{\bgm},\bfX, \sigma^2) \pi\left(\sigma^2\right) d\bbeta d\sigma^{2}     \\
 && \quad \propto \frac{1}{\gn^{\rho_{\bgm}/2}} \int \int_{\mathcal{H}} \frac{1}{\sigma^{n+\rho_{\bgm}+2}} \exp\left\{-\frac{1}{2\sigma^{2}} \left( \left\| \bfy - \bXgm \Bgamma \right\|^{2} + \frac{1}{\gn} \Bgamma^{\prime} \bXgm^{\prime} \bXgm \Bgamma  \right) \right\}
d\Bgamma d\sigma^{2}\\
 && \quad \propto  \frac{1}{\gn^{\rho_{\bgm}/2}} \int \int_{\mathcal{H}} \frac{1}{\sigma^{n+\rho_{\bgm}+2}} \exp\left[-\frac{1}{2\sigma^{2}} \left\{ \| \bfy\|^{2} + \left(1+\frac{1}{\gn} \right) \Bgamma^{\prime} \bXgm^{\prime} \bXgm \Bgamma  \right. \right. \\
 && \hspace{1.5 in} \left. \left. - 2 \bfy^{\prime}\bXgm \left(\bXgm^{\prime} \bXgm \right)^{+} \bXgm^{\prime} \bXgm \Bgamma \right\} \right]
d\Bgamma d\sigma^{2}.
\end{eqnarray*}
Here, the first expression follows from the fact that for a singular normal prior of $\Bgamma$, density exists on a hyperplane $\mathcal{H}$ such that for any vector ${\bf u}\in \mathcal{H}$, ${\bf N}^{\prime} {\bf u}={\bf 0}$, where ${\bf N}$ satisfies ${\bf N}^{\prime} \left(\bXgm^{\prime} \bXgm \right)^{+}={\bf 0}$ and $ {\bf N}^{\prime}{\bf N}= \I$ (see, e.g., \cite{Mardia_book}) and $({\bf A}^{+})^{+}={\bf A}$ for any matrix ${\bf A}$. Further, the second expression is due to an application of the rank cancellation law on the equality
$\bXgm^{\prime}\bXgm \left(\bXgm^{\prime} \bXgm \right)^{+} \bXgm^{\prime} \bXgm  =\bXgm^{\prime}\bXgm$ (see, e.g., \cite{RB_book}).

After integrating out $\Bgamma$ over $\mathcal{H}$, $m_{\bgm}(\bfy \mid \bfX)$ is obtained to be proportional to

   {\centering
   $  (1+\gn )^{-\rho_{\bgm}/2}
    \int  \displaystyle\frac{1}{\sigma^{n+2}} \exp\left\{-\frac{1}{2\sigma^{2}} \left( \| \bfy \|^{2}  - \frac{\gn}{1+\gn} \bfy^{\prime}\bXgm \left(\bXgm^{\prime} \bXgm \right)^{+}  \bXgm^{\prime}\bfy \right) \right\}
 d\sigma^{2}$
 \par}

\noindent as ${\bf A}^{+} {\bf A} {\bf A}^{+} = {\bf A}^{+}$ for any matrix ${\bf A}$. Integrating out $\sigma^{2}$, we get the desired result.
\end{proof}

\subsubsection{Proof of Lemma \ref{lm:1}} \label{sec:Proof_lm1}
\begin{proof}[Proof of (A2)]
The random vector $\bfX$ is distributed as a scaled sub-Gaussian distribution with positive definite scale matrix $\Sg$ and sub-Gaussian parameter $\kappa$, if $\bfZ=\Sg^{-1/2}\bfX$ is a sub-Gaussian random vector with sub-Gaussian parameter $\kappa$ and $\var(\bfZ)=\I$. Therefore, for any norm one vector ${\bf e}$, ${\bf e}^{\prime} \bfZ$ is a sub-Gaussian random variable with sub-Gaussian parameter less than or equal to $\kappa$ (see, e.g., \cite{Vershynin_notes}).

For any $i\in \{1,\ldots,n\}$, $\mu_{i} = \bbeta_{\bgmo}^{\prime} \bfX_{\bgmo,i} $, where $\bfX_{\bgmo,i} = {\bf Q}_{\bgmo} \bfX_{i}$ and ${\bf Q}_{\bgmo}$ is the $|\bgmo|\times p$ sub-matrix of the identity matrix taking the rows corresponding to $\bgmo$. Thus, $\textstyle\big(\bbeta_{\bgmo}^{\prime}  {\bf Q}_{\bgmo} \Sg {\bf Q}_{\bgmo}^{\prime}  \bbeta_{\bgmo} \big)^{-1/2} \mu_{i}= \textstyle\big\|   \Sg^{1/2} {\bf Q}_{\bgmo}^{\prime}  \bbeta_{\bgmo} \big\|^{-1}\bbeta_{\bgmo}^{\prime}  {\bf Q}_{\bgmo} \Sg^{1/2} \bfz_{i} $, which is sub-Gaussian. Consequently, $\textstyle\big\|   \Sg^{1/2} {\bf Q}_{\bgmo}^{\prime}  \bbeta_{\bgmo} \big\|^{-2} \mu_{i}^{2}$ is sub-exponential with sub-exponential parameter at most $2\kappa^{2}$.

As $\bfX_{i}$s are centered, $E(\mu_{i})={\bf 0}$. Therefore, $E(\mu_{i}^{2}) =\var(\mu_{i})= \big\|   \Sg^{1/2} {\bf Q}_{\bgmo}^{\prime}  \bbeta_{\bgmo} \big\|^{2} \leq  \tau_{\max} \|\bbeta_{\bgmo}\|^{2} \leq \tau_{\max} M^{2} |\bgmo|$.
Using a Bernstein type inequality (see, e.g., \cite{ALPT_2010}), we have
\begin{eqnarray*}
    P\left( \frac{\bmu^{\prime} \bmu}{n|\bgmo|}> M_{0}\right) &\leq &
    P\left[ \frac{\sum_{i} \left\{\mu_{i}^{2} - E(\mu_{i}^{2} )\right\} }{ \bbeta_{\bgmo}^{\prime}  {\bf Q}_{\bgmo} \Sg {\bf Q}_{\bgmo}^{\prime}  \bbeta_{\bgmo}  }> \frac{n|\bgmo|( M_{0} -\tau_{\max} M^{2})}{\bbeta_{\bgmo}^{\prime}  {\bf Q}_{\bgmo} \Sg {\bf Q}_{\bgmo}^{\prime}  \bbeta_{\bgmo} }\right]\\
    & \leq & P\left( \frac{\sum_{i} \left\{\mu_{i}^{2} - E(\mu_{i}^{2} )\right\} }{ \bbeta_{\bgmo}^{\prime}  {\bf Q}_{\bgmo} \Sg {\bf Q}_{\bgmo}^{\prime}  \bbeta_{\bgmo}  }> \frac{n|\bgmo| ( M_{0} - \tau_{\max} M^{2})}{\tau_{\max} \bbeta_{\bgmo}^{\prime}   \bbeta_{\bgmo} }\right)\\
    &\leq & 2 \exp \left\{-c n \zeta_{n} \left(\zeta_{n}\wedge 1 \right) \right\},
\end{eqnarray*}
where $\zeta_{n} = |\bgmo| ( M_{0} - \tau_{\max} M^{2}) /\left(\kappa^{2} \tau_{\max}^{2}  \| \bbeta_{\bgmo}\|^{2}  \right)$.
Assumption (A2) follows trivially if we choose $M_{0}> \tau_{\max} M^{2}$.
\end{proof}

\begin{proof}[Proof of (A3)]
Fix any constant $c^{\star}>0$, and let $\mM_{2}^{\prime}= \left\{ \bgm :   |\bgm|= \pg\leq K ,~ \bgmo \nsubseteq \bgm \right\} $, where $K=\lfloor c^{\star} n^{\xi} \rfloor$.
Let us partition $\mM_{2}^{\prime}$ into $K$ disjoint sets $\mM_{2}^{\prime}= \mM_{2,1} \cup \cdots \cup \mM_{2,K}$ such that $\mM_{2,k}$ contains models of dimension $k$ only.
For each $k$, we split $\mM_{2,k}$ again into $L=\binom{|\bgm_{0}|}{0}+ \cdots +\binom{|\bgm_{0}|}{k\wedge |\bgmo|-1} $ disjoint sets, with $\mM_{2,k}= \cup_{\bl} \mM_{2,k,\bl}$, where the index set $\bl$ specifies a proper subset of $\bgm_{0}$. For a fixed $\bl$, all models in $\mM_{2,k,\bl}$ has a fixed set of covariates common with $\bgmo$, given by $\bl \cap \bgmo$.

Consider any $\bXgm\in \mM_{2,k,\bl}$ for some fixed $k$ and $\bl$. We first partition $\bfX_{\bgm_{0}}$ as $ [\bfX_{1,0,\bl} : \bfX_{2,0,\bl}]$, where $\bfX_{1,0,\bl}$ contains the columns of $\bfX_{\bgm_{0}}$ which are common with the columns of $\bXgm$, and $\bfX_{2,0,\bl}$ contains the remaining columns. Then it is easy to see that
    \begin{eqnarray*}
\bfX_{\bgm_{0}}^{\prime} (\I - \Pgm)\bfX_{\bgm_{0}}  = \begin{bmatrix} {\bf 0} & {\bf 0}  \\ {\bf 0} & \bfX_{2,0,\bl}^{\prime} (\I - \Pgm)\bfX_{2,0,\bl} \end{bmatrix}
\end{eqnarray*}
as the column space of $\bfX_{1,0,\bl}$ is contained in that of $\bXgm$, which implies $(\I - \Pgm)\bfX_{1,0,\bl}={\bf 0}$.

Correspondingly, writing $\bbeta_{\bgmo}=[\bbeta_{1,0,\bl}^{\prime} : \bbeta_{2,0,\bl}^{\prime} ]^{\prime}$ and $\bmu_{2,{\bf l}}= \bfX_{2,0,\bl} \bbeta_{2,0,\bl}$ we get

{\centering
$ \bmu^{\prime} (\I - \Pgm) \bmu = \bbeta_{2,0,\bl}^{\prime}  \bfX_{2,0,\bl}^{\prime} (\I - \Pgm)\bfX_{2,0,\bl}\bbeta_{2,0,\bl} =
 \bmu_{2,\bl}^{\prime} (\I - \Pgm)\bmu_{2,\bl} := \Delta_{\bgm,2,\bl}.$
\par}

Next, define ${\bf Q}_{2,0,\bl}$ as the submatrix of $\I$ considering the rows corresponding to $\bgmo\setminus \bl$.
Then, as before one can show that $\| \Sg^{1/2} {\bf Q}_{2,0,\bl}^{\prime} \bbeta_{2,0,\bl}\|^{-1} \mu_{2,\bl,i}$ for $i=1,\ldots,n$ are i.i.d. sub-Gaussian with zero mean and sub-Gaussian parameter at most $\kappa$. Then by Hanson–Wright inequality (see, e.g., \cite{RV_2013}), given $(k,\bl)$ and $\bXgm$, for any $c_{0}>0$ we have
\begin{multline*}
P\left[ \frac{ \left| \Delta_{\bgm,2,\bl} - E\left\{ \Delta_{\bgm,2,\bl} \mid \bXgm\right\} \right|}{\| \Sg^{1/2} {\bf Q}_{2,0,\bl}^{\prime} \bbeta_{2,0,\bl}\|^{2}} > c_{0} \mid  \bXgm\right]
 \leq   2\exp\left\{ - \frac{c_{0}^{2} c}{\kappa^{4} (n-|\bgm|)} \right\}
\leq
2\exp\left\{ - \frac{c_{0}^{2} c}{\kappa^{4} n} \right\},
\end{multline*}
for some constant $c>0$.
Next, observe that $\mathrm{var}( \bmu_{2,\bl}) =  \bbeta_{2,0,\bl}^{\prime} \Sg_{2,0,\bl} \bbeta_{2,0,\bl}$, where $\Sg_{2,0,\bl} = {\bf Q}_{2,0,\bl} \Sg {\bf Q}_{2,0,\bl}^{\prime}$. Using Cauchy's interlacing theorem (see, e.g., \cite{Hwang_2004}), we get
    \begin{multline*}
    E\left( \Delta_{\bgm,2,\bl} \mid \bXgm\right) =
    \mathrm{trace} \left\{(\I - \Pgm)\mathrm{var}( \bmu_{2,\bl}) \right\} \\
    \geq  \tau_{\min} (n-|\bgm|) \|\bbeta_{2,0,\bl}\|^{2}
     \geq  \tau_{\min} (n-K) \|\bbeta_{2,0,\bl}\|^{2}.
\end{multline*}
Further, by proposition

{\centering
$\tau_{\min} m^{2}  \leq   \tau_{\min} \|\bbeta_{2,0,\bl}\|^{2} \leq \| \Sg^{1/2} {\bf Q}_{2,0,\bl}^{\prime} \bbeta_{2,0,\bl}\|^{2} = \bbeta_{2,0,\bl}^{\prime} \Sg_{2,0,\bl} \bbeta_{2,0,\bl} \leq \tau_{\max}\|\bbeta_{2,0,\bl}\|^{2}. $
\par}

\noindent Therefore, given $(k,\bl)$ and $\bXgm$, we get
\begin{flalign*}
  & P\left[ \frac{1}{n} \bmu^{\prime} (\I - \Pgm) \bmu \leq \delta  \mid \bXgm \right] \leq P\left[ \frac{\bmu^{\prime} (\I - \Pgm) \bmu}{ \| \Sg^{1/2} {\bf Q}_{2,0,\bl}^{\prime} \bbeta_{2,0,\bl}\|^{2}}  \leq \frac{n\delta}{\tau_{\min} m^{2} }  \mid \bXgm \right]\\
  &\quad = P\left[  \frac{\Delta_{\bgm,2,\bl} - E\left( \Delta_{\bgm,2,\bl} \mid \bXgm\right) }{\| \Sg^{1/2} {\bf Q}_{2,0,\bl}^{\prime} \bbeta_{2,0,\bl}\|^{2}}  \leq \frac{n\delta}{\tau_{\min} m^{2}} - \frac{E\left( \Delta_{\bgm,2,\bl} \mid \bXgm\right) }{\| \Sg^{1/2} {\bf Q}_{2,0,\bl}^{\prime} \bbeta_{2,0,\bl}\|^{2}}   \mid \bXgm  \right] \\
 &\quad\leq P\left[  \frac{\Delta_{\bgm,2,\bl} -E\left( \Delta_{\bgm,2,\bl} \mid \bXgm\right) }{\| \Sg^{1/2} {\bf Q}_{2,0,\bl}^{\prime} \bbeta_{2,0,\bl}\|^{2}} \leq   \frac{n\delta}{\tau_{\min} m^{2}} -  \frac{\tau_{\min} (n-K)}{\tau_{\max} }  \mid \bXgm  \right].
\end{flalign*}
One can choose $\delta$ in such a way that $\tau_{\min}^{-1} M^{-2} n\delta -  \tau_{\max}^{-1}\tau_{\min} (n-K)= - nc_{0}$, for some $c_{0}>0$.
Combining these, given $(k,\bl)$ and $\bXgm$, we get
    \begin{eqnarray*}
    P\left[ \frac{1}{n} \bmu^{\prime} (\I - \Pgm) \bmu \leq \delta  \mid \bXgm \right] \leq  P\left[  \frac{\Delta_{\bgm,2,\bl} -E\left( \Delta_{\bgm,2,\bl} \mid \bXgm\right) }{\| \Sg^{1/2} {\bf Q}_{2,0,\bl}^{\prime} \bbeta_{2,0,\bl}\|^{2}} \leq  - n c_{0} \right]
    \leq   2\exp\left\{ - \frac{n c_{0}^{2} c}{ \kappa^{4}  } \right\}.
\end{eqnarray*}

Observe that the upper bound of the probability is free of $\bXgm$, and therefore the unconditional probability attains the same upper bound. Finally, observe that
\vspace{-.1 in}
    \begin{eqnarray*}
        P\left( \inf_{k}\inf_{\bl}\frac{1}{n} \bmu^{\prime} (\I - \Pgm) \bmu \leq \delta  \right) &\leq &
        \sum_{k} \sum_{\bl}  P\left( \frac{1}{n} \bmu^{\prime} (\I - \Pgm) \bmu \leq \delta  \right) \\
        &\leq & 2\exp\left\{ - \frac{n c_{0}^{2} c}{ \kappa^{4} } \right\} \sum_{k=1}^{K} \binom{p}{k}
        ~\leq ~ Kp^{K} 2\exp\left\{  - \frac{n c_{0}^{2} c}{ \kappa^{4} } \right\}.
\end{eqnarray*}
As $K\leq c^{\star} n^{\xi}$, $p=n^{t}$ and $|\bgmo|=n^{b}$, taking log of the last expression, we get

{\centering
$ C_{1} + C_{2} \log n + C_{3} n^{\xi} \log n  - C_{4} n \leq -C_{5} n,  $
\par}

\noindent $C_{i}>0$ for $i=1,\ldots,5$ are constants and $C_{5}$ is suitably chosen depending on the other constants. The result follows by an application of the Borel-Cantelli lemma.
\end{proof}

\subsection{Supplement to the results in Section \ref{sec:GSIS}}

\subsubsection{Proof of Lemma \ref{lm:ProjMat}}
\begin{proof}
 \noindent{\it Part (a)}
Define $\bfZ=(\I-\bfP_{k})\bfX_{(-k)}$. Clearly, $\bfX_{k}^{\prime} \bfZ={\bf 0}$, i.e., the columns of $\bfZ$ belong to the null space of $\bfX_{k}$. Now, it is easy to see that the column spaces of $\bfX$ and $[\bfX_{k}: \bfZ]$ are the same. Therefore, $\bfP$ is also the projection matrix of $[\bfX_{k}: \bfZ]$, which is equal to $\bfP_{k}+\bfP(\bfZ)$, where $\bfP(\bfZ)$ is the projection matrix of $\bfZ$.

\vskip5pt
\noindent {\it Part (b)} First, we state a few properties of singular values as follows:
\vspace{-.1 in}
\begin{enumerate}[(i)]
\itemsep-0.25em
    \item Singular values of any matrix ${\bf W}$ is same as that of $-{\bf W}$.
    \item Let $\sigma_{\max}({\bf W})$ be the highest singular value of a matrix ${\bf W}$. Then, for any two matrices ${\bf A}$ and ${\bf B}$, $\sigma_{\max}({\bf A}+{\bf B}) \leq \sigma_{\max}({\bf A})+ \sigma_{\max}({\bf B})  $ and $\sigma_{\max}({\bf AB}) \leq \sigma_{\max}({\bf A})\times \sigma_{\max}({\bf B})$.
    \item Let ${\bf W}$ be a symmetric matrix, then $\sigma_{\max} ({\bf W}) = \left|\lambda_{\max} ({\bf W}) \right|\vee \left|\lambda_{\min} ({\bf W}) \right| $.
\end{enumerate}
\vspace{-.1 in}
Proofs of these results can be found in \cite{Bhatia_book}.

Writing $\bfX_{(-k)}={\bf W}$, observe that 
\vspace{-.1 in}
    \begin{align*}
 &\max_{k} \lambda_{\max} \left( \bfP_{k}\bfP_{(-k)}\bfP_{k}  \right)\\
 &=  \max_{k} \lambda_{\max} \left( \bfX_{k} \left( \bfX_{k}^{\prime} \bfX_{k} \right)^{-1} \bfX_{k}^{\prime} {\bf W} \left( {\bf W}^{\prime} {\bf W} \right)^{-1} {\bf W}^{\prime} \bfX_{k} \left( \bfX_{k}^{\prime} \bfX_{k} \right)^{-1} \bfX_{k}^{\prime}  \right)\\
    &\leq  \max_{k} \lambda_{\max}\left( \left( \bfX_{k}^{\prime} \bfX_{k} \right)^{-1} \right) \max_{k}  \lambda_{\max} \left( \bfX_{k} \left( \bfX_{k}^{\prime} \bfX_{k} \right)^{-1} \bfX_{k}^{\prime} {\bf W}  {\bf W}^{\prime} \bfX_{k} \left( \bfX_{k}^{\prime} \bfX_{k} \right)^{-1} \bfX_{k}^{\prime}  \right) \\
    &\leq  \max_{k} \left\{ \lambda_{\max}\left( \left( \bfX_{k}^{\prime} \bfX_{k} \right)^{-1} \right) \right\}^{2}
 \max_{k} \lambda_{\max} \left( \bfX_{k}^{\prime} {\bf W}  {\bf W}^{\prime} \bfX_{k}
 \right) = \frac{\eta_{n}^{2}}{\tau_{\min}^{\star ~2}}  .
\end{align*}
The first inequality above follows from the spectral decomposition of $\big( \bfX_{k}^{\prime} \bfX_{k} \big)^{-1}$, while the second inequality is due to the additional fact that the non-zero eigenvalues of ${\bf AB}$ and ${\bf BA}$ are same.
Therefore, $\max_{k} \sigma_{\max} (\bfP_{k} \bfP_{(-k)} ) = \max_{k} \sigma_{\max} (\bfP_{(-k)} \bfP_{k} ) \leq \eta_{n} /\tau_{\min}^{\star}$.
\end{proof}

\subsubsection{Proof of Lemma \ref{lm:GroupProb}}
\begin{proof}
[Proof of part (a)] Let $k\in \mA$, then $\bfX_{(k)}=\left[ \bfX_{G_{k} \cap \bgm_{0}}: \bfX_{G_{k} \setminus \bgm_{0}}  \right].$
By part (a) of Lemma \ref{lm:ProjMat}, we can write
$  \| \bfP({\bf X}_{(k)}) \bmu\|^{2} = \bmu^{\prime} \bfP(\bfX_{G_{k} \cap \bgm_{0}}) \bmu +\bmu^{\prime} \boldsymbol{\Psi} \bmu \geq  \bmu^{\prime} \bfP(\bfX_{G_{k} \cap \bgm_{0}})\bmu$ as $\boldsymbol{\Psi}$ is positive semi-definite.

Denoting the collection of all covariates absent in $G_{k}$ by $ G_{(-k)}$, we split $\bmu$ as
\vspace{-.1 in}
\begin{eqnarray*}
\bmu = \bfX_{\bgmo} \bbeta_{\bgmo} =
\bfX_{ G_{k} \cap \bgmo} \bbeta_{ G_{k} \cap \bgmo}+\bfX_{ G_{(-k)} \cap \bgmo} \bbeta_{ G_{(-k)} \cap \bgmo} = \bmu_{ G_{k} \cap \bgmo} + \bmu_{ G_{(-k)}\cap \bgmo}
\end{eqnarray*}
Applying the above decomposition of $\bmu$ on the expression of $\| \bfP(\bfX_{G_{k} \cap \bgm_{0}}) \bmu\|^{2}$, we get
\vspace{-.1 in}
\begin{eqnarray*}
    \| \bfP(\bfX_{G_{k} \cap \bgm_{0}}) \bmu\|^{2} &\geq & \bmu_{ G_{k} \cap \bgmo}^{\prime} \bfP(\bfX_{G_{k} \cap \bgm_{0}})\bmu_{ G_{k} \cap \bgmo} + \zeta_{n} \\
    &\geq &  \bmu_{ G_{k} \cap \bgmo}^{\prime} \bmu_{ G_{k} \cap \bgmo} + \zeta_{n}  ~ \geq ~ n |G_{k} \cap \bgmo| \beta_{\min}^{\star}\tau_{\min}^{\star} + \zeta_{n}
\end{eqnarray*}
as $n^{-1}\lambda_{\min} \big(\bfX_{\bgmo}^{\prime}\bfX_{\bgmo}\big)\geq \tau_{\min}^{\star}$ and an application of Cauchy's Interlacing Theorem.
Using Lemma \ref{lm:ProjMat} (b), we have
\vspace{-.1 in}
\begin{eqnarray*}
    \zeta_{n} &=& \bmu_{ G_{(-k)} \cap \bgmo}^{\prime} \bfP(\bfX_{G_{k} \cap \bgm_{0}})\bmu_{ G_{(-k)} \cap \bgmo} + 2\bmu_{ G_{k} \cap \bgmo}^{\prime} \bfP(\bfX_{G_{k} \cap \bgm_{0}})\bmu_{ G_{(-k)} \cap \bgmo} \\
    &\geq &  2\bbeta_{ G_{k} \cap \bgmo}^{\prime} \bfX_{G_{k} \cap \bgm_{0}}^{\prime} \bfX_{ G_{(-k)} \cap \bgmo} \bbeta_{ G_{(-k)} \cap \bgmo} \\
    &\geq & - 2n \eta_{n}  \| \bbeta_{ G_{k} \cap \bgmo}^{\prime} \| \| \bbeta_{ G_{(-k)} \cap \bgmo}^{\prime} \| \gtrsim - 2n  \sqrt{|G_{k} \cap \bgmo| |\bgmo|} \eta_{n} \beta_{\min}^{\star~2} .
\end{eqnarray*}
Here, the second last inequality is a consequence of the Cauchy-Schwartz inequality and the fact that for any matrix ${\bf A}$ and any vector ${\bf x}$, we have $\|{\bf Ax}\|\leq \|{\bf A}\|\|{\bf x}\|$, where $\|{\bf A}\|=\sigma_{\max}({\bf A}) $.

Therefore, by assumption (B2\*), we have
\vspace{-.1 in}
\begin{eqnarray*}
 \| \bfP({\bf X}_{(k)}) \bmu\|^{2} &\geq & n \beta_{\min}^{\star} \tau_{\min}^{\star} |G_{k} \cap \bgmo| \left\{ 1- 2 \eta_{n} \sqrt{|\bgmo|} \beta_{\min}^{\star} \tau_{\min}^{\star~-1}\right\}\\
 &=& n |G_{k} \cap \bgmo| \beta_{\min}^{\star}\tau_{\min}^{\star} \{1+o(1)\}
\end{eqnarray*}
\vskip5pt
\noindent {\it Proof of part (b)}
For an inactive set, we have $G_{k} \cap \bgmo=\phi$. Therefore, by part (b) of Lemma \ref{lm:ProjMat}

\vspace{-.1 in}
{\centering
    $\displaystyle\| \bfP(\bfX_{(k)}) \bmu \|^{2} = \bmu^{\prime} \bfP_{\bgmo}\bfP(\bfX_{(k)}) \bfP_{\bgmo} \bmu \leq \frac{\eta_{n}^{2}}{\tau_{\min}^{\star~2}} \|\bmu\|^{2} \lesssim \frac{\eta_{n}^{2}}{\tau_{\min}^{\star~2}} n |\bgmo|. $
\par}
\end{proof}

\subsection{Supplement to the results in Section \ref{sec:GiVSA}}
\label{sec:Proof_Mixing}

Proofs of Lemma \ref{lm:PostRatioLowerbound} and Lemmas \ref{res:1}-\ref{res:3} are provided in this section.
\vspace{-.1 in}
\subsubsection{Proof of Lemma \ref{lm:PostRatioLowerbound}}
\begin{proof}
We write

\vspace{-.1 in}
{\centering
$\min_{\bgm} P(M_{\bgm}|\bfy) = \displaystyle P(M_{\bgm_{0}}|\bfy) \min_{\bgm} \frac{P(M_{\bgm}|\bfy)}{P(M_{\bgm_{0}}|\bfy)}$
\par}
\vskip5pt

\noindent and derive an asymptotic lower bound for each component of the expression stated in the RHS.

Observe that under assumptions (A2$^{\prime}$) and (A3$^{\prime}$) (see Remark \ref{rm:VSC}), Theorem \ref{thm:VSC} can be proved in an exactly similar way. The probability statements will now be with respect to the randomness of $\bfy$ only.

From the proof of Theorem \ref{thm:VSC} (see equations (\ref{eqn:ProbBoundM2}), (\ref{eqn:ProbBoundM1}) and (\ref{eqn:ProbBoundM3})), we have
\vspace{-.1 in}
    \begin{multline*}
P(M_{\bgm_{0}}|\bfy) \geq  \min \left\{ c \exp \left\{ - p ^{(1+\epsilon)^{3}} k_n^{-1} g_{n}^{-1/2} |\bgm_{0}|\right\} , \left\{1+(pk_{n} \sqrt{1+g_{n}})^{|\bgmo|} \left(  1+ 3^{-1}\delta \right)^{-n/2} \right\}^{-1},\right. \\
 \left.  \Big[1+ \left\{  p ^{(1+\epsilon)^{3}}(k_{n}\sqrt{1+g_{n}})^{-1}\right\}^{n^{\xi}-|\bgm_{0}|} \Big]^{-1} \right\},
\end{multline*}
with probability at least $1- p^{1-\epsilon}/(p-1)$. By the choices of $k_{n}$ and $g_{n}$ stated in Theorem \ref{thm:Mixing_rate}, all the above terms converges to one with probability tending to one. Therefore, $P(M_{\bgm_{0}}|\bfy)>1/2$ for sufficiently large $n$.

To get a lower bound of the second part, recall that
\begin{align}
 \min_{\bgm} \frac{P(M_{\bgm}|\bfy)}{P(M_{\bgm_{0}}|\bfy)} \geq \hspace{.75 in} & \notag \\
  \min_{\bgm} (k_{n} \sqrt{1+g_n})^{|\bgm_{0}|-|\bgm|} &\frac{\binom{p}{|\bgm_{0}|}}{\binom{p}{|\bgm|}} \times \min_{\bgm} \left\{ 1+ \frac{\bfy^{\prime}(\Pgmo- \Pgm) \bfy}{g_{n}^{-1} \bfy^{\prime} \bfy+ \bfy^{\prime}(\I- \Pgmo) \bfy}\right\}^{-n/2}. \label{eqn:minPostProb}
\end{align}
Now, simple algebra shows that
\begin{eqnarray*}
    \binom{p}{|\bgm_{0}|} \min_{\bgm}\binom{p}{|\bgm|}^{-1} \geq  \binom{p}{|\bgm_{0}|}  \min\left\{ \min_{\bgm: |\bgm|\leq |\bgmo|}\binom{p}{|\bgm|}^{-1} , \min_{\bgm: |\bgm|> |\bgmo|}\binom{p}{|\bgm|}^{-1}  \right\} \hspace{.5 in}\\
 \geq    \min\left\{ \min_{\bgm: |\bgm|\leq |\bgmo|}\left(\frac{p-|\bgm|}{|\bgmo|}\right)^{|\bgmo|-|\bgm|} , \min_{\bgm: |\bgm|>|\bgmo|}  \left(\frac{|\bgmo|}{p-|\bgm|}\right)^{|\bgm|-|\bgmo|} \right\} .
\end{eqnarray*}
Further, the first part of RHS of (\ref{eqn:minPostProb}) is a decreasing function of $|\bgm|$.
Therefore,
\begin{equation}
\min_{\bgm\in \mM} (k_{n} \sqrt{1+g_n})^{|\bgm_{0}|-|\bgm|} \binom{p}{|\bgm_{0}|}\binom{p}{|\bgm|}^{-1} \geq
\left\{ \frac{|\bgm_{0}|}{p k_{n} \sqrt{1+\gn}} \right\}^{\ps -|\bgmo|} \notag 
\end{equation}

Next, consider the denominator of the second part of (\ref{eqn:minPostProb}). As shown in the proof of Theorem \ref{thm:VSC}, both $g_{n}^{-1} \bfy^{\prime} \bfy$ and $n^{-1}\bfy^{\prime} (\I-\Pgmo) \bfy$ converges almost surely to $0$ and $\sigma^{2}$, respectively. Using similar concentration inequalities for the $\chi^{2}$ distribution, it can be shown that the denominator is bounded below by $n\sigma^{2} (1-\epsilon)$ for any $\epsilon>0$ with probability at least $1-\exp\{ -c n \epsilon^{2} \}$.

Finally, consider the numerator $\bfy^{\prime}(\Pgmo- \Pgm) \bfy$. For $\bgm \in \mM_{2} $, we have

{\centering
    $\bfy^{\prime}(\Pgmo- \Pgm) \bfy \leq \bmu^{\prime}(\I - \Pgm) \bmu + 2  \sqrt{\bmu^{\prime}\bmu \bvep^{\prime} \bvep}+ \bvep^{\prime} \Pgmo \bvep \leq  c n |\bgm_{0}|$
    \par}

\noindent with probability at least $1- \exp\{ -cn\}$.
If $\bgm \in \mM_{1}$, we have $\bfy^{\prime}(\Pgmo- \Pgm) \bfy \leq 0$.

Combining the above facts, we get

{\centering
$ \displaystyle\min_{\bgm} \frac{P(M_{\bgm}|\bfy)}{P(M_{\bgm_{0}}|\bfy)} \geq  \displaystyle\left\{ \frac{|\bgm_{0}|}{p k_{n} \sqrt{1+\gn}} \right\}^{\ps -|\bgmo|} \times \left(1+c |\bgm_{0}| \right)^{-n/2}$.
 \par}
\end{proof}
\vspace{-.1 in}
\subsubsection{Proof of Lemma  \ref{res:1}}\label{sec:Proof_Mixing1}
\begin{proof}
{\bf Case 1: Supermodels.}
In this case, $\bgmp=\bgm \setminus \{j\}$ with $j=\argmin_{l\in \bgm \setminus \bgmo} \|\bfP(\bfx_{(l)}) \bmu  \|$. Therefore, $\bgmp$ is also a supermodel of smaller dimension and we have
    \begin{eqnarray*}
\frac{P(M_{\bgm}|\bfy)}{ P(M_{\bgmp}|\bfy)} &=& \frac{|\bgm|}{(p-|\bgm|+1) k_{n} \sqrt{1+g_{n}}}\left\{ \frac{g_{n}^{-1} \bfy^{\prime} \bfy + \bvep^{\prime} (\I - \Pgmp) \bvep}{g_{n}^{-1} \bfy^{\prime} \bfy + \bvep^{\prime} (\I - \Pgm) \bvep} \right\}^{n/2} \\
    &=& \frac{1}{u_{n}} \left\{ 1 + \frac{ \bvep^{\prime} (\Pgm - \Pgmp) \bvep}{g_{n}^{-1} \bfy^{\prime} \bfy + \bvep^{\prime} (\I - \Pgm) \bvep} \right\}^{n/2}
    \leq  \frac{1}{u_{n}} \left\{ 1 + \frac{ \bvep^{\prime} (\Pgm - \Pgmp) \bvep}{\bvep^{\prime} (\I - \Pgm) \bvep} \right\}^{n/2},
\end{eqnarray*}
where $u_{n}=(p-|\bgm|+1) k_{n} \sqrt{1+g_{n}} /|\bgm|$.

Observe that $ U_{n,\bgm,j}= \sigma^{-2} \bvep^{\prime} (\Pgm - \Pgmp) \bvep  $ and $ V_{n,\bgm,j} = \sigma^{-2} \bvep^{\prime} (\I - \Pgm) \bvep$ are independently distributed $\chi^{2}$ distributions with degrees of freedom $ 1$ and $(n-|\bgm|)$, respectively.
Therefore, for any $\epsilon>0$, by \citet[Lemma 1]{LM2000}
    \begin{eqnarray*}
    & P\left[ \sup_{\bgm} \sigma^{-2} \bvep^{\prime} (\Pgm - \Pgmp) \bvep \geq (2+\epsilon) (\ps -|\bgmo|+1) \log p \right] \hspace{2 in}  \\
    & \leq  \sum_{\bgm} P\left[ \sigma^{-2} \bvep^{\prime} (\Pgm - \Pgmp) \bvep \geq (2+\epsilon) (\ps-|\bgmo|+1) \log p \right] \\
    &  \leq  \displaystyle\sum_{|\bgm|-|\bgmo|=1}^{\ps-|\bgmo|} (p-|\bgmo|)^{|\bgm|-|\bgmo|} \exp\{-(\ps-|\bgmo|+1) \log p \}
\rightarrow 0.
\end{eqnarray*}
Further, for any $\alpha \in (0,1)$, there exists a $c>0$ such that
\begin{eqnarray*}
    P\left( \inf_{\bgm} \sigma^{-2} \bvep^{\prime} (\I - \Pgm) \bvep \leq \alpha n \right)
    &\leq  \sum_{\bgm} P\left( \sigma^{-2} \bvep^{\prime} (\I - \Pgm) \bvep \leq \alpha n  \right) \hspace{1  in} \\
    & \leq  \displaystyle\sum_{|\bgm|-|\bgmo|=1}^{\ps-|\bgmo|} (p-|\bgmo|)^{|\bgm|-|\bgmo|} \exp\{-c n \}
\rightarrow 0 \hspace{.5  in}
\end{eqnarray*}
as $\ps=n^{s}$ with $s<1-b$.
Combining these we get
\begin{eqnarray*}
\frac{P(M_{\bgm}|\bfy)}{ P(M_{\bgmp}|\bfy)} &\leq&  \frac{1}{u_{n}}\left(1+ \frac{(2+\epsilon) \ps  \log p }{\alpha n} \right)^{n/2}\\
&\leq& \frac{1}{u_{n}} \exp\left\{ (2+\epsilon) \ps (2\alpha)^{-1} \log p  \right\}\leq  \frac{\ps p^{(1+\epsilon)\ps}}{(p-\ps)k_{n}\sqrt{1+g_{n}}},
\end{eqnarray*}
using the inequality $\log(1+x)\leq x$ for any $x>0$ and $\alpha \in (0,1)$.
As as $\epsilon>0$ is arbitrary and $k_{n}^{2} g_{n} \gtrsim p^{2(1+\lambda)\ps}$ for some $\lambda>0$, the result follows.

\paragraph{Case 2: Saturated non-supermodels.}
In this case too, $\bgmp=\bgm \setminus \{j\}$ with $j= \argmin_{l \in \bgm\setminus \bgmo}  \|  (\bfP_{\bgm} - \bfP_{\bgm\setminus \{l\}}) \bmu\| $. Therefore, $\bgmp$ is also a non-supermodel of smaller dimension, and so
\begin{eqnarray*}
    \frac{P(M_{\bgm}|\bfy)}{ P(M_{\bgmp}|\bfy)} &=& \frac{|\bgm|}{(p-|\bgm|+1) k_{n} \sqrt{1+g_{n}}}\left\{ \frac{g_{n}^{-1} \bfy^{\prime} \bfy + \bfy^{\prime} (\I - \Pgmp) \bfy}{g_{n}^{-1} \bfy^{\prime} \bfy + \bfy^{\prime} (\I - \Pgm) \bfy} \right\}^{n/2} \\
    &=& \frac{1}{u_{n}} \left\{ 1 + \frac{ \bfy^{\prime} (\Pgm - \Pgmp) \bfy}{g_{n}^{-1} \bfy^{\prime} \bfy + \bfy^{\prime} (\I - \Pgm) \bfy} \right\}^{n/2}
\end{eqnarray*}
where $u_{n}=(p-|\bgm|+1) k_{n} \sqrt{1+g_{n}} /|\bgm|$.

Consider the numerator $ \bfy^{\prime} (\Pgm - \Pgmp) \bfy =\bmu^{\prime} (\Pgm - \Pgmp) \bmu + \bmu^{\prime} (\Pgm - \Pgmp) \bvep + \bvep^{\prime} (\Pgm - \Pgmp) \bvep$.
As before, we can show that $\bvep^{\prime}  (\Pgm - \Pgmp) \bvep \leq \sigma^{2} (2+\epsilon) (\ps+1) \log p$, for any $\epsilon>0$, with probability tending to one.

Next, we provide an upper bound to $\bmu^{\prime} (\Pgm - \Pgmp) \bmu$.
Let $\hat{\bbeta} (\bgm)$ be as follows

{\centering
$\hat{\bbeta} (\bgmp) =\mathrm{argmin}_{\bbeta\in \mathbb{R}^{|\bgmp|}} \| \bfX_{\bgmp} \bbeta_{\bgmp} - \bmu_{\bgmo\setminus \bgm} \|^{2}, $
\par}

\noindent where $\bmu_{\bgmo\setminus \bgm}=\bfX_{\bgmo\setminus \bgm}\bbeta_{\bgmo\setminus \bgm} $,
and $\hat{\bbeta}_{\Tilde{\bgm}} (\bgmp)$ be the sub-vector of $\hat{\bbeta} (\bgmp)$ taking the co-ordinates corresponding to $\Tilde{\bgm}$, where $\Tilde{\bgm}\subset \bgmp$.

For any $l \in \bgm\setminus \bgmo$, we have
\vspace{-.05 in}
\begin{eqnarray}
    \bmu^{\prime} (\Pgm - \bfP_{\bgm\setminus\{l\}}) \bmu
    &=& \bmu_{\bgmo\setminus \bgm}^{\prime} (\Pgm - \bfP_{\bgm\setminus\{l\}}) \bmu_{\bgmo\setminus \bgm} \notag \\
    &=& \| (\I- \bfP_{\bgm\setminus\{l\}}) \bmu_{\bgmo\setminus \bgm} \|^{2} -  \| (\I- \bfP_{\bgm}) \bmu_{\bgmo\setminus \bgm}  \|^{2} \notag \\
 &=& \left\| \bfX_{\bgm\setminus\{l\}} \hat{\bbeta} (\bgm\setminus\{l\} ) - \bmu_{\bgmo\setminus \bgm} \right\|^{2} - \left\| \bfX_{\bgm} \hat{\bbeta} (\bgm ) - \bmu_{\bgmo\setminus \bgm} \right\|^{2}. \label{eqn:1}
\end{eqnarray}
By definition, $\hat{\bbeta} (\bgm \setminus \{l\})$ is the minimizer of $\| \bfX_{\bgm\setminus\{l\}} \bbeta   - \bmu_{\bgmo\setminus \bgm} \|^{2}$. Hence, the first quantity in (\ref{eqn:1}) is less than, or equal to $\| \bfX_{\bgm\setminus\{l\}} \bbeta   - \bmu_{\bgmo\setminus \bgm} \|^{2}$ for any choice of $\bbeta$, in particular if $\bbeta =  \hat{\bbeta}_{\bgm\setminus \{l\}}(\bgm)$. Therefore, we have
\vspace{-.05 in}
\begin{eqnarray*}
    \left\| \bfX_{\bgm\setminus\{l\}} \hat{\bbeta} (\bgm \setminus \{l\})  - \bmu_{\bgmo\setminus \bgm} \right\|^{2}
    & \leq & \left\| \bfX_{\bgm\setminus\{l\}} \hat{\bbeta}_{\bgm\setminus \{l\}}(\bgm) - \bmu_{\bgmo\setminus \bgm} \right\|^{2} \\
    & = & \left\| \bfX_{\bgm} \hat{\bbeta}_{\bgm}(\bgm) - \bmu_{\bgmo\setminus \bgm} - \bfx_{(l)} \hat{\bbeta}_{l}(\bgm)\right\|^{2}\\
    & \leq & \left\| \bfX_{\bgm} \hat{\bbeta}_{\bgm}(\bgm) - \bmu_{\bgmo\setminus \bgm} \right\|^{2} + \left\| \bfx_{(l)} \hat{\bbeta}_{l}(\bgm)\right\|^{2}
\end{eqnarray*}
as $\bfx_{(l)}^{\prime} (\bfX_{\bgm} \hat{\bbeta}_{\bgm}(\bgm) - \bmu_{\bgmo\setminus \bgm} ) = 0$ for any $l\in \bgm$. From (\ref{eqn:1}), it follows that

{\centering
$    \bmu^{\prime} (\Pgm - \bfP_{\bgm\setminus\{l\}}) \bmu
     \leq   \big\| \bfx_{(l)} \hat{\beta}_{l}(\bgm) \big\|^{2} = n |\hat{\beta}_{l} (\bgm)|^{2}$
\par}

\noindent as each $\bfx_{(l)}/\sqrt{n}$ is normalized to have norm $1$ for any $l\in \bgm$.
As $j$ minimizes $\bmu^{\prime} (\I - \bfP_{\bgm \setminus\{j\}})\bmu$,
we have
\vspace{-.05 in}
\begin{eqnarray*}
    \bmu^{\prime} (\Pgm - \bfP_{\bgm\setminus\{j\}}) \bmu &\leq & \frac{1}{|\bgm\setminus \bgmo|} \sum_{ l\in \bgm\setminus \bgmo } \bmu^{\prime} (\Pgm - \bfP_{\bgm\setminus\{l\}}) \bmu \\
    &\leq & \frac{n}{|\bgm\setminus \bgmo|} \sum_{ l\in \bgm\setminus \bgmo }  |\hat{\beta}_{l} (\bgm)|^{2} = \frac{n}{\ps -| \bgmo|} \| \hat{\bbeta} (\bgm)\|^{2},
\end{eqnarray*}
given $\bgm$ is saturated.

Finally,

{\centering
$ \| \hat{\bbeta} (\bgm)\|^{2} = \big\| \left( \bfX_{\bgm}^{\prime} \bfX_{\bgm} \right)^{-1} \bfX_{\bgm}^{\prime} \bmu_{\bgmo\setminus \bgm}  \big\|^{2} \leq n^{-1} \tau_{\min}^{\star ~-1} \|\bmu\|^{2} 
.$
\par}

Combining the above, we get $\bmu^{\prime} (\bfP_{\bgm} - \bfP_{\bgmp}) \bmu \leq  (\ps -|\bgmo|)^{-1} \|\bmu\|^{2} \tau_{\min}^{\star ~-1}   $ uniformly over the saturated non-supermodels $\bgm$.

Next, consider the part $\bmu^{\prime} (\bfP_{\bgm} - \bfP_{\bgmp})\bvep$. From the above two upper bounds and using Cauchy-Schwarz's inequality, we have

{\centering
$\left|\bmu^{\prime} (\bfP_{\bgm} - \bfP_{\bgmp})\bvep \right| \lesssim  \|\bmu\| \sqrt{( \ps  -|\bgmo|)^{-1}  \tau_{\min}^{\star ~-1} ( \ps +1 )\log p} \lesssim  \|\bmu\| \sqrt{\log p}.  $
\par}

\noindent Combining the above inequalities, for any $\epsilon>0$, we have

{\centering
$ \bfy^{\prime} (\Pgm - \Pgmp) \bfy  \leq \|\bmu\|^{2} \tau_{\min}^{\star ~-1} p_{n}^{\star~-1} +c \|\bmu\| \sqrt{\log p}  + \sigma^{2} (2+\epsilon) (\ps +1) \log p.$
\par}

Next, consider the denominator

{\centering
$  \bfy^{\prime} (\I - \Pgm) \bfy =  \bmu^{\prime} (\I - \Pgm) \bmu + 2 \bmu^{\prime} (\I - \Pgm) \bvep + \bvep^{\prime} (\I - \Pgm) \bvep. $
\par}

\noindent By the separability assumption (C2), we have

{\centering
$\bmu^{\prime} (\I - \Pgm) \bmu \geq   (1-\alpha_{0}) \|\bmu\|^{2} $.
\par}

Recall that for any $\epsilon >0$, it was shown in the proof of Theorem \ref{thm:VSC} (see equation (\ref{eqn:m_I-P_e})) that
$ P(\sup_{\bgm: |\bgm|=p_{n}^{\star}}  |\bmu^{\prime} (\I - \Pgm) \bvep| >\epsilon n ) \rightarrow 0 $
as $p_{n}^{\star} \sim n^{s}$ and $s<1-b$.
Therefore,
$\bmu^{\prime} (\I - \Pgm) \bmu + 2 \bmu^{\prime} (\I - \Pgm) \bvep \geq c\|\bmu\|^{2}$ for some constant $c>0$.

Further, as in the previous case, $ \bvep^{\prime} (\I - \Pgm) \bvep \geq \sigma^{2}\alpha n$ for $\alpha\in (0,1)$ with probability tending to one, uniformly over $\bgm$ in the class of saturated models.

Combining all these, we state that for some $c>0$,

{\centering
$  \bfy^{\prime} (\I - \Pgm) \bfy \geq c \|\bmu\|^{2}. $
\par}

Therefore, from the bounds for numerator and the denominator we get
$$ \displaystyle\frac{P(M_{\bgm}|\bfy)}{ P(M_{\bgmp}|\bfy)} \leq \frac{1}{u_{n}}  \left[1+ \frac{\|\bmu\|^{2} \tau_{\min}^{\star ~-1} p_{n}^{\star~-1} +c \|\bmu\| \sqrt{\log p}  + \sigma^{2} (2+\epsilon) (\ps +1) \log p}{c \|\bmu\|^{2}+ \sigma^{2} \alpha n} \right]^{n/2}. $$
Next, we consider two cases separately: (i) $\|\bmu\|^{2} \gtrsim p_{n}^{\star 2} \log p$ and (ii) $\|\bmu\|^{2} \lesssim p_{n}^{\star 2} \log p $.

In case (i), there exists some $c>0$ such that

{\centering
$   \displaystyle\frac{P(M_{\bgm}|\bfy)}{ P(M_{\bgmp}|\bfy)} \leq \frac{1}{u_{n}}  \left( 1+ c p_{n}^{\star~-1} \right)^{n/2} \lesssim \frac{1}{u_{n}}\exp\{  c p_{n}^{\star~-1} n\}. $
\par}

\noindent As $ \exp\{{\ps}^{-1} n\}= o\left( k_{n} \sqrt{g_n}\right)$, the result follows.

In case (ii), for any $\epsilon>0$, we have

{\centering
$\displaystyle\frac{P(M_{\bgm}|\bfy)}{ P(M_{\bgmp}|\bfy)} \leq \frac{1}{u_{n}}  \left\{  1+ 2(1+\epsilon) \sigma^{2} (c\|\bmu\|^{2}+\sigma^{2}\alpha n)^{-1}  p_{n}^{\star} \log p  \right\}^{n/2}  \lesssim \frac{p^{\ps}}{u_{n}}. $
\par}

\noindent As $\epsilon>0$ is arbitrary and $k_{n}^{2} g_{n} \gtrsim p^{2(1+\lambda)\ps}$ for some $\lambda>0$, the result follows.

\paragraph{Case 3: Unsaturated and over-explored non-supermodels.}
If $\bgm$ is an over-explored and unsaturated non-supermodel, then among the active groups the best covariate $l$ is included, which is not presently included in the model. Thus, $\mG(\bgm)=\bgmp= \bgm\cup\{ l\}$ with $ l = \argmax_{j \in \bgmo\setminus \bgm} \|\bfP_{\bgm\cup \{j\} } \bmu \| $.

Let $\{l\} \in G^{\star}$ for the active group $G^{\star}$ and $\bar{G}^{\star}$ is the collection of all covariates not in $G^{\star}$. Then, we have
\vspace{-.05 in}
\begin{eqnarray*}
    \frac{P(M_{\bgm}|\bfy)}{ P(M_{\bgmp}|\bfy)} &=& \frac{(p-|\bgm|+1) k_{n} \sqrt{1+g_{n}}}{|\bgm|}\left\{ \frac{g_{n}^{-1} \bfy^{\prime} \bfy + \bfy^{\prime} (\I - \Pgmp) \bfy}{g_{n}^{-1} \bfy^{\prime} \bfy + \bfy^{\prime} (\I - \Pgm) \bfy} \right\}^{n/2} \\
    &=& u_{n} \left\{ 1 - \frac{ \bfy^{\prime} (\Pgmp - \Pgm) \bfy}{g_{n}^{-1} \bfy^{\prime} \bfy + \bfy^{\prime} (\I - \Pgm) \bfy} \right\}^{n/2}.
\end{eqnarray*}
Next, we provide a lower bound to $ \bfy^{\prime} (\Pgmp - \Pgm) \bfy$.
From \citet[Lemma 6]{YWJ_2016}, we have
\begin{small}
\vspace{-.1 in}
\begin{eqnarray*}
    \bmu^{\prime} (\Pgmp - \Pgm) \bmu &=& \bmu^{\prime} (\I - \Pgm) \bfx_{(l)} \left[ \bfx_{(l)}^{\prime} (\I- \Pgm) \bfx_{(l)} \right]^{-1} \bfx_{(l)}^{\prime}(\I- \Pgm) \bmu \\
    &\geq & \frac{1}{n}\bmu^{\prime} (\I - \Pgm) \bfx_{(l)}  \bfx_{(l)}^{\prime}(\I- \Pgm) \bmu,
\end{eqnarray*}
\end{small}
as $\|\bfx_{(l)}\|^{2}=n$.
Now, writing $\bmu = \bmu_{G^{\star}} + \bmu_{\bar{G^{\star}}}$ we get
\vspace{-.05 in}
\begin{eqnarray}
&&    \textstyle\sum_{j\in (\bgmo\setminus \bgm)\cap G^{\star}} \frac{1}{n}\bmu^{\prime} (\I - \Pgm) \bfx_{(j)}  \bfx_{(j)}^{\prime}(\I- \Pgm) \bmu \notag \\
&& \hspace{1.25 in} = \frac{1}{n}\bmu^{\prime} (\I - \Pgm) \bfX_{(\bgmo\setminus \bgm)\cap G^{\star}}\bfX_{(\bgmo\setminus \bgm)\cap G^{\star}}^{\prime}(\I- \Pgm) \bmu \notag \\
&& \hspace{1.25 in} = \frac{1}{n}\bmu_{G^{\star}}^{\prime} (\I - \Pgm) \bfX_{\bgmo\cap G^{\star}} \bfX_{\bgmo\cap G^{\star}}^{\prime} (\I- \Pgm)  \bmu_{G^{\star}} \notag \\
&& \hspace{1.25 in} \qquad + \frac{1}{n}\bmu_{\bar{G^{\star}}}^{\prime} (\I - \Pgm) \bfX_{\bgmo\cap G^{\star}} \bfX_{\bgmo\cap G^{\star}}^{\prime} (\I- \Pgm) \bmu_{\bar{G^{\star}}} \notag \\
&& \hspace{1.25 in} \qquad + \frac{2}{n} \bmu_{\bar{G^{\star}}}^{\prime} (\I - \Pgm) \bfX_{\bgmo\cap G^{\star}} \bfX_{\bgmo\cap G^{\star}}^{\prime}(\I- \Pgm) \bmu_{G^{\star}},~  \label{eqn:3}
\end{eqnarray}
as $(\I-\Pgm)\bfx_{(j)}={\bf 0}$ for any $j\in\bgm$.
Now, the first component in the above expression
\vspace{-.05 in}
\begin{eqnarray*}
&& \bmu_{G^{\star}}^{\prime} (\I - \Pgm) \bfX_{\bgmo\cap G^{\star}} \bfX_{\bgmo\cap G^{\star}}^{\prime} (\I- \Pgm)  \bmu_{G^{\star}} \\
&& \hspace{.2 in}\geq  \lambda_{\min}^{2} \left( \bfX_{(\bgmo\cap G^{\star})\setminus \bgm}^{\prime} (\I - \Pgm) \bfX_{(\bgmo\cap G^{\star})\setminus \bgm}  \right) \left\| \bbeta_{(\bgmo\cap G^{\star})\setminus \bgm}\right\|^{2}\\
&&  \hspace{.2 in} \geq  \beta_{\min}^{\star~2} n^{2} \tau_{\min}^{\star~2} |(\bgmo\cap G^{\star})\setminus \bgm|.
\end{eqnarray*}
To see the last inequality, let $\bfZ=\left[\bfX_{\bgm}: \bfX_{(\bgmo\cap G^{\star})\setminus \bgm} \right]$ and partition $\left(\bfZ^{\prime} \bfZ \right)^{-1}$ corresponding to $\bfX_{\bgm}$ and the rest. Then, the $(2,2)$-th block would be the inverse of $\bfX_{(\bgmo\cap G^{\star})\setminus \bgm}^{\prime} (\I - \Pgm) \bfX_{(\bgmo\cap G^{\star})\setminus \bgm}$. Now,
by Cauchy's interlacing theorem and assumption (C1) the inequality follows.

Next, consider the third component of (\ref{eqn:3}) as follows:
\vspace{-.05 in}
\begin{eqnarray*}
&&    \left| \bmu_{\bar{G^{\star}}}^{\prime} (\I - \Pgm) \bfX_{\bgmo\cap G^{\star}} \bfX_{\bgmo\cap G^{\star}}^{\prime}(\I- \Pgm) \bmu_{G^{\star}} \right|\\
    &&  \qquad \qquad \leq  \left\| \bfX_{\bgmo\cap G^{\star}}^{\prime}(\I- \Pgm) \bmu_{\bar{G^{\star}}} \right\|\left\| \bfX_{\bgmo\cap G^{\star}}^{\prime}(\I- \Pgm) \bmu_{G^{\star}} \right\| \\
    && \qquad \qquad \leq  \lambda_{\max}\left( \bfX^{\prime}_{(\bgmo\cap G^{\star})\setminus \bgm}\bfX_{(\bgmo\cap G^{\star})\setminus \bgm} \right) \|\bbeta_{(\bgmo\cap G^{\star})\setminus \bgm}\|
    \left\| \bfX_{\bgmo\cap G^{\star}}^{\prime}(\I- \Pgm) \bmu_{\bar{G^{\star}}} \right\|\\
    &&  \qquad \qquad  \lesssim \lambda_{\max}\left( \bfX^{\prime}_{\bgmo}\bfX_{\bgmo} \right)  \sqrt{|(\bgmo\cap G^{\star})\setminus \bgm | } \left\| \bfX_{\bgmo\cap G^{\star}}^{\prime}(\I- \Pgm) \bmu_{\bar{G^{\star}}} \right\| .
\end{eqnarray*}
Now,
 $   \big\| \bfX_{\bgmo\cap G^{\star}}^{\prime}(\I- \Pgm) \bmu_{\bar{G^{\star}}} \big\| =    \big\| \bfX_{(\bgmo\cap G^{\star})\setminus \bgm}^{\prime}(\I- \Pgm) \bmu_{\bar{G^{\star}}\setminus \bgm} \big\|
    \leq  2 \big\| \bfX_{(\bgmo\cap G^{\star})\setminus \bgm}^{\prime}\bmu_{\bar{G^{\star}}\setminus \bgm}  \big\|.$
Also, by assumption (B2\*), we have

{\centering
$\big\| \bfX_{(\bgmo\cap G^{\star})\setminus \bgm}^{\prime}\bmu_{\bar{G^{\star}}\setminus \bgm}  \big\| \leq \sigma_{\max} \big(\bfX_{(\bgmo\cap G^{\star})\setminus \bgm}^{\prime}\bfX_{(\bgmo\cap \bar{G^{\star}})\setminus \bgm} \big) \big\| \bbeta_{(\bgmo\cap \bar{G^{\star}})\setminus \bgm} \big\|  \lesssim  n\eta_{n} \sqrt{|(\bgmo\cap\bar{G^{\star}})\setminus \bgm |}.$
\par}

Combining the above and the fact that $\lambda_{\max} \left(n^{-1} \bfX_{\bgmo}^{\prime} \bfX_{\bgmo} \right) \leq \tau_{\max}^{\star}$, we get

{\centering
$\big| \bmu_{\bar{G^{\star}}}^{\prime} (\I - \Pgm) \bfX_{\bgmo\cap G^{\star}} \bfX_{\bgmo\cap G^{\star}}^{\prime}(\I- \Pgm) \bmu_{G^{\star}} \big| \lesssim  n^{2} \eta_{n} \sqrt{|(\bgmo\cap \bar{G^{\star}})\setminus \bgm | |(\bgmo\cap G^{\star})\setminus \bgm |}
.$
\par}

\noindent Finally, observe that the last term in (\ref{eqn:3}):
~$ \bmu_{\bar{G^{\star}}}^{\prime} (\I - \Pgm) \bfX_{\bgmo\cap G^{\star}} \bfX_{\bgmo\cap G^{\star}}^{\prime} (\I- \Pgm) \bmu_{\bar{G^{\star}}} \geq 0
$.

As $ |\bgmo|\eta_{n}^{2}\rightarrow 0$, we have

{\centering
 $\sum_{j\in (\bgmo\setminus \bgm)\cap G^{\star}} n^{-1}\bmu^{\prime} (\I - \Pgm) \bfx_{(j)}  \bfx_{(j)}^{\prime}(\I- \Pgm) \bmu \gtrsim  n  |(\bgmo\cap G^{\star})\setminus \bgm|.$
\par}

Now, as $\bgmp$ is so chosen that $\bgmp=\bgm\cup\{l\}$ with

{\centering
$ l = \argmax_{j \in \bgmo\setminus \bgm} \|\bfP_{\bgm\cup \{j\} } \bmu \|= \argmax_{j \in \bgmo\setminus \bgm} \|(\bfP_{\bgm\cup \{j\} } - \bfP_{\bgm}) \bmu \| ,$
\par}

\noindent we have

{\centering
$\bmu(\bfP_{\bgm\cup \{l\} } - \bfP_{\bgm}) \bmu \geq \displaystyle\frac{1}{|(\bgmo\cap G^{\star})\setminus \bgm|} \sum_{j\in (\bgmo\setminus \bgm)\cap G^{\star}} \bmu(\bfP_{\bgm\cup \{j\} } - \bfP_{\bgm})\bmu \gtrsim  n .$
\par}

The above lower bound does not depend on the particular choice of group $G^{\star}$ as long as the group is active.

Next, recall that $\bvep(\bfP_{\bgm\cup \{l\} } - \bfP_{\bgm}) \bvep \lesssim \ps \log p$, uniformly over all $\bgm$ as $|\bgm|<\ps$, with probability tending to one.
By an application of the Cauchy-Schwarz inequality, we get

{\centering
$
    \bmu(\bfP_{\bgm\cup \{j\} } - \bfP_{\bgm})\bvep \lesssim \sqrt{n|\bgmo| \ps \log p}, $
\par}

\noindent with probability tending to one.
As $|\bgmo|p_{n}^{\star} \log p/n \rightarrow 0$, for any $\epsilon>0$, we have
\vspace{-.05 in}
\begin{equation}
    \bfy^{\prime} (\Pgmp - \Pgm) \bfy \geq c_{0}^{2} (1-\epsilon) n \tau_{\min}^{\star~2},  \label{eqn:num_bound}
\end{equation}
with probability tending to one.


Next, consider the denominator $g_{n}^{-1} \bfy^{\prime} \bfy + \bfy^{\prime} (\I - \Pgm) \bfy$. Recall that $g_{n}^{-1} \bfy^{\prime} \bfy <\epsilon$ for any $\epsilon>0$, with probability at least $1-\exp\{-cn\}$.

Also, $ \bfy^{\prime} (\I - \Pgm) \bfy \leq \|\bmu\|^{2} + 2 \|\bmu\|\|\bvep\|+ \| \bvep\|^{2} \lesssim n|\bgmo|$.
Combining this and equation (\ref{eqn:num_bound}), we get
    \vspace{-.05 in}
\begin{eqnarray*}
    \frac{P(M_{\bgm}|\bfy)}{ P(M_{\bgmp}|\bfy)}  \leq u_{n} \left( 1- \frac{c}{|\bgmo|}\right)^{n/2} \lesssim \exp\{-cn^{1-b}\}
\end{eqnarray*}
by the choices of $k_{n}$ and $g_{n}$ provided in (i$^{\prime}$) and (ii$^{\prime}$) of Theorem \ref{thm:Mixing_rate}.


\paragraph{\bf Case 4: Unsaturated and under-explored on-supermodels.}
If $\bgm$ is an under-explored and unsaturated non-supermodel, then among the active groups with no representative in $\bgm$, the best  active covariate $l$ in the best group $k^{\star}$ is included. Thus, $\mG(\bgm)=\bgmp= \bgm\cup\{ l\}$ where
$ l = \argmax_{j \in G_{k^{\star}}\cap \bgmo  } \|\bfP_{\bgm\cup \{j\} } \bmu \|$ and $k^{\star} = \argmax_{k} \| \bfP(\bfX_{k})\bmu \|   $.

The proof of this case is similar to that of case 3. The only difference is that in this case $(\bgmo\cap G^{\star})\setminus \bgm=\bgmo\cap G^{\star}$, as the group $G^{\star}$ is not yet explored. However, in can be easily verified that this little deviation does not make any difference in the upper bound of $P(M_{\bgm}|\bfy)/ P(M_{\bgmp}|\bfy)$.
\end{proof}

\subsubsection{Proof of Lemma \ref{res:2}}\label{sec:Proof_Mixing2}
\begin{proof}
{\bf Case I: Non-supermodels.} Let $\bgm$ be a non-supermodel, all it's precedents are non-supermodels. Let $\bgms$ be a precedent of $\bgm$.

Any influential covariate, once included, is never discarded. So, the number of saturated models in $\bgms\to\bgm$ can at most be $(|\bgm\cap\bgmo|-|\bgms\cap \bgmo|+1) \leq |\bgmo|$. Further, there must be at least one unsaturated model in between two consecutive saturated models, as an `R' move is taken at a saturated model. Combining all these facts, if the length of the path from $\bgms\to \bgm$ is $k$, and $u$ is the number of intermediate saturated models, then $u\leq \lceil(k+1)/2\rceil \wedge |\bgmo| $.

Further, to see the number of possible models $\bgms$ such that $|\mT_{\bgms, \bgm}|=k$, note that
if $\bgmp$ is saturated, then there are at most $p-|\bgmo| ~(\leq p)$ possible components leading to $\mG(\bgmp)$. On the contrary, if $\bgmp$ is unsaturated, then there are $ |\bgmo|$ possible components leading to $\mG(\bgmp)$.
Therefore,
    \vspace{-.05 in}
\begin{eqnarray*}
\sup_{\bgm} \sum_{\bgms: \bgms\in \Lambda(\bgm), |\bgms \setminus \bgm|=k}
&& \hspace{-.1 in}\frac{P(M_{\bgms}|\bfy)}{ P(M_{\bgm}|\bfy)} \\
&\leq &  \sum_{u=0}^{\lceil(k+1)/2\rceil \wedge |\bgmo|} \left( \frac{p}{pw_{n}} \right)^{u} \left( |\bgmo| \exp\{-c n^{1-b} \} \right)^{k-u} \\
&\leq & \left( \frac{|\bgmo|^{-1}}{w_{n}} \exp\{c n^{1-b} \} \right)^{\lceil(k+1)/2\rceil \wedge |\bgmo|} \hspace{-.1 in} \times \left( |\bgmo| \exp\{-c n^{1-b} \} \right)^{k}
\\
& \leq & \left( \frac{1}{w_{n}}  \right)^{\lceil(k+1)/2\rceil \wedge |\bgmo|} \hspace{-.1 in}\left( |\bgmo| \exp\{-c n^{1-b} \} \right)^{\lfloor(k-1)/2\rfloor \vee (k-|\bgmo|) }
\end{eqnarray*}
Now, as $k$ can vary from $1, \ldots, p_{n}^{\star}$, we have the following
    \vspace{-.05 in}
    \begin{eqnarray*}
&& \frac{1}{P(M_{\bgm}|\bfy)}\sum_{\bar{\bgm} \in \Lambda (\bgm) } P(M_{\bar{\bgm}}|\bfy) \\
&& \qquad \leq \sum_{k=1}^{p_{n}^{\star}} \left( \frac{1}{w_{n}}  \right)^{\lceil(k+1)/2\rceil \wedge |\bgmo|} \left( |\bgmo| \exp\{-c n^{1-b} \} \right)^{\lfloor(k-1)/2\rfloor \vee (k-|\bgmo|) } \\
&& \qquad = \sum_{k=1}^{2|\bgmo|} \left( \frac{1}{w_{n}}  \right)^{\lceil(k+1)/2\rceil} \left( \frac{|\bgmo|}{ \exp\{c n^{1-b} \}} \right)^{\lfloor(k-1)/2\rfloor  } \\
&& \hspace{2 in}+\left( \frac{1}{w_{n}}  \right)^{ |\bgmo|}\sum_{k=2|\bgmo|+1}^{p_{n}^{\star}}  \left( \frac{|\bgmo|}{ \exp\{c n^{1-b} \}} \right)^{(k-|\bgmo|) } \\ && \qquad \leq 2 \mbox{ as } n\rightarrow\infty.
\end{eqnarray*}
\vspace{-.2 in}

\paragraph{Case II: Supermodels.}
When $\bgm$ is a supermodel, we write $ \Lambda(\bgm)= \Lambda_{1} (\bgm) + \Lambda_{2}(\bgm)$, where $\Lambda_{1} (\bgm)  = \{ \bgms : \bgm \in \mT_{\bgms,\bgmo} ; ~\bgmo\subseteq \bgms\}$ and $\Lambda_{2} (\bgm)  = \{ \bgms : \bgm \in \mT_{\bgms,\bgmo} ; ~\bgmo\nsubseteq \bgms\}$. Let $\bgms\in \Lambda_{1} (\bgm)$ and $ |\bgms \setminus \bgm|=k$, then as per the canonical path construction
$\bgm=\mG(\cdots \mG(\bgms))=\mG^{k}(\bgms)$, and all the intermediate models are supermodels of $\bgmo$. Thus, using Lemma \ref{res:1}, we get
    \vspace{-.05 in}
\begin{eqnarray*}
    \sup_{\bgm} \sum_{\bgms: \bgms\in \Lambda_{1}(\bgm), |\bgms \setminus \bgm|=k}
\frac{P(M_{\bgms}|\bfy)}{ P(M_{\bgm}|\bfy)}  \leq (p-|\bgm|)_{k} \left(\frac{1}{pw_{n}} \right)^{k} \leq w_{n}^{-k}.
\end{eqnarray*}
Therefore, as $w_{n}\rightarrow\infty$
    \vspace{-.05 in}
\begin{eqnarray*}
    \frac{1}{P(M_{\bgm}|\bfy)}\sum_{\bar{\bgm} \in \Lambda_{1}(\bgm) } P(M_{\bar{\bgm}} |\bfy) \leq \sum_{k=0}^{p-|\bgm|} w_{n}^{-k} \rightarrow 1.
\end{eqnarray*}
Next, let $\bgms\in \Lambda_{2}(\bgm)$. Then the chain $\bgms\to \bgm$ can be split into two parts $\bgms \to \bgmp$ and $\bgmp\to \bgm$, where $\bgmp$ is the first state in the path of $\bgms\to \bgm$, when the chain enters the class of supermodels. As per the canonical path ensemble once the chain enters the class of supermodels, it remains their. Let the length of the path $\bgms\to \bgm$ be $k$, then the length of $\bgmp\to\bgm $ can be any number $u$ between $0,1,\ldots,k-1$. Therefore, by Case I and the previous part of Case II, we have
    \vspace{-.05 in}
\begin{eqnarray*}
&& \sup_{\bgm} \sum_{\bgms: \bgms\in \Lambda_{2}(\bgm), |\bgms \setminus \bgm|=k}
\frac{P(M_{\bgms}|\bfy)}{ P(M_{\bgm}|\bfy)} \\
&& \quad =  \sum_{u=0}^{k-1} \left( \frac{1}{w_{n}}  \right)^{\lceil(k-u+1)/2\rceil \wedge |\bgmo|} \left( |\bgmo| \exp\{-c n^{1-b} \} \right)^{\lfloor(k-u-1)/2\rfloor \vee (k-u-|\bgmo|) }
 \left( \frac{1}{w_{n}}\right)^{u}\\
&& \quad =  \sum_{u=0}^{k-1} \left( \frac{1}{w_{n}}  \right)^{\lceil(k+u+1)/2\rceil \wedge (|\bgmo|+u)} \left( |\bgmo| \exp\{-c n^{1-b} \} \right)^{\lfloor(k-u-1)/2\rfloor \vee (k-u-|\bgmo|) }\\
&& \quad \leq  k \left( \frac{1}{w_{n}}  \right)^{\lceil(k+1)/2\rceil \wedge |\bgmo|  }.
\end{eqnarray*}
Thus,
    \vspace{-.05 in}
\begin{eqnarray*}
     \frac{1}{P(M_{\bgm}|\bfy) } \sum_{\bar{\bgm} \in \Lambda_{2}(\bgm) } P(M_{\bar{\bgm}}|\bfy) &\leq & \sum_{k=1}^{p_{n}^{\star}} k w_{n}^{-\{\lceil(k+1)/2\rceil \wedge |\bgmo|\} } \\
     &\leq & \sum_{k=1}^{2|\bgmo|} k w_{n}^{-\lceil(k+1)/2\rceil  } + w_{n}^{-|\bgmo| } \sum_{k=2|\bgmo|+1}^{p_{n}^{\star}} k \\
     &\leq & 2\left(1-\frac{1}{\sqrt{w_n}} \right)^{-2} + p_{n}^{\star 2} w_{n}^{-|\bgmo| }  \leq 5 \mbox{ as } n\rightarrow\infty.
     \end{eqnarray*}
\end{proof}

\vspace{-.15 in}
\subsubsection{Proof of Lemma \ref{res:3}}\label{sec:Proof_Mixing3}
\begin{proof}
{\bf Case I:} Let $\bgm$ be a supermodel, or a saturated non-supermodel. In this case, in the canonical path ensemble, an `R' move is taken. Recall from (\ref{eqn:P_MH}) and (\ref{eqn:P_MH2}) that in case of an `R' move, we have
$$P_{\mathrm{MH}}(\bgm \to \bgmp) =\frac{1}{3} \min \displaystyle\left\{ \frac{1}{|\bgm|} , \frac{  P(M_{\bgmp}| \bfy) }{P(M_{\bgm}| \bfy) } \times \sum_{k=1}^{\kn} \frac{q_{k} I(j \in G_{k}) }{p_{k}-|G_{k} \cap \bgmp|} \right\} . $$
As $\inf_{k\in \mI} q_{k} \gtrsim q_{\mI}^{\star}$ and maximum group size is $p_g^{\star} \lesssim n^{t^{\star}}$ with $t^{\star}<1$ (by Lemma \ref{res:1}), we have $P_{\mathrm{MH}}(\bgm \to \bgmp) \gtrsim \min\left\{ p_{n}^{\star-1}, pw_{n} q_{\mI}^{\star}/p_g^{\star} \right\} $.

{\bf Case II:} Let $\bgm$ be a unsaturated non-supermodel. Then, an `A' move is adapted. Recall from (\ref{eqn:P_MH}) and (\ref{eqn:P_MH2}) that in case of an `A' move, we have
$$P_{\mathrm{MH}}(\bgm \to \bgmp) =\frac{1}{3} \displaystyle\min\left\{\sum_{k=1}^{\kn} \frac{q_{k} I(l \in G_{k}) }{p_{k}-|G_{k} \cap \bgm|} , \frac{ (|\bgm|+1) P(M_{\bgmp}| \bfy) }{P(M_{\bgm}| \bfy) }  \right\} . $$
As $\inf_{k\in \mA} q_{k} \gtrsim q_{\mA}^{\star}$, by Lemma \ref{res:1}, we have $P_{\mathrm{MH}}(\bgm \to \bgmp) \gtrsim \min\left\{ q_{\mA}^{\star} \pg^{\star-1}, e^{cn^{1-b}} \right\}=q_{\mA}^{\star} \pg^{\star-1}$.
\end{proof}


\bibliographystyle{apalike}
\bibliography{ref}


\end{document}